\documentclass[12pt,a4paper]{iopart}
\pdfoutput=1
\usepackage[utf8]{inputenc}
\usepackage{ulem}

\makeatletter
\providecommand*\@nameundef[1]{\expandafter\let\csname #1\endcsname\@undefined}
\@nameundef{equation*}
\@nameundef{endequation*}
\makeatother
\usepackage{amsmath}
\usepackage{amssymb}
\usepackage{amsfonts}
\usepackage{mathtools}
\usepackage{lmodern}
\usepackage[T1]{fontenc}
\usepackage{bbm}
\DeclareMathAlphabet{\mathbfi}{OML}{cmm}{b}{it}

\usepackage{upgreek}

\usepackage[scr=boondoxo,scrscaled=1.05]{mathalfa}

\usepackage[unicode=true,bookmarksopen=true,colorlinks=true,urlcolor=blue,anchorcolor=blue,citecolor=blue,filecolor=blue,linkcolor=blue,menucolor=blue,
linktocpage=true,pdfa=true]{hyperref}

\let\originalleft\left
\let\originalright\right
\renewcommand{\left}{\mathopen{}\mathclose\bgroup\originalleft}
\renewcommand{\right}{\aftergroup\egroup\originalright}

\makeatletter
\newenvironment{equations}[1][]{\subequations\ifx\relax#1\relax\else\label{#1}\fi\align\ignorespaces}{\endalign\ignorespacesafterend\endsubequations}
\def\@spliteq#1{\begin{equation}\begin{split}#1\end{split}\end{equation}}
\def\splitequation{\collect@body\@spliteq}

\makeatother

\renewcommand{\vec}[1]{{\ifnum9<1#1\mathbf{#1}\else\ifcat\noexpand#1\relax\boldsymbol{#1}\else\mathbfi{#1}\fi\fi}}
\newcommand{\mathe}{\mathrm{e}}
\newcommand{\mathi}{\mathrm{i}}
\newcommand{\total}{\mathop{}\!\mathrm{d}}
\newcommand{\laplace}{\mathop{}\!\bigtriangleup}
\newcommand{\abs}[1]{{\left\lvert{#1}\right\rvert}}
\newcommand{\eqend}[1]{\,#1}
\newcommand{\bigo}[1]{\mathcal{O}({#1})}
\newcommand{\bra}[1]{\left\langle{#1}\right\rvert}
\newcommand{\ket}[1]{\left\lvert{#1}\right\rangle}
\newcommand{\expect}[1]{\left\langle{#1}\right\rangle}
\newcommand{\bnabla}{\boldsymbol{\nabla}}
\DeclareMathOperator{\lie}{\mathcal{L}}

\usepackage[numbers,sort&compress]{natbib}
\allowdisplaybreaks
\usepackage{xspace}
\newcommand{\ie}{\textit{i.\,e.}\xspace}
\newcommand{\eg}{\textit{e.\,g.}\xspace}

\begin{document}

\title{Synchronous coordinates and gauge-invariant observables in cosmological spacetimes}

\author{Markus B.\ Fr\"ob and William C.\ C.\ Lima\footnote{Author to whom any correspondence should be addressed.}}
\address{Institut f\"ur Theoretische Physik, Universit\"at Leipzig,\\ Br\"uderstra\ss e 16, 04103 Leipzig, Germany}
\ead{\href{mailto:mfroeb@itp.uni-leipzig.de}{mfroeb@itp.uni-leipzig.de}, \href{mailto:williamcclima@gmail.com}{williamcclima@gmail.com}}

\begin{abstract}
We consider the relational approach to construct gauge-invariant observables in cosmological perturbation theory using synchronous coordinates. We construct dynamical synchronous coordinates as non-local scalar functionals of the metric perturbation in the fully non-linear theory in an arbitrary gauge. We show that the observables defined in this dynamical coordinate system are gauge-independent, and that the full perturbed metric has the expected form in these coordinates. Our construction generalises the familiar synchronous gauge in linearised gravity, widely used in cosmological perturbation theory, to the non-linear theory. We also work out the expressions for the gauge-invariant Einstein equation, sourced either by an ideal fluid or a scalar field up to second order in perturbation theory, and give explicit expressions for the Hubble rate --- as measured by synchronous observers or by observers co-moving with the matter field --- up to that order. Finally, we consider quantised linear perturbations around Minkowski and de~Sitter backgrounds, and compute the two-point function of the gauge-invariant metric perturbation in synchronous coordinates, starting with two-point function in a general linear covariant gauge. Although the gauge-fixed two-point function contains gauge modes, we show that the resulting gauge-invariant two-point function only contains the physical tensor modes and it is thus positive, \ie, it has a spectral representation.

\end{abstract}

\maketitle

%%%%%%%%%%%%%%%%%%%%%%%%%%%%%%%%%%%%%%%%%%%%%%%%%%%%%%%%%%%%%%%%%%%%%%%%%%%%%%%%%%%%%%%%%%%%%%%%%%%%%%%%%%%%%%%%%%%%%%%%%%%%%%%%%%%%%%%%%%%%%%%%%%%%%
\section{Introduction}                                                                                                                              %
\label{sec:introduction}                                                                                                                            %
%%%%%%%%%%%%%%%%%%%%%%%%%%%%%%%%%%%%%%%%%%%%%%%%%%%%%%%%%%%%%%%%%%%%%%%%%%%%%%%%%%%%%%%%%%%%%%%%%%%%%%%%%%%%%%%%%%%%%%%%%%%%%%%%%%%%%%%%%%%%%%%%%%%%%

A foundational problem in any theory of gravity is the construction of observables corresponding as closely as possible to experimental measurements. Since diffeomorphisms are symmetries of gravity, any local field defined at a fixed point of the background manifold is not invariant, and can therefore not correspond to an observable. It is of course possible to construct non-local observables by integrating scalar densities over the manifold, which are by construction invariant under diffeomorphisms; such observables comprise in particular topological invariants such as the Euler--Poincaré and Pontryagin characteristics. Other observables which also contain information about correlations can be obtained by considering quantities held at a fixed geodesic distance and then integrating over the manifold, or the average distance of spheres of a fixed size, which are used in causal dynamical triangulation approaches to quantum gravity~\cite{klitgaard_loll_prd_2018a,klitgaard_loll_prd_2018b,loll_silva_prd_2023}. However, all of these only give information about quantities averaged over the full manifold, while experiments also provide us with data measured in compact regions.

To obtain localised quantities that are invariant under diffeomorphisms, one has to use the relational formalism (see, \eg, Refs.~\cite{komar_pr_1958,bergmann_pr_1961,bergmann_rmp_1961,rovelli_cqg_1991,montesinos_grg_2001,rovelli_prd_2002a} or~\cite{tambornino_sigma_2012,goeller_hoehn_kirklin_arxiv_2022} for reviews), where observables are defined by the state of fields with respect to a dynamical coordinate system, which itself is constructed from (a subset of) the fields of the theory, the clocks or reference fields. Relational observables have a long history, and have been used in canonical and Loop Quantum Gravity (see, \eg, Refs.~\cite{dittrich_cqg_2006,dittrich_grg_2007,dittrich_tambornino_cqg_2007,giesel_ijmpa_2008,giesel_et_al_cqg_2010a,giesel_et_al_cqg_2010b,giesel_thiemann_cqg_2015,giesel_et_al_cqg_2018}), Group Field Theory~\cite{gielen_2018} and the Asymptotic Safety Programme~\cite{baldazzi_falls_ferrero_2022}.\footnote{In the context of quantum gravity as an effective quantum field theory, Ware, Saotome and Akhoury~\cite{waresaotomeakhoury2013} and Donnelly, Giddings and Perkins~\cite{donnelly_giddings_prd_2016a,donnelly_giddings_prd_2016b,giddings_perkins_arxiv_2022} have proposed to extended the concept of dressed observables in gauge theories~\cite{dirac_cjp_1955,kibble1968,kulishfaddeev1970,steinmann1984,baganlavellemcmullan2000a,baganlavellemcmullan2000b,mitraratabolesharatchandra2006} to perturbative quantum gravity. Their observables can also be reformulated in the relational approach, see, \eg, Ref.~\cite{goeller_hoehn_kirklin_arxiv_2022}.}

The question remains how one has to choose the dynamical coordinate system. In the earliest works, it was proposed to use curvature scalars, which clearly only works if the spacetime is sufficiently inhomogeneous such that one can discriminate points by the values of these scalars. This is a problem in perturbative quantum gravity on highly symmetric backgrounds, since points related by a symmetry transformation cannot be distinguished by curvature scalars, which take the same value on the whole orbit of any such transformation. A way out is to explicitly add material reference systems such as dust~\cite{brown_kuchar_prd_1995,brown_marolf_prd_1996,bicak_kuchar_prd_1997}, but these change the dynamics of the theory~\cite{giesel_et_al_prd_2020,giesel_li_singh_prd_2021}. Another possibility is to construct the required dynamical coordinate scalars from the gauge-dependent parts of the metric, which has the advantage that it works for an arbitrary spacetime. Without explicitly using the relational formalism, this is the way that suitable observables at linear order have been constructed in cosmology long time ago, such as the Bardeen variables~\cite{bardeen_prd_1980} and the Sasaki--Mukhanov variable~\cite{sasaki_ptp_1986,mukhanov_zetf_1988}. However, the extension to higher orders is difficult and often not systematic; see, \eg, Refs.~\cite{bruni_matarrese_mollerach_sonego_cqg_1997,matarrese_mollerach_bruni_prd_1998,nakamura_ptp_2003,malik_jcap_2005,nakamura_ptp_2007,finelli_marozzi_vacca_venturi_prd_2006,pons_salisbury_sundermeyer_prd_2009,pons_salisbury_sundermeyer_mpla_2009,gasperini_marozzi_veneziano_jcap_2009} for second- and higher-order constructions.

A practical and systematic way to construct dynamical coordinate systems for highly symmetric spacetimes was determined recently in Ref.~\cite{brunetti_etal_jhep_2016}. Their idea was to choose the dynamical coordinates as solutions of some set of scalar (differential) equations which are identically fulfilled on the background, which then can be explicitly solved order by order in perturbation theory around the highly symmetric background. The equations defining the dynamical coordinates must be chosen such as to reflect the experimental situation that should be modelled. It was shown in Ref.~\cite{brunetti_etal_jhep_2016} that for a cosmological background one can recover the Bardeen variables by making a suitable choice of time coordinate (depending on the perturbed inflaton and the scalar curvature of the constant-inflaton hypersurfaces), and using the Laplacian of the constant-inflaton hypersurfaces to define the spatial coordinates. This method has been further extended by us to other backgrounds~\cite{froeb_cqg_2018,froeb_cqg_2019,froeb_lima_cqg_2018}, and we have shown how to impose causality in this framework, such that the invariant observables only depend on the dynamics of the coordinates in their past light cone. However, the so-defined dynamical coordinates (which we call ``generalised harmonic coordinates'') do not have a straightforward interpretation, unlike, \eg, the geodesic light cone (GLC) coordinates~\cite{gasperini_et_al_jcap_2011,fanizza_etal_jcap_2013,fanizza_etal_jcap_2015,nugier_mg_2015,fleury_nugier_fanizza_jcap_2016,fanizza_etal_jcap_2021,mitsou_etal_cqg_2021,froeb_lima_jcap_2022,fanizza_marozzi_medeiros_jcap_2023}. In GLC coordinates, the dynamical coordinate scalars are constructed from the proper time of a geodesic observer, a null coordinate labelling the observer's past light cone and angles that determine the incoming directions of photons. GLC coordinates have found widespread use in theoretical cosmology (see, \eg, Refs.~\cite{bonvin_durrer_prd_2011,bertacca_maartens_clarkson_jcap_2014,didio_durrer_marozzi_montanari_jcap_2014,biern_yoo_jcap_2017,fanizza_yoo_biern_jcap_2018,fanizza_gasperini_marozzi_veneziano_jcap_2020,magi_yoo_jcap_2022}), even if their use is in most cases only implicit and not mentioned explicitly. They are very well adapted to compute important observables such as galaxy number counts and clustering or luminosity distances to high orders in classical perturbation theory, and clearly facilitate a direct comparison with observations. Unfortunately, they do not manifestly display the translational symmetry of constant-time hypersurfaces due to the use of null coordinates, which makes the quantisation awkward and complicated to the extent that not even the tree-level two-point function of the metric perturbation is known. For the same reason, numerical simulations almost always use synchronous gauge~\cite{cmbfast_zaldarriaga_seljak_bertschinger_apj_1998, camb_lewis_challinor_lasenby_jcap_2000, cmbeasy_doran_jcap_2005, class_lesgourgues_arxiv_2011, foidl_rindler-daller_prd_2022}. We remark that the difference between the results obtained using GLC and synchronous coordinates should be small in non-relativistic situations, where all velocities are small compared to the speed of light $c$. Although this calculation has not yet been performed, we believe it would be interesting to show in detail that the expansion of GLC coordinates for large $c$ and short distances (such that the time that light needs to travel is short in comparison to the characteristic time scales of local dynamics) would give back the synchronous coordinates.

To remedy the situation, in this article we construct the dynamical synchronous coordinate system and the corresponding gauge-invariant relational observables in perturbation theory. These coordinates are defined by the proper time of a congruence of free-falling observers, and spatial coordinates that are orthogonal to the observers' four-velocity. At linear order, the relational observables constructed in this dynamical coordinate system agree with fields in the well-known synchronous gauge~\cite{lifshitz_grg_2017}. Our results thus show how to extend this to arbitrary orders in perturbation theory. The paper is organised as follows: After a more detailed introduction to relational observables in Sec.~\ref{sec:relational_observables}, we explicitly construct the dynamical coordinates and the relationally defined gauge-invariant metric to second order in Sec.~\ref{sec:synchronous_coord}. We then determine the perturbed Einstein equation for both an ideal fluid and an inflaton field as source in Sec.~\ref{sec:perturbed_einstein_eq}, and as a further example of relational gauge-invariant observables we construct the invariant Hubble rate (the local expansion rate of the universe) in Sec.~\ref{sec:hubble_rate}. Lastly, in Sec.~\ref{sec:linearised_qg} we quantise the metric perturbations around Minkowski and de Sitter backgrounds, and show explicitly that the correlator of the gauge-invariant relational metric perturbation in synchronous coordinates is independent of the gauge fixing. It turns out that for both backgrounds, the invariant correlator only contains the propagating tensor modes, and thus captures exactly the dynamical content of the theory. We conclude in Sec.~\ref{sec:conclusions}, and leave a detailed comparison between the perturbed Einstein equation for the ideal fluid and the inflaton field to~\ref{sec:comparison_fluid_field}.

\paragraph{Conventions:} We use the ``$+++$'' convention of~\cite{misner_thorne_wheeler_book} for the metric and curvature tensors, work in a $n$-dimensional spacetime and set $c = \hbar = 1$ and $\kappa^2 \equiv 16 \pi G_\mathrm{N}$.

%%%%%%%%%%%%%%%%%%%%%%%%%%%%%%%%%%%%%%%%%%%%%%%%%%%%%%%%%%%%%%%%%%%%%%%%%%%%%%%%%%%%%%%%%%%%%%%%%%%%%%%%%%%%%%%%%%%%%%%%%%%%%%%%%%%%%%%%%%%%%%%%%%%%%
\section{Relational observables}                                                                                                                    %
\label{sec:relational_observables}                                                                                                                  %
%%%%%%%%%%%%%%%%%%%%%%%%%%%%%%%%%%%%%%%%%%%%%%%%%%%%%%%%%%%%%%%%%%%%%%%%%%%%%%%%%%%%%%%%%%%%%%%%%%%%%%%%%%%%%%%%%%%%%%%%%%%%%%%%%%%%%%%%%%%%%%%%%%%%%

As mentioned in the introduction, the idea of relational observables in gravity (or in any theory invariant under diffeomorphisms) assumes that one can fix points by the values that dynamical fields assume at those points. Diffeomorphism-invariant quantities --- observables --- can then be obtained by measuring any tensor field with respect to these dynamical fields.

To make these ideas more concrete, let us consider a gravitational system formed by a spacetime with metric $g_{\mu\nu}$ (covered by coordinates $x^\mu$) and matter fields that we will collectively denote by $\psi$. Let us assume that we can find $n$ scalar fields $X^{(\mu)}[g,\psi]$ that are functionals of the spacetime metric and matter fields.\footnote{We keep the index $\mu$ in $X^{(\mu)}$ within parentheses to stress that these are a collection of scalar fields.} We further assume that the map $\mathcal{X}\colon x^\mu \to X^{(\mu)}[g(x),\psi(x)]$ is invertible, and thus can view the dynamical fields $X^{(\mu)}[g,\phi]$ as a field-dependent reference frame or dynamical coordinate system. We can then measure any tensor field with components $T^{\mu_1 \cdots \mu_m}_{\nu_1 \cdots \nu_p}(x)$ in this dynamical coordinate system via the pullback resp.\ pushforward induced by this map:
\begin{equation}
\label{eq:invariant_tensor}
\mathscr{T}^{\mu_1 \cdots \mu_m}_{\nu_1 \cdots \nu_p}(X) \equiv \frac{\partial X^{(\mu_1)}}{\partial x^{\alpha_1}} \cdots \frac{\partial X^{(\mu_m)}}{\partial x^{\alpha_m}} \frac{\partial x^{\beta_1}}{\partial X^{(\nu_1)}} \cdots \frac{\partial x^{\beta_p}}{\partial X^{(\nu_p)}} T^{\alpha_1 \cdots \alpha_m}_{\beta_1 \cdots \beta_p}[x(X)] \eqend{,}
\end{equation}
where we write $x(X) = \mathcal{X}^{-1}(X^{\mu})$ with the inverse of the map $\mathcal{X}$.

To convince ourselves that Eq.~\eqref{eq:invariant_tensor} produces fields that are invariant under diffeomorphisms, let us consider the simpler case of a scalar field $S(x)$. In that case, Eq.~\eqref{eq:invariant_tensor} reduces to
\begin{equation}
\label{eq:invariant_scalar}
\mathscr{S}(X) = S[x(X)] = (S \circ \mathcal{X}^{-1})(X) \eqend{.}
\end{equation}
Now, consider an arbitrary diffeomorphism $f \colon x^\mu \to x'^\mu = f^\mu(x)$ of compact support, \ie, such that $f^\mu(x) \neq x^\mu$ only in a compact region. We will call such diffeomorphisms localised. Since $S$ and the $X^{(\mu)}$ are scalar fields, they transform under $f$ as
\begin{equation}
S \to S \circ f^{-1} \eqend{,} \quad X \to X \circ f^{-1} \ \Rightarrow \ \mathcal{X} \to \mathcal{X} \circ f^{-1} \eqend{,}
\end{equation}
which means that
\begin{equations}
\delta X^{(\mu)}(x) &= X'^{(\mu)}(x) - X^{(\mu)}(x) = X^{(\mu)}[f^{-1}(x)] - X^{(\mu)}(x) \eqend{,} \\
\delta S(x) &= S'(x) - S(x) = S[f^{-1}(x)] - S(x) \eqend{,}
\end{equations}
\ie, the scalar field $S$ and the field-dependent frame $X^{(\mu)}$ both change under diffeomorphisms when evaluated at a point of the background manifold. However, for the quantity $\mathscr{S}$~\eqref{eq:invariant_scalar} we obtain
\begin{equation}
\mathscr{S} \to S \circ f^{-1} \circ ( \mathcal{X} \circ f^{-1} )^{-1} = S \circ \mathcal{X}^{-1} = \mathscr{S} \eqend{,}
\end{equation}
which shows that it is defined in such a way that the changes of the field-dependent frame and the scalar field compensate each other, leaving the final expression invariant.

For many situations, such as the ones found in cosmology, it is a good approximation to assume that the spacetime metric and the matter fields can be written as a fixed background plus perturbations. Hence, let us write the full metric $\tilde{g}_{\mu\nu}$ and matter fields $\tilde{\psi}$ as
\begin{equation}
\label{eq:perturbed_fields}
\tilde{g}_{\mu\nu} = g_{\mu\nu} + \kappa h_{\mu\nu} \quad\text{and}\quad \tilde{\psi} = \psi + \kappa \psi^{(1)} \eqend{.}
\end{equation}
In this case, it is reasonable to assume that we can write the field-dependent frame as a power series in the perturbations. Furthermore, we assume that the coordinates $X^{(\mu)}$ are chosen in such a way that they agree with the background coordinates $x^\mu$ at zeroth order. Hence, we write
\begin{equation}
\label{eq:X_power_series}
X^{(\mu)}(x) = x^\mu + \kappa X^{(\mu)}_{(1)}(x) + \kappa^2 X^{(\mu)}_{(2)}(x) + \bigo{\kappa^3} \eqend{.}
\end{equation} 
The power series for the inverse of $X^{(\mu)}$ can be obtained from Eq.~\eqref{eq:X_power_series} by first isolating the zero-order terms, substituting it into the argument of the power-series coefficients and then expanding the result up to the desired order. The result up to second order reads
\begin{equation}
\label{eq:X_inverse_power_series}
x^\mu(X) = X^{(\mu)} - \kappa X^{(\mu)}_{(1)} - \kappa^2 \left[ X^{(\mu)}_{(2)}(X) - X^{(\nu)}_{(1)}(X) \partial_\nu X^{(\mu)}_{(1)}(X) \right] + \bigo{\kappa^3} \eqend{.}
\end{equation}
As an illustration of the perturbative expansion for observables defined by Eq.~\eqref{eq:invariant_tensor}, let us consider the example of the full metric. We can produce an invariant metric tensor $\mathscr{g}_{\mu\nu}$ by defining
\begin{equation}
\label{eq:invariant_metric}
\mathscr{g}_{\mu\nu}(X) \equiv \frac{\partial x^\alpha(X)}{\partial X^{(\mu)}} \frac{\partial x^\beta(X)}{\partial X^{(\nu)}} \tilde{g}_{\alpha\beta}[x(X)] \eqend{.}
\end{equation}
Substituting Eqs.~\eqref{eq:perturbed_fields} and~\eqref{eq:X_inverse_power_series} into Eq.~\eqref{eq:invariant_metric}, we obtain the following perturbative expansion of the invariant metric up to second order:
\begin{equation}
\mathscr{g}_{\mu\nu} = g_{\mu\nu} + \kappa \mathscr{g}_{\mu\nu}^{(1)} + \kappa^2 \mathscr{g}_{\mu\nu}^{(2)} + \bigo{\kappa^3}
\end{equation}
with
\begin{equations}[eq:invariant_metric_corrections]
\mathscr{g}_{\mu\nu}^{(1)} &\equiv h_{\mu\nu} - X^{(\rho)}_{(1)} \partial_\rho g_{\mu\nu} - g_{\mu\rho} \partial_\nu X^{(\rho)}_{(1)} - g_{\nu\rho} \partial_\mu X^{(\rho)}_{(1)} = h_{\mu\nu} - \lie_{X_{(1)}} g_{\mu\nu} \eqend{,} \label{eq:invariant_metric_corrections_1} \\
\mathscr{g}_{\mu\nu}^{(2)} &= - \lie_{X_{(2)}} g_{\mu\nu} - \lie_{X_{(1)}} h_{\mu\nu} + \frac{1}{2} \lie_{X_{(1)}}^2 g_{\mu\nu} + \frac{1}{2} \lie_{X^{(\rho)}_{(1)} \partial_\rho X_{(1)}} g_{\mu\nu} \eqend{,} \label{eq:invariant_metric_corrections_2a}
\end{equations}
where $\lie_X$ is the Lie derivative with respect to $X^\mu$, and the explicit expression for the second order $\mathscr{g}_{\mu\nu}^{(2)}$ is given in Eq.~\eqref{eq:invariant_metric_corrections_2b}. We can also define a invariant metric perturbation by taking
\begin{equation}
\label{eq:invariant_metric_perturbation}
\mathscr{h}_{\mu\nu}(X) \equiv \kappa^{-1} \left[ \mathscr{g}_{\mu\nu}(X) - g_{\mu\nu}(X) \right] \eqend{,}
\end{equation}
which leads to
\begin{equation}
\label{eq:invariant_metric_perturbation_expansion}
\mathscr{h}_{\mu\nu} = \mathscr{g}^{(1)}_{\mu\nu} + \kappa \mathscr{g}^{(2)}_{\mu\nu} + \bigo{\kappa^2} \eqend{.}
\end{equation}

For latter use, we also record the perturbative expansion for observables constructed from scalar and co-vector fields. We consider the perturbed co-vector and scalar fields $\tilde{W}_\mu$ and $\tilde{S}$, and assume they can be expanded as
\begin{equations}
\tilde{W}_\mu(x) &= W_\mu(x) + \kappa W^{(1)}_\mu(x) + \kappa^2 W^{(2)}_\mu(x) + \bigo{\kappa^3} \eqend{,} \\
\tilde{S}(x) &= S(x) + \kappa S^{(1)}(x) + \kappa^2 S^{(2)}(x) + \bigo{\kappa^3} \eqend{.}
\end{equations}
The co-vector observable is defined as
\begin{equation}
\label{eq:covector}
\mathscr{W}_\mu(X) \equiv \frac{\partial x^\alpha(X)}{\partial X^{(\mu)}} \tilde{W}_\alpha[x(X)] \eqend{.}
\end{equation}
Its perturbative expansion reads
\begin{equation}
\mathscr{W}_\mu = \mathscr{W}_\mu + \kappa \mathscr{W}^{(1)}_\mu + \kappa^2 \mathscr{W}^{(2)}_\mu + \bigo{\kappa^3}
\end{equation}
with
\begin{equations}[eq:expansion_invariant_covector]
\mathscr{W}^{(1)}_\mu &\equiv W^{(1)}_\mu - X^{(\rho)}_{(1)} \partial_\rho W_\mu - \partial_\mu X^{(\rho)}_{(1)} W_\rho = W^{(1)}_\mu - \lie_{X_{(1)}} W_\mu \eqend{,} \\
\mathscr{W}^{(2)}_\mu &\equiv W^{(2)}_\mu - \lie_{X_{(2)}} W_\mu - \lie_{X_{(1)}} W^{(1)}_\mu + \frac{1}{2} \lie_{X_{(1)}}^2 W_\mu + \frac{1}{2} \lie_{X^{(\rho)}_{(1)} \partial_\rho X_{(1)}} W_\mu \eqend{,}
\end{equations}
where the explicit expression for the second order $\mathscr{W}^{(2)}_\mu$ is given in Eq.~\eqref{eq:expansion_invariant_covector_2}.
The scalar observable was defined in Eq.~\eqref{eq:invariant_scalar} and its perturbative expansion is given by
\begin{equation}
\mathscr{S} = \mathscr{S} + \kappa \mathscr{S}^{(1)} + \kappa^2 \mathscr{S}^{(2)} + \bigo{\kappa^3} \eqend{,}
\end{equation}
with
\begin{equations}[eq:expansion_invariant_scalar]
\mathscr{S}^{(1)} &\equiv S^{(1)} - X^{(\sigma)}_{(1)} \partial_\sigma S = S^{(1)} - \lie_{X_{(1)}} S \eqend{,} \\
\begin{split}
\mathscr{S}^{(2)} &\equiv S^{(2)} - X^{(\sigma)}_{(1)} \partial_\sigma S^{(1)} + \frac{1}{2} X^{(\rho)}_{(1)} X^{(\sigma)}_{(1)} \partial_\rho \partial_\sigma S + X^{(\rho)}_{(1)} \partial_\rho X^{(\sigma)}_{(1)} \partial_\sigma S - X^{(\sigma)}_{(2)} \partial_\sigma S \\
&= S^{(2)} - \lie_{X_{(2)}} S - \lie_{X_{(1)}} S^{(1)} + \frac{1}{2} \lie_{X_{(1)}}^2 S + \frac{1}{2} \lie_{X^{(\rho)}_{(1)} \partial_\rho X_{(1)}} S \eqend{.}
\end{split}
\end{equations}

We shall also write down the expression for the covariant derivative $\bnabla_\mu$ compatible with the invariant metric $\mathscr{g}_{\mu\nu}$ as observables often involve derivatives of tensors. Furthermore, since we are interested in perturbation theory around a given background metric $g_{\mu\nu}$, it is convenient to express $\bnabla_\mu$ in terms of the background covariant derivative $\nabla_\mu$. A simple way to find this expression is to recall that the action of any two covariant derivatives differs by a tensor field~\cite{wald_gr_book}. Hence, for example in the case of a co-vector field $W_\nu$, we have
\begin{equation}
\label{eq:inv_derivative}
\bnabla_\mu W_\nu = \nabla_\mu W_\nu - \mathscr{C}_{\mu\nu}^\rho W_\rho \eqend{,}
\end{equation}
where the tensor $\mathscr{C}_{\mu\nu}^\rho$ reads
\begin{equation}
\label{eq:C_invariant_tensor}
\mathscr{C}_{\mu\nu}^\rho = \frac{1}{2} \mathscr{g}^{\rho\sigma} \left( \nabla_\mu \mathscr{g}_{\nu\sigma} + \nabla_\nu \mathscr{g}_{\mu\sigma} - \nabla_\sigma \mathscr{g}_{\mu\nu} \right) \eqend{.}
\end{equation}
In Eq.~\eqref{eq:C_invariant_tensor}, $\mathscr{g}^{\mu\nu}$ denotes the invariant inverse metric, which is defined from the inverse metric by
\begin{equation}
\label{eq:invariant_inverse_metric}
\mathscr{g}^{\mu\nu}(X) \equiv \frac{\partial X^{(\mu)}}{\partial x^\alpha} \frac{\partial X^{(\nu)}}{\partial x^\beta} \tilde{g}^{\alpha\beta}[x(X)] \eqend{,}
\end{equation}
and fulfills $\mathscr{g}^{\mu\nu} \mathscr{g}_{\nu\rho} = \delta^\mu_\rho$.

At this point is natural to ask how one defines the field-dependent frame $X^{(\mu)}$. Ultimately, one has to choose it such that it describes the system in which the measurement is performed, and, thus, it is an integral part of the definition of the observables one is considering. Nevertheless, as discussed in the introduction, there are different ways to model $X^{(\mu)}$ as scalar fields. For the case of perturbation theory around highly symmetric backgrounds, as is needed for cosmological perturbation theory, one can choose $X^{(\mu)}$ as solutions of some set of scalar differential equations~\cite{brunetti_etal_jhep_2016}
\begin{equation}
\label{eq:X_equations}
D^{(\mu)}_{\tilde{g}, \tilde{\psi}}(X) = 0 \eqend{,}
\end{equation}
where $D^{(\mu)}_{\tilde{g}, \tilde{\psi}}$ are (possibly non-linear) differential operators involving the perturbed metric and/or matter fields. The differential operator then needs to be chosen such as to reflect the experimental situation. Since the $X^{(\mu)}$ are solutions of differential equations with coefficients involving $\tilde{g}_{\mu\nu}$ and $\tilde{\psi}$, they will be (in general non-local) functionals of the metric and matter field perturbations. To obtain a sensible solution of Eqs.~\eqref{eq:X_equations}, we require them (i) to reduce to $D^{(\mu)}_{g, \psi}(x) = 0$ at the background level and (ii) to be causal, \ie, the $X^{(\mu)}(x)$ should only depend on perturbations within the past lightcone of the observation point $x$. Condition (i) realises our assumption that the field-dependent and background frames coincide in the absence of perturbations, see Eq.~\eqref{eq:X_power_series}, while condition (ii) avoids the observables to display unphysical effects coming from arbitrary large spacelike separations.

It is interesting to compare the relational approach discussed here with the more familiar procedure of gauge fixing the metric perturbation $h_{\mu\nu}$. We recall that since the metric perturbation is arbitrary, we are free to perform coordinate transformations that leave the background metric unchanged. On the other hand, the field-dependent frame $X^{(\mu)}$ is completely fixed as the solution of Eqs.~\eqref{eq:X_equations} (with some given boundary condition). Hence, a simple way to fix the background coordinates $x^\mu$ in the full spacetime is to impose that $x^\mu = X^{(\mu)}(x)$. Fixing $x^\mu$ this way amounts to take
\begin{equation}
\label{eq:gauge_condition}
D^{(\mu)}_{\tilde{g}, \tilde{\psi}}(x) = 0 \eqend{,}
\end{equation}
which are (possibly non-linear) gauge-fixing conditions on the perturbed metric and matter fields. We see from Eq.~\eqref{eq:invariant_metric_perturbation} that when the perturbations satisfy the gauge condition~\eqref{eq:gauge_condition} such that $x^\mu$ and $X^{(\mu)}(x)$ coincide, the invariant metric perturbation corresponds to the gauge-fixed metric perturbation. It is clear now why Eq.~\eqref{eq:invariant_tensor} produces tensor field components that are invariant under diffeomorphisms: it corresponds to the gauge-fixed components of the tensor field in the gauge defined by $X^{(\mu)}$, but expressed in terms of an arbitrary metric. This is convenient if we need to fix different gauges for the full metric $\tilde{g}_{\mu\nu}$ and the invariant metric $\mathscr{g}_{\mu\nu}$: for example, if it is easier to solve the equation of motion for $h_{\mu\nu}$ in a gauge different from Eq.~\eqref{eq:gauge_condition}, but the observables of interest are measured in the frame defined by Eq.~\eqref{eq:X_equations}. The main advantage in formulating the observables in the relational framework, however, is that it allows us to easily obtain observables in the non-linear regime. Moreover, by reformulating known gauge conditions in the linear theory as conditions on the dynamical frame (and then extending these to the non-linear theory), one obtains a clear interpretation of these conditions, namely to which experimental setup they correspond. Examples of field-dependent reference frames satisfying requirements (i) and (ii) have been worked out in the case of harmonic coordinates, where $D^{(\mu)}_{\tilde{g}, \tilde{\psi}}$ corresponds to the Laplace-Beltrami operator of the perturbed geometry~\cite{froeb_cqg_2018,froeb_lima_cqg_2018}, co-moving coordinates in de~Sitter spacetime~\cite{lima_cqg_2021} and geodesic lightcone coordinates~\cite{froeb_lima_jcap_2022}.

%%%%%%%%%%%%%%%%%%%%%%%%%%%%%%%%%%%%%%%%%%%%%%%%%%%%%%%%%%%%%%%%%%%%%%%%%%%%%%%%%%%%%%%%%%%%%%%%%%%%%%%%%%%%%%%%%%%%%%%%%%%%%%%%%%%%%%%%%%%%%%%%%%%%%
\section{Synchronous coordinates}                                                                                                                   %
\label{sec:synchronous_coord}                                                                                                                       %
%%%%%%%%%%%%%%%%%%%%%%%%%%%%%%%%%%%%%%%%%%%%%%%%%%%%%%%%%%%%%%%%%%%%%%%%%%%%%%%%%%%%%%%%%%%%%%%%%%%%%%%%%%%%%%%%%%%%%%%%%%%%%%%%%%%%%%%%%%%%%%%%%%%%%

\subsection{Synchronous coordinates on perturbed cosmological spacetimes}
\label{subsec:synchronous_coord}
%%%%%%%%%%%%%%%%%%%%%%%%%%%%%%%%%%%%%%%%%%%%%%%%%%%%%%%%%%%%%%%%%%%%%%%%%%%%%%%%%%%%%%%%%%%%%%%%%%%%%%%%%%%%%%%%%%%%%%%%%%%%%%%%%%%%%%%%%%%%%%%%%%%%%

Let us consider a cosmological spacetime that can be written as a FLRW metric $g_{\mu\nu}$ plus perturbations as in Eq.~\eqref{eq:perturbed_fields}. For flat spatial sections, we can write the FLRW metric line element as
\begin{equation}
\label{eq:FLRW_line_element}
\total s^2 = - \total t^2 + a^2(t) \total \vec{x}^2 \eqend{,}
\end{equation}
where $a(t)$ is the scale factor depending on cosmic time $t$, and the $x^i$ are Cartesian coordinates. The background coordinates are synchronous, \ie, the four-velocity $u_\mu \equiv - \partial_\mu t$ of co-moving observers is orthogonal to the spatial coordinate basis:
\begin{equation}
\label{eq:synchronous_condition_background}
u^\mu \partial_\mu x^i = 0 \eqend{.}
\end{equation}
For later use, let us compute the Christoffel symbols of the background metric in these coordinates. Noting that
\begin{equation}
\label{eq:metric_coord_derivative}
\partial_\alpha g_{\mu\nu} = - 2 H u_\alpha \left( u_\mu u_\nu + g_{\mu\nu} \right)
\end{equation}
with the Hubble parameter $H \equiv a^{-1} \dot a$, we obtain
\begin{equation}
\label{eq:christoffel_symbols}
\Gamma^\rho_{\mu\nu} = H \left( g_{\mu\nu} u^\rho - u_\mu \delta_\nu^\rho - u_\nu \delta_\mu^\rho - u_\mu u_\nu u^\rho \right) \eqend{.}
\end{equation}

We now want to construct synchronous coordinates in the perturbed spacetime. Hence, we pick a congruence of free-falling observers with proper time $\tilde{t}$, and use this proper time to foliate the perturbed spacetime. The normal to the spatial hypersurfaces in this foliation are the observers' four-velocities, which satisfy
\begin{equation}
\label{eq:perturbed_four_velocity}
\tilde{u}_\mu \equiv - \partial_\mu \tilde{t} \quad\text{with}\quad \tilde{u}^\mu \tilde{u}_\mu = -1 \eqend{.}
\end{equation}
Each hypersurface of the foliation can then be covered with coordinates $\tilde{x}^i$ fulfilling
\begin{equation}
\label{eq:synchronous_condition_perturbed}
\tilde{u}^\mu \partial_\mu \tilde{x}^i = 0 \eqend{,}
\end{equation}
\ie, they are orthogonal to the observers' four-velocity. Eqs.~\eqref{eq:perturbed_four_velocity} and~\eqref{eq:synchronous_condition_perturbed} form a set of $n$ scalar differential equations whose solutions give us dynamical synchronous coordinates in the perturbed spacetime, as we wrote in Eq.~\eqref{eq:X_equations}. In the following, we also write $\tilde{t} = X^{(0)}$ and $\tilde{x}^i = X^{(i)}$. Since these equations are fulfilled on the background~\eqref{eq:synchronous_condition_background}, we can expand the solution in perturbation theory and obtain up to second order
\begin{equations}[eq:power_series_coordinates]
\tilde{t} &= t + \kappa t_{(1)} + \kappa^2 t_{(2)} + \bigo{\kappa^3} \eqend{,} \\
\tilde{x}^i &= x^i + \kappa x^i_{(1)} + \kappa^2 x^i_{(2)} + \bigo{\kappa^3} \eqend{.}
\end{equations}
The substitution of the perturbed metric~\eqref{eq:perturbed_fields} and the expansion~\eqref{eq:power_series_coordinates} into Eqs.~\eqref{eq:perturbed_four_velocity} and~\eqref{eq:synchronous_condition_perturbed} results in
\begin{splitequation}
\label{eq:expansion_four_vel_norm}
-1 = \tilde{g}^{\mu\nu} \tilde{u}_\mu \tilde{u}_\nu &= -1 + \kappa \left( 2 u^\mu u_\mu^{(1)} - h^{\mu\nu} u_\mu u_\nu \right) \\
&\quad+ \kappa^2 \left( 2 u^\mu u_\mu^{(2)} + u^\mu_{(1)} u_\mu^{(1)} - 2 h^{\mu\nu} u_\mu u_\nu^{(1)} + h^{\mu\rho} h_\rho{}^\mu u_\mu u_\nu\right) + \bigo{\kappa^3}
\end{splitequation}
and
\begin{splitequation}
\label{eq:expansion_synchronous_condition}
0 = \tilde{u}^\mu \partial_\mu \tilde{x}^i &= \kappa \left( u^\mu \partial_\mu x^i_{(1)} + u^\mu_{(1)} \partial_\mu x^i - h^{\mu\nu} u_\mu \partial_\nu x^i \right) \\
&\quad+ \kappa^2 \Big( u^\mu \partial_\mu x^i_{(2)} + u^\mu_{(2)} \partial_\mu x^i + u^\mu_{(1)} \partial_\mu x^i_{(1)} \\
&\qquad- h^{\mu\nu} u^{(1)}_\mu \partial_\nu x^i - h^{\mu\nu} u_\mu \partial_\nu x^i_{(1)} + h^{\mu\rho} h_\rho{}^\nu u_\mu \partial_\nu x^i \Big) + \bigo{\kappa^3} \eqend{.}
\end{splitequation}
We thus obtain the following equations for the four-velocity and spatial coordinates at first and second order in perturbation theory:
\begin{equations}[eq:four_vel_first_second_order]
u^\mu u^{(1)}_\mu &= \frac{1}{2} h^{\mu\nu} u_\mu u_\nu \eqend{,} \\
u^\mu u^{(2)}_\mu &= - \frac{1}{2} \left( u^\mu_{(1)} u_\mu^{(1)} - 2 h^{\mu\nu} u_\mu u_\nu^{(1)} + h^{\mu\rho} h_\rho{}^\nu u_\mu u_\nu \right)
\end{equations}
and
\begin{equations}[eq:spatial_coord_first_second_order]
u^\mu \partial_\mu x^i_{(1)} &= - u^\mu_{(1)} \partial_\mu x^i + h^{\mu\nu} u_\mu \partial_\nu x^i \\
\begin{split}
u^\mu \partial_\mu x^i_{(2)} &= - u^\mu_{(2)} \partial_\mu x^i - u^\mu_{(1)} \partial_\mu x^i_{(1)} + h^{\mu\nu} u^{(1)}_\mu \partial_\nu x^i + h^{\mu\nu} u_\mu \partial_\nu x^i_{(1)} \\
&\quad- h^{\mu\rho} h_\rho{}^\nu u_\mu \partial_\nu x^i \eqend{.}
\end{split}
\end{equations}

To obtain explicit differential equations for the coordinate corrections produced by the metric perturbation, we use the background coordinate basis~\eqref{eq:FLRW_line_element}. At first order, this gives
\begin{equations}[eq:coord_correction_first_order]
\partial_t t_{(1)} &= - \frac{1}{2} h_{tt} \eqend{,} \\
\partial_t x^i_{(1)} &= \partial^i t_{(1)} + h_t{}^i \eqend{.}
\end{equations}
These equations can be integrated after initial conditions for the perturbed coordinates have been specified. Here we will assume that the metric perturbation is either of compact support or falls off fast enough at past infinity. In the absence of metric perturbations, the background and perturbed coordinates should agree. Thus, the solution for Eqs.~\eqref{eq:coord_correction_first_order} is
\begin{equations}[eq:coord_correction_first_order_integral]
t_{(1)}(t,\vec{x}) &= - \frac{1}{2} \int_{-\infty}^t h_{tt}(s,\vec{x}) \total s \eqend{,} \label{eq:coord_correction_first_order_integral_t} \\
\begin{split}
x^i_{(1)}(t,\vec{x}) &= \int_{-\infty}^t \left( \partial^i t_{(1)} + h_t{}^i \right)(s,\vec{x}) \total s \\
&= - \frac{1}{2} \partial^i \int_{-\infty}^t \int_{-\infty}^s h_{tt}(s',\vec{x}) \total s' \total s + \int_{-\infty}^t h_t{}^i(s,\vec{x}) \total s \eqend{,} \label{eq:coord_correction_first_order_integral_x}
\end{split}
\end{equations}
which agrees with the results of~\cite{giesel_et_al_cqg_2018}. The differential equations for the second-order correction to the coordinates are
\begin{equations}[eq:coord_correction_second_order]
\partial_t t_{(2)} &= \frac{1}{2} \left( \partial^\mu t_{(1)} \partial_\mu t_{(1)} + 2 h_t{}^\mu \partial_\mu t_{(1)} + h_t{}^\mu h_{t\mu} \right) \eqend{,} \\
\partial_t x^i_{(2)} &= a^{-2} \partial_i t_{(2)} + \partial^\mu t_{(1)} \partial_\mu x^i_{(1)} - h_{i\mu} \partial^\mu t_{(1)} + h_t{}^\mu \partial_\mu x^i_{(1)} - h_{t\mu} h^{i\mu} \eqend{.}
\end{equations}
Using the same initial conditions as at first order, we obtain the solutions
\begin{equations}[eq:coord_correction_second_order_integral]
t_{(2)}(t,\vec{x}) &= \frac{1}{2} \int_{-\infty}^t \left( \partial^\mu t_{(1)} \partial_\mu t_{(1)} + 2 h_t{}^\mu \partial_\mu t_{(1)} + h_t{}^\mu h_{t\mu} \right)(s,\vec{x}) \total s \eqend{,} \label{eq:coord_correction_second_order_integral_t} \\
\begin{split}
x^i_{(2)}(t,\vec{x}) &= \int_{-\infty}^t \Big( a^{-2} \partial_i t_{(2)} + \partial^\mu t_{(1)} \partial_\mu x^i_{(1)} - h^{i\mu} \partial_\mu t_{(1)} + h_t{}^\mu \partial_\mu x^i_{(1)} \\
&\qquad\qquad- h_{t\mu} h^{i\mu} \Big)(s,\vec{x}) \total s \label{eq:coord_correction_second_order_integral_x} \eqend{.}
\end{split}
\end{equations}

As explained in the introduction, the diffeomorphism invariance of the full gravity theory results in a gauge symmetry for the metric perturbation, which is thus gauge dependent. For infinitesimal and localised diffeomorphisms that preserve the background, $x^\mu \to x^\mu - \kappa \xi^\mu(x)$, the perturbed metric changes by
\begin{equation}
\label{eq:metric_perturbation_transformation}
\delta_\xi h_{\mu\nu} = \lie_\xi \tilde{g}_{\mu\nu} = \partial_\mu \xi_\nu + \partial_\nu \xi_\mu - 2 \Gamma^\rho_{\mu\nu} \xi_\rho + \kappa \left( \xi^\rho \partial_\rho h_{\mu\nu} - h_{\mu\rho} \partial_\nu \xi^\rho - h_{\nu\rho} \partial_\mu \xi^\rho \right) \eqend{,}
\end{equation}
where $\lie_V$ is the Lie derivative with respect to $V^\mu$. Given the transformation of the metric perturbation, we can now check that the perturbed coordinates $X^{(\mu)} = (\tilde{t},\tilde{\vec{x}})$ indeed transform as scalar fields. At first order, we have that the correction to the time coordinate~\eqref{eq:coord_correction_first_order_integral_t} changes by
\begin{splitequation}
\delta_\xi t_{(1)}(t,\vec{x}) &= - \frac{1}{2} \int_{-\infty}^t \delta_\xi h_{tt}(s,\vec{x}) \total s = - \int_{-\infty}^t \left( \partial_t \xi_t - \Gamma_{tt}^\rho \xi_\rho \right)(s,\vec{x}) \total s + \bigo{\kappa} \\
&= \int_{-\infty}^t \left( \partial_t \xi^t \right)(s,\vec{x}) \total s + \bigo{\kappa} = \xi^t(t,\vec{x}) - \lim_{s \to - \infty} \xi^t(s,\vec{x}) + \bigo{\kappa} \\
&= \xi^t(t,\vec{x}) + \bigo{\kappa} \eqend{,}
\end{splitequation}
where we have used Eqs.~\eqref{eq:christoffel_symbols} and~\eqref{eq:metric_perturbation_transformation} and the assumption that the diffeomorphism is localised, and thus vanishes at past infinity. The change in the first-order correction to the spatial coordinates~\eqref{eq:coord_correction_first_order_integral_x} can be computed analogously, and the final result can be written as
\begin{equation}
\delta_\xi t_{(1)} = \xi^\mu \partial_\mu t \quad\text{and}\quad \delta_\xi x^i_{(1)} = \xi^\mu \partial_\mu x^i \eqend{.}
\end{equation}
To compute the change produced by diffeomorphisms on the second-order correction to the coordinates, we first notice that the change of the metric perturbation leaves a next-order term at first order. Indeed, for the first-order correction to the time coordinate, we find in total
\begin{splitequation}
\label{eq:coord_correction_deltaxi_t1}
\delta_\xi t_{(1)}(t,\vec{x}) &= - \frac{1}{2} \int_{-\infty}^t \delta_\xi h_{tt}(s,\vec{x}) \total s \\
&= \xi^t(t,\vec{x}) - \frac{\kappa}{2} \int_{-\infty}^t \left[ \xi^\rho \partial_\rho h_{tt} + 2 \partial_t \xi^\rho h_{t\rho} \right](s,\vec{x}) \total s \eqend{.}
\end{splitequation}
The change of the second-order correction~\eqref{eq:coord_correction_second_order_integral_t} to the time coordinate is computed in Eq.~\eqref{eq:coord_correction_deltaxi_t2} and reads
\begin{equation}
\delta_\xi t_{(2)}(t,\vec{x}) = \xi^\mu \partial_\mu t_{(1)}(t,\vec{x}) + \frac{1}{2} \int_{-\infty}^t \left[ \xi^\mu \partial_\mu h_{tt} + 2 \partial_t \xi^\mu h_{t\mu} \right](s,\vec{x}) \total s + \bigo{\kappa} \eqend{,}
\end{equation}
and we see that the integral term exactly cancels the one from~\eqref{eq:coord_correction_deltaxi_t1}. The computation of the change in the spatial coordinates again leads to similar results. In conclusion, we have shown that the perturbed synchronous coordinates change by
\begin{equations}
\delta_\xi \tilde{t} &= \kappa \delta_\xi t_{(1)} + \kappa^2 \delta_\xi t_{(2)} + \bigo{\kappa^3} = \kappa \xi^\mu \partial_\mu \tilde{t} + \bigo{\kappa^3} \eqend{,} \\
\delta_\xi \tilde{x}^i &= \kappa \delta_\xi x_{(1)}^i + \kappa^2 \delta_\xi x^i_{(2)} + \bigo{\kappa^3} = \kappa \xi^\mu \partial_\mu \tilde{x}^i + \bigo{\kappa^3}
\end{equations}
under infinitesimal and localised diffeomorphisms that preserve the background. Hence, we have verified that $X^{(\mu)} = (\tilde{t},\tilde{\vec{x}})$ transform as scalar fields up to second order. As discussed earlier, this is a key assumption in the construction of our relational observables.

\subsection{Gauge-invariant metric}
\label{subsec:gauge_invariant_metric}
%%%%%%%%%%%%%%%%%%%%%%%%%%%%%%%%%%%%%%%%%%%%%%%%%%%%%%%%%%%%%%%%%%%%%%%%%%%%%%%%%%%%%%%%%%%%%%%%%%%%%%%%%%%%%%%%%%%%%%%%%%%%%%%%%%%%%%%%%%%%%%%%%%%%%

In Eqs.~\eqref{eq:invariant_metric}--\eqref{eq:invariant_metric_perturbation} we have already defined the gauge-invariant metric $\mathscr{g}_{\mu\nu}$, its perturbative expansion and the gauge-invariant metric perturbation $\mathscr{h}_{\mu\nu}$, respectively. Our aim now is to check that the gauge-invariant metric perturbation in the dynamical synchronous coordinate system $X^{(\mu)} = (\tilde{t}, \tilde{\vec{x}})$ satisfies the synchronous condition $\mathscr{h}_{t\nu} = 0$, compare the discussion at the end of Sec.~\ref{sec:relational_observables}. We first show this explicitly, up to second order in perturbation theory. The correction to the invariant metric at first order~\eqref{eq:invariant_metric_corrections_1} is explicitly given by $\mathscr{g}^{(1)}_{t\nu} = h_{t\nu} + \partial_\nu t_{(1)} - g_{t\nu} \partial_t t_{(1)} - g_{i\nu} \partial_t x^i_{(1)}$, and using Eqs.~\eqref{eq:coord_correction_first_order} we have $g_{i\nu} \partial_t x^i_{(1)} = g_{\nu\mu} \left( \partial^\mu t_{(1)} + h_t{}^\mu \right) - g_{t\nu} \left( \partial^t t_{(1)} + h_t{}^t \right)$ such that
\begin{splitequation}
\mathscr{g}^{(1)}_{t\nu} &= h_{t\nu} + \partial_\nu t_{(1)} - g_{t\nu} \partial_t t_{(1)} - g_{\nu\mu} \left( \partial^\mu t_{(1)} + h_t{}^\mu \right) + g_{t\nu} \left( \partial^t t_{(1)} + h_t{}^t \right) \\
&= - g_{t\nu} \left( 2 \partial_t t_{(1)} + h_{tt} \right) = 0
\end{splitequation}
as required. For the second-order correction, Eq.~\eqref{eq:invariant_metric_corrections_2a} or Eq.~\eqref{eq:invariant_metric_corrections_2b}, the analogous computation is done in Eq.~\eqref{eq:synchronous_metric_second_order} with the result $\mathscr{g}_{t\nu}^{(2)} = 0$. Hence, we have shown that
\begin{equation}
\label{eq:synchronous_metric_perturbation}
\mathscr{h}_{t\nu} = \bigo{\kappa^2} \eqend{.}
\end{equation}

We can also obtain the result~\eqref{eq:synchronous_metric_perturbation} by inspecting the form of the invariant inverse perturbed metric defined in Eq.~\eqref{eq:invariant_inverse_metric}. From that definition and Eqs.~\eqref{eq:perturbed_four_velocity} and~\eqref{eq:synchronous_condition_perturbed}, we have that
\begin{equation}
\mathscr{g}^{tt} = \frac{\partial \tilde{t}}{\partial x^\mu} \frac{\partial \tilde{t}}{\partial x^\nu} \tilde{g}^{\mu\nu} = \tilde{u}^\mu \tilde{u}_\mu = -1
\end{equation} 
and that
\begin{equation}
\mathscr{g}^{ti} = \frac{\partial \tilde{t}}{\partial x^\mu} \frac{\partial \tilde{x}^i}{\partial x^\nu} \tilde{g}^{\mu\nu} = \tilde{u}^\mu \partial_\mu \tilde{x}^i = 0 \eqend{.}
\end{equation}
The invariant inverse perturbed metric is thus block-diagonal, which means that the invariant perturbed metric $\mathscr{g}_{\mu\nu}$ is also block-diagonal with $\mathscr{g}_{tt} = -1$ and $\mathscr{g}_{ti} = 0$ since $\mathscr{g}^{\mu\nu} \mathscr{g}_{\nu\rho} = \delta^\mu_\rho$. This results in Eq.~\eqref{eq:synchronous_metric_perturbation}, not only at second but at all orders in perturbation theory.

%%%%%%%%%%%%%%%%%%%%%%%%%%%%%%%%%%%%%%%%%%%%%%%%%%%%%%%%%%%%%%%%%%%%%%%%%%%%%%%%%%%%%%%%%%%%%%%%%%%%%%%%%%%%%%%%%%%%%%%%%%%%%%%%%%%%%%%%%%%%%%%%%%%%%
\section{Perturbed Einstein equation}                         									                                                                  %
\label{sec:perturbed_einstein_eq}                                                                                                                   %
%%%%%%%%%%%%%%%%%%%%%%%%%%%%%%%%%%%%%%%%%%%%%%%%%%%%%%%%%%%%%%%%%%%%%%%%%%%%%%%%%%%%%%%%%%%%%%%%%%%%%%%%%%%%%%%%%%%%%%%%%%%%%%%%%%%%%%%%%%%%%%%%%%%%%

We now consider the Einstein equation for the perturbed metric when sourced by matter perturbations around a FLRW background. Our aim here is to use the relational approach to obtain the gauge-invariant part of these equations in synchronous coordinates, up to second order in the metric and matter perturbations. For the matter we will consider two popular models in the literature, namely, the ideal fluid and the scalar field with a potential.

Thus, let us consider the Einstein equation for the perturbed spacetime in the form
\begin{equation}
\label{eq:perturbed_Einstein_eq}
\tilde{E}_{\mu\nu} \equiv 2 \tilde{G}_{\mu\nu} - \kappa^2 \tilde{T}_{\mu\nu} = 0 \eqend{,}
\end{equation}
where $\tilde{G}_{\mu\nu}$ is the perturbed Einstein tensor and $\tilde{T}_{\mu\nu}$ is the perturbed stress tensor for the matter. We can then expand both the Einstein and the stress tensors up to second order in $\kappa$:
\begin{equations}
\tilde{G}_{\mu\nu} &= G_{\mu\nu} + \kappa G_{\mu\nu}^{(1)} + \kappa^2 G_{\mu\nu}^{(2)} + \bigo{\kappa^3} \eqend{,} \\
\tilde{T}_{\mu\nu} &= T_{\mu\nu} + \kappa T_{\mu\nu}^{(1)} + \kappa^2 T_{\mu\nu}^{(2)} + \bigo{\kappa^3} \eqend{.}
\end{equations}
Assuming the FLRW metric~\eqref{eq:FLRW_line_element} for the background, we obtain for the perturbed Einstein tensor up to first order the following expressions:
\begin{equations}[eq:perturbed_einstein_tensor]
G_{\mu\nu} &= - \frac{1}{2} (n-2) (n-1-2\epsilon) H^2 g_{\mu\nu} + (n-2) H^2 \epsilon u_\mu u_\nu \eqend{,} \\
\begin{split}
G_{\mu\nu}^{(1)} &= \nabla^\alpha \nabla_{(\mu} h_{\nu)\alpha} - \frac{1}{2} \nabla^2 h_{\mu\nu} - \frac{1}{2} g_{\mu\nu} \nabla^\alpha \nabla^\beta h_{\alpha\beta} - \frac{1}{2} \nabla_\mu \nabla_\nu h + \frac{1}{2} g_{\mu\nu} \nabla^2 h \\
&\quad- \frac{1}{2} (n-1) (n-2\epsilon) H^2 h_{\mu\nu} + \frac{1}{2} H^2 [ (n-1-\epsilon) h + (n-2) \epsilon u^\alpha u^\beta h_{\alpha\beta} ] g_{\mu\nu} \eqend{,}
\end{split}
\end{equations}
where $\nabla^2 \equiv \nabla^\rho \nabla_\rho$, $h \equiv g^{\mu\nu} h_{\mu\nu}$ and $\epsilon \equiv - H^{-2} \dot{H}$ is the deceleration parameter; the second-order expression $G_{\mu\nu}^{(2)}$ is given in Eq.~\eqref{eq:perturbed_einstein_tensor_2}. We will leave the matter stress tensor unspecified for the time being. The perturbative expansion of the Einstein equation~\eqref{eq:perturbed_Einstein_eq} then reads
\begin{equation}
\tilde{E}_{\mu\nu} = E_{\mu\nu} + \kappa E^{(1)}_{\mu\nu} + \kappa^2 E^{(2)}_{\mu\nu} + \bigo{\kappa^3} \eqend{,}
\end{equation}
with $E_{\mu\nu} = 0$ being the Einstein equation for the background and
\begin{equations}[eq:expansion_einstein_eq]
E^{(1)}_{\mu\nu} &= 2 G^{(1)}_{\mu\nu} - \kappa^2 T^{(1)}_{\mu\nu} \eqend{,} \\
E^{(2)}_{\mu\nu} &= 2 G^{(2)}_{\mu\nu} - \kappa^2 T^{(2)}_{\mu\nu} \eqend{.}
\end{equations}

To obtain the gauge-invariant part of the Einstein equation, we perform the transformation~\eqref{eq:invariant_tensor} for the tensor $\tilde{E}_{\mu\nu}$. This results in
\begin{splitequation}
\label{eq:invariant_Einstein_eq}
\mathscr{E}_{\mu\nu}(X) &\equiv \frac{\partial x^\alpha(X)}{\partial X^{(\mu)}} \frac{\partial x^\beta(X)}{\partial X^{(\nu)}} \tilde{E}_{\alpha\beta}[x(X)] \\
&= \kappa E^{(1)}_{\mu\nu} + \kappa^2 \left( E^{(2)}_{\mu\nu} - X^{(\rho)}_{(1)} \partial_\rho E^{(1)}_{\mu\nu} - \partial_\mu X^{(\rho)}_{(1)} E^{(1)}_{\rho\nu} - \partial_\nu X^{(\rho)}_{(1)} E^{(1)}_{\rho\mu} \right) + \bigo{\kappa^3} \eqend{,}
\end{splitequation}
where we have used Eq.~\eqref{eq:X_inverse_power_series}, and that the background metric satisfies the background Einstein equation $E_{\mu\nu} = 0$. We see that expressing the Einstein equation in a relational way just corresponds to a rearrangement of the terms in the perturbative series, \ie, if we have a solution $\tilde{E}_{\mu\nu} = 0$ order by order in perturbation theory, also $\mathscr{E}_{\mu\nu} = 0$ order by order. In this sense, the Einstein equation is already gauge-invariant, in contrast to a general field (which does not vanish). However, the right-hand side of Eqs.~\eqref{eq:invariant_Einstein_eq} is still expressed in terms of the gauge-dependent metric and matter field perturbations. To express Eqs.~\eqref{eq:invariant_Einstein_eq} in an explicitly gauge-invariant way, we need to replace the gauge-dependent fields by their gauge-independent parts. We will do exactly this in the following.

For later convenience we also perform the $(3+1)$-decomposition of the metric tensor and write the induced background spatial metric as
\begin{equation}
\label{eq:induced_spatial_metric}
\bar{g}_{\mu\nu} \equiv u_\mu u_\nu + g_{\mu\nu} \eqend{.}
\end{equation}
We denote the covariant derivative associated with $\bar{g}_{\mu\nu}$ as $\bar{\partial}_\alpha$ and the projected part of a vector $v_\mu$ as
\begin{equation}
\label{eq:spatial_projection}
\bar{v}_\mu \equiv \bar{g}_\mu{}^\nu v_\nu \eqend{.}
\end{equation}

\subsection{Ideal fluid model}
\label{sec:ideal_fluid}
%%%%%%%%%%%%%%%%%%%%%%%%%%%%%%%%%%%%%%%%%%%%%%%%%%%%%%%%%%%%%%%%%%%%%%%%%%%%%%%%%%%%%%%%%%%%%%%%%%%%%%%%%%%%%%%%%%%%%%%%%%%%%%%%%%%%%%%%%%%%%%%%%%%%%

In the case of an ideal fluid, the perturbed stress tensor reads
\begin{equation}
\tilde{T}_{\mu\nu} = (\tilde{\rho} + \tilde{p}) \tilde{V}_\mu \tilde{V}_\nu + \tilde{p} \tilde{g}_{\mu\nu} \eqend{,}
\end{equation}
where $\tilde{\rho}$ is the perturbed energy density, $\tilde{p} = f(\tilde{\rho})$ is the perturbed pressure, $f$ gives the equation of state and $\tilde{V}^\mu$ is the perturbed four-velocity of the fluid. To perform the perturbative expansion for the fluid, it is convenient express these fields as
\begin{equations}[eq:fluid_perturbation]
\tilde{\rho} &= \rho \left( 1 + \kappa d \right) \eqend{,} \\
\tilde{p} &= f\left[ \rho \left( 1 + \kappa d \right) \right] = p + \kappa c^2_\mathrm{s} \rho d + \frac{1}{2} \kappa^2 \rho^2 \frac{\total c_\text{s}^2}{\total \rho} d^2 + \bigo{\kappa^3} \eqend{,} \\
\tilde{V}_\mu & = u_\mu + \kappa v_\mu \eqend{,}
\end{equations}
where $\rho$ and $p$ are the background energy density and pressure of the fluid, $d$ is the fractional density perturbation, $c^2_\mathrm{s} \equiv f'(\rho)$ is the square of the speed of sound in the fluid and $v_\mu$ is the four-velocity perturbation. The normalisation of the fluid's four-velocity constrains $v_t$, the time component of the four-velocity perturbation. Indeed, we obtain
\begin{splitequation}
\label{eq:norm_fluid_four_vel}
-1 &= \tilde{g}^{\mu\nu} \tilde{V}_\mu \tilde{V}_\nu \\
&= -1 + \kappa \left( 2 u^\mu v_\mu - h^{\mu\nu} u_\mu u_\nu \right) + \kappa^2 \left( v^\mu v_\mu - 2 h^{\mu\nu} u_\mu v_\nu + h^{\mu\rho} h_\rho{}^\nu u_\mu u_\nu \right) + \bigo{\kappa^3} \eqend{.}
\end{splitequation}
As a result, we have to express $v_t$ as a power series,
\begin{equation}
\label{eq:time_component_perturbed_four_vel}
v_t = v^{(1)}_t + \kappa v^{(2)}_t + \bigo{ \kappa^2 } \eqend{,}
\end{equation}
and then substitute this expansion back into Eq.~\eqref{eq:norm_fluid_four_vel}. The result is
\begin{equations}[eq:time_component_4_vel]
v_t^{(1)} &= \frac{1}{2} h_{tt} \eqend{,} \\
v_t^{(2)} &= - \frac{1}{2} \left( v^i v_i - 2 h_t{}^i v_i + h_t{}^i h_{ti} - \frac{1}{4} h_{tt}^2 \right) \eqend{.}
\end{equations}
The perturbative expansion for the fluid's stress tensor results in 
\begin{equations}
T_{\mu\nu} &= \rho u_\mu u_\nu + p \bar{g}_{\mu\nu} \eqend{,} \\
T_{\mu\nu}^{(1)} &= \rho d u_\mu u_\nu + (\rho + p) ( v_\mu u_\nu + u_\mu v_\nu ) + c_\text{s}^2 \rho d \bar{g}_{\mu\nu} + p h_{\mu\nu} \eqend{,} \\
\begin{split}
T_{\mu\nu}^{(2)} &= (\rho + p) v_\mu v_\nu + \rho (1 + c_\text{s}^2) d ( u_\mu v_\nu + u_\nu v_\mu ) + \frac{1}{2} \rho^2 \frac{\total c_\text{s}^2}{\total \rho} d^2 \bar{g}_{\mu\nu} + c_\text{s}^2 \rho d h_{\mu\nu} \eqend{,}
\end{split}
\end{equations}
where we have used Eqs.~\eqref{eq:fluid_perturbation}, and we recall that $\bar{g}_{\mu\nu}$ is the background spatial metric defined in Eq.~\eqref{eq:induced_spatial_metric}.

To work out the expansion of the fluid's equation of motion, we now consider the divergence
\begin{equation}
\label{eq:divergence_stress_tensor}
\tilde{F}_\nu \equiv \tilde{\nabla}^\mu \tilde{T}_{\mu\nu} \eqend{.}
\end{equation}
Its perturbative expansion reads
\begin{equation}
\tilde{F}_\nu = F_\nu + \kappa F^{(1)}_\nu + \kappa^2 F^{(2)}_\nu + \bigo{\kappa^3} \eqend{,}
\end{equation}
where the first-order correction $F^{(1)}_\nu$ reads
\begin{equation}
\label{eq:eom_matter_1st_order}
F^{(1)}_\nu \equiv \nabla^\mu T_{\mu\nu}^{(1)} - h^{\rho\mu} \nabla_\rho T_{\mu\nu} - \left( \nabla_\mu h^{\mu\rho} - \frac{1}{2} \nabla^\rho h \right) T_{\rho\nu} - \frac{1}{2} \nabla_\nu h^{\mu\rho} T_{\mu\rho} \eqend{,}
\end{equation}
and the second-order correction $F^{(2)}_\nu$ is given in Eq.~\eqref{eq:eom_matter_2nd_order}.

As mentioned earlier, to make the gauge invariance of Eq.~\eqref{eq:invariant_Einstein_eq} explicit, we have to express it in terms of the gauge-invariant parts of the perturbed fields. The invariant metric perturbation was defined in Eq.~\eqref{eq:invariant_metric_perturbation}. For the matter perturbations, it is convenient to start with the invariant energy density for the fluid. The perturbative expansion of a scalar observable was given in Eqs.~\eqref{eq:expansion_invariant_scalar}, and in the case of the invariant energy density it reads
\begin{splitequation}
\label{eq:invariant_rho}
\uprho(X) &\equiv \tilde{\rho}[x(X)] = \rho \left( 1 + \kappa \mathscr{d}(X) \right) \\
&= \rho + \kappa \left( \rho d - \dot{\rho} t_{(1)} \right) - \kappa^2 \big[ \dot{\rho} t_{(1)} d + \rho t_{(1)} \dot{d} + \rho x^i_{(1)} \partial_i d - \frac{1}{2} t_{(1)}^2 \ddot{\rho} \\
&\quad- t_{(1)} \partial_t t_{(1)} \dot{\rho} - x^i_{(1)} \partial_i t_{(1)} \dot{\rho} + t_{(2)} \dot{\rho} \big] + \bigo{\kappa^3} \\
&= \rho + \kappa \uprho_{(1)} + \kappa^2 \uprho_{(2)} + \bigo{\kappa^3} \eqend{.}
\end{splitequation}
Above, $\mathscr{d}$ denotes the invariant fractional density perturbation and we have defined the expansion of the invariant density perturbation
\begin{equations}[eq:invariant_rho_1st_2nd_orders]
\uprho_{(1)} &\equiv \rho d - \dot{\rho} t_{(1)} \eqend{,} \\
\uprho_{(2)} &\equiv - \dot{\rho} t_{(1)} d - \rho t_{(1)} \dot{d} - \rho x^i_{(1)} \partial_i d + \frac{1}{2} \ddot{\rho} t_{(1)}^2 - \frac{1}{2} \dot{\rho} t_{(1)} h_{tt} + \dot{\rho} x^i_{(1)} \partial_i t_{(1)} - \dot{\rho} t_{(2)} \eqend{,}
\end{equations}
where we have used Eqs.~\eqref{eq:coord_correction_first_order} to eliminate the time derivative of the coordinate corrections. The invariant fractional density perturbation then reads
\begin{equation}
\label{eq:invariant_d}
\mathscr{d} = \frac{1}{\rho} \left( \uprho_{(1)} + \kappa \uprho_{(2)} \right) + \bigo{\kappa^2} \eqend{.}
\end{equation}
The invariant scalar corresponding to the pressure in the fluid is obtained from the equation of state of the fluid evaluated for the invariant energy density defined in Eq.~\eqref{eq:invariant_rho}. The result is
\begin{splitequation}
\label{eq:invariant_p}
\mathscr{p}(X) &\equiv \tilde{p}[x(X)] = f[\uprho(X)] = f\left[ \rho \left( 1 + \kappa \mathscr{d} \right) \right] + \bigo{\kappa^3} \\
&= p + \kappa c^2_\mathrm{s} \rho \mathscr{d} + \frac{1}{2} \kappa^2 \frac{\total c^2_\mathrm{s}}{\total \rho} \left( \rho \mathscr{d} \right)^2 + \bigo{\kappa^3} \eqend{.}
\end{splitequation}

We define the invariant four-velocity as the co-vector observable
\begin{equation}
\label{eq:inv_4_vel_fluid}
\mathscr{V}_\mu(X) \equiv \frac{\partial x^\nu(X)}{\partial X^{(\mu)}} \tilde{V}_\nu[x(X)] = u_\mu + \kappa \mathscr{V}^{(1)}_\mu + \kappa^2 \mathscr{V}^{(2)}_\mu + \bigo{\kappa^3} \eqend{.}
\end{equation}
The perturbative expansion of a co-vector observable was given in Eqs.~\eqref{eq:expansion_invariant_covector}. After eliminating the time derivatives of the coordinate corrections using Eqs.~\eqref{eq:coord_correction_first_order} and~\eqref{eq:coord_correction_second_order}, we obtain the following expressions for the first- and second-order corrections to the temporal and spatial components of the invariant four-velocity:
\begin{equations}[eq:invariant_4_vel_corrections_components]
\mathscr{V}_t^{(1)} &= v^{(1)}_t - \frac{1}{2} h_{tt} \eqend{,} \label{eq:invariant_4_vel_corrections_components_1_t} \\
\mathscr{V}_i^{(1)} &= v_i + \partial_i t_{(1)} \eqend{,} \label{eq:invariant_4_vel_corrections_components_1_i} \\
\mathscr{V}_t^{(2)} &= v_t^{(2)} - \left( \partial^i t_{(1)} + h_t{}^i \right) v_i - \frac{1}{8} h_{tt}^2 + \frac{1}{2} h_t{}^i h_{ti} - \frac{1}{2} \partial^i t_{(1)} \partial_i t_{(1)} \eqend{,} \label{eq:invariant_4_vel_corrections_components_2_t} \\
\begin{split}
\mathscr{V}_i^{(2)} &= - t_{(1)} \partial_t v_i - x_{(1)}^j \partial_j v_i - \partial_i t_{(1)} v_t^{(1)} - \partial_i x_{(1)}^j v_j + \partial_i t_{(2)} + \frac{1}{2} \partial_i \left( t_{(1)} h_{tt} \right) \\
&\quad- \partial_i \left( x_{(1)}^j \partial_j t_{(1)} \right) \eqend{,} \label{eq:invariant_4_vel_corrections_components_2_i}
\end{split}
\end{equations}
and we recall that the power series for the temporal components of the four-velocity perturbation was given in Eqs.~\eqref{eq:time_component_4_vel}. We then define the invariant four-velocity perturbation as
\begin{equation}
\label{eq:invariant_4_vel_perturbation}
\mathscr{v}_\mu(X) \equiv \kappa^{-1} \left[ \mathscr{V}_\mu(X) - u_\mu \right] = \mathscr{V}^{(1)}_\mu + \kappa \mathscr{V}^{(2)}_\mu + \bigo{\kappa^2} \eqend{.}
\end{equation}
In passing, we note that the normalisation of the 4-velocity $\mathscr{V}_\mu$ with the invariant metric $\mathscr{g}_{\mu\nu}$ implies that
\begin{equations}[eq:time_components_invariant_4_vel]
\mathscr{V}^{(1)}_t &= \frac{1}{2} \mathscr{h}_{tt} = 0 \eqend{,} \label{eq:time_components_invariant_4_vel_1} \\
\mathscr{V}^{(2)}_t &= - \frac{1}{2} \left( \mathscr{V}^i_{(1)} \mathscr{V}_i^{(1)} + 2 \mathscr{h}^{ti} \mathscr{V}_i^{(1)} - \mathscr{h}^{ti} \mathscr{h}_{ti} - \frac{1}{4} \mathscr{h}_{tt}^2 \right) = - \frac{1}{2} \mathscr{V}^i_{(1)} \mathscr{V}_i^{(1)} \eqend{.} \label{eq:time_components_invariant_4_vel_2}
\end{equations}
Furthermore, Eqs.~\eqref{eq:time_components_invariant_4_vel} can also be obtained by substituting Eqs.~\eqref{eq:time_component_4_vel} into Eqs.~\eqref{eq:invariant_4_vel_corrections_components_1_t} and~\eqref{eq:invariant_4_vel_corrections_components_2_t}.

Finally, we can obtain the relational version of the fluid equation of motion by transforming the divergence~\eqref{eq:divergence_stress_tensor} as
\begin{splitequation}
\label{eq:invariant_eom_matter}
\mathscr{F}_\mu(X) &\equiv \frac{\partial x^\nu(X)}{\partial X^{(\mu)}} \tilde{F}_\nu[x(X)] \\
&= \kappa F^{(1)}_\mu + \kappa^2 \left( F^{(2)}_\mu - X_{(1)}^{(\sigma)} \partial_\sigma F^{(1)}_\mu - \partial_\mu X_{(1)}^{(\sigma)} F^{(1)}_\sigma \right) + \bigo{\kappa^3} \eqend{,}
\end{splitequation}
where again we have used Eq.~\eqref{eq:X_inverse_power_series} to obtain the perturbative expansion and that the equation of motion is satisfied on the background, $F_\mu = 0$. We see again that as for the Einstein equation~\eqref{eq:invariant_Einstein_eq} the fluid equation of motion is already gauge-invariant, in the sense that the relational version just corresponds to a rearrangement of the terms in the perturbative series, such that $\mathscr{F}_\mu = 0$ and $\tilde{F}_\mu = 0$ are equivalent.

\subsubsection{Background}
%%%%%%%%%%%%%%%%%%%%%%%%%%%%%%%%%%%%%%%%%%%%%%%%%%%%%%%%%%%%%

At the background level, the trace and the time-time and space-space components of the Einstein equation give the Friedmann equations
\begin{equations}[eq:friedmann_fluid]
(n-1) (n-2) H^2 = \kappa^2 \rho \eqend{,} \label{eq:rho_eq} \\
- (n-2) (n-1-2\epsilon) H^2 = \kappa^2 p \eqend{.}
\end{equations}
The equation of motion of the fluid is obtained by taking the divergence of its stress tensor, assuming that it is in a homogeneous and isotropic state such that $\rho$ and $p$ only depend on time. The result is
\begin{equation}
\label{eq:dotrho_rhop}
\dot{\rho} + (n-1) H (\rho + p) = 0 \eqend{,}
\end{equation}
and we recall that we are given an equation of state $p = f(\rho)$ which determines the pressure in terms of the energy density.

To express the speed of sound in the fluid in terms of the background geometry parameters, we take a time derivative of Eqs.~\eqref{eq:friedmann_fluid} and use the fact that $\dot{p} = c_\text{s}^2 \dot{\rho}$ to replace $\dot{p}$, while $\dot{\rho}$ is given by Eq.~\eqref{eq:dotrho_rhop}. The result is
\begin{equation}
\label{eq:speed_sound}
c_\text{s}^2 = - 1 + 2 \frac{\epsilon-\delta}{n-1} \eqend{,}
\end{equation} 
where $\delta \equiv \dot{\epsilon}/(2\epsilon H)$ is the second slow-roll parameter. In what follows, we will also need the derivative of $c_\text{s}^2$ with respect to the background energy density $\rho$. That derivative can be easily related to the background geometry parameters by taking the time derivative of $c_\text{s}^2$~\eqref{eq:speed_sound}:
\begin{equation}
\frac{\total c_\text{s}^2}{\total t} = \frac{\total c_\text{s}^2}{\total \rho} \dot{\rho} \eqend{.}
\end{equation}
Then, by taking the time derivative of Eq.~\eqref{eq:speed_sound} and the Friedmann equation for $\rho$~\eqref{eq:rho_eq}, we obtain
\begin{equation}
\frac{\total c_\text{s}^2}{\total \rho} = - \kappa^2 \frac{2 \delta}{(n-1)^2 (n-2) H^2} \left( 1 - \frac{\dot{\delta}}{2 \epsilon \delta H} \right) \eqend{.}
\end{equation}
For later use, we also calculate
\begin{equation}
\label{eq:nu}
\nu \equiv \frac{p}{\rho} = - 1 + \frac{2 \epsilon}{n-1} \quad\text{and}\quad \dot{\nu} = \frac{4 \epsilon\delta H}{n-1} \eqend{.}
\end{equation}

\subsubsection{First order}
%%%%%%%%%%%%%%%%%%%%%%%%%%%%%%%%%%%%%%%%%%%%%%%%%%%%%%%%%%%%%

As we can see from Eqs.~\eqref{eq:invariant_Einstein_eq}, the assumption that the Einstein equation is satisfied by the background implies that the first-order correction to the Einstein equation $E^{(1)}_{\mu\nu}$ is gauge invariant, which is an instance of the Stewart--Walker lemma~\cite{stewart_walker_prsla_1974}. Therefore, we can simply replace all the perturbed fields by their gauge-invariant parts, since we know that the gauge-dependent parts will cancel out. This amounts to taking
\begin{equation}
\label{eq:substitutions_fluid_1st_order}
h_{\mu\nu} \to \mathscr{h}_{\mu\nu} \eqend{,} \quad v_i \to \mathscr{v}_i \eqend{,} \quad v_t^{(1)} \to 0 \quad\text{and}\quad d \to \mathscr{d} \eqend{,}
\end{equation}
and we recall that $u^\mu \mathscr{h}_{\mu\nu} = 0$ at all orders~\eqref{eq:synchronous_metric_perturbation} and that $\mathscr{v}^{(1)}_t = 0$~\eqref{eq:time_components_invariant_4_vel_1}. After performing these substitutions, we can write down the constraint and dynamical equations for the gauge-invariant part of the perturbations at linear order. (Of course, one can also substitute the full expressions and check explicitly that the gauge-dependent parts cancel.)

The constraint equations are obtained by computing the contraction
\begin{equation}
\label{eq:constraint_eq_linear}
\Tilde{E}_\mu \equiv \tilde{E}_{\mu\nu} \tilde{u}^\nu = \kappa E_{\mu\nu}^{(1)} u^\nu + \bigo{\kappa^2} \equiv \kappa E_\mu^{(1)} + \bigo{\kappa^2} \eqend{.}
\end{equation}
This results in
\begin{splitequation}
\label{eq:constraint_eq_fluid_1st_order}
&\partial_t \bar{\partial}^\alpha \mathscr{h}_{\alpha\mu} - \partial_t \bar{\partial}_\mu \mathscr{h} + 2 (n-2) \epsilon H^2 \bar{\mathscr{v}}_\mu \\
&\quad- u_\mu \left[ (n-2) H \partial_t \mathscr{h} - \bar{\partial}_\alpha \bar{\partial}^\alpha \mathscr{h} + \bar{\partial}_\alpha \bar{\partial}_\beta \mathscr{h}^{\alpha\beta} - (n-1) (n-2) H^2 \mathscr{d} \right] = 0 \eqend{,}
\end{splitequation}
where we have defined $\mathscr{h} \equiv g^{\mu\nu} \mathscr{h}_{\mu\nu}$. The temporal and spatial components of Eq.~\eqref{eq:constraint_eq_fluid_1st_order} are
\begin{equations}[eq:constraint_eq_timespace_fluid_1st_order]
- (n-2) H \partial_t \mathscr{h} + a^{-2} \laplace \mathscr{h} + \partial_i \partial_j \mathscr{h}^{ij} - (n-1) (n-2) H^2 \mathscr{d} = 0 \eqend{,} \label{eq:constraint_eq_time_fluid_1st_order} \\
\partial_t \partial^j \mathscr{h}_{ij} - \partial_i \partial_t \mathscr{h} + 2 (n-2) \epsilon H^2 \mathscr{v}_i = 0 \eqend{,} \label{eq:constraint_eq_space_fluid_1st_order}
\end{equations}
where the indices were raised with the induced background metric $\bar{g}_{\mu\nu}$, and $\laplace$ is the usual Laplace operator in Euclidean space.

On the other hand, the dynamical equation is obtained by projecting both indices of $E^{(1)}_{\mu\nu}$ with the induced background spatial metric. The trace of $E^{(1)}_{\mu\nu}$ with respect to the background metric $g_{\mu\nu}$ gives
\begin{splitequation}
\label{eq:dynamical_eq_fluid_1st_order_trace}
- \partial_t^2 \mathscr{h} - n H \partial_t \mathscr{h} + \bar{\partial}_\alpha \bar{\partial}^\alpha \mathscr{h} - \bar{\partial}_\alpha \bar{\partial}_\beta \mathscr{h}^{\alpha\beta} + (n-1) (n-2\epsilon+2\delta) H^2 \mathscr{d} = 0 \eqend{,}
\end{splitequation}
and we can use this equation to eliminate $\partial_t^2 \mathscr{h}$ from the dynamical equation, which then reads
%\begin{splitequation}
%\label{eq:dynamical_eq_fluid_1st_order}
%&\left[ \partial_t^2 + (n-5) H \partial_t - 2 (n-3-\epsilon) H^2 - \bar{\partial}_\alpha \bar{\partial}^\alpha \right] \mathscr{h}_{\mu\nu} + 2 \bar{\partial}^\alpha \bar{\partial}_{(\mu} \mathscr{h}_{\nu)\alpha} - \bar{\partial}_\mu \bar{\partial}_\nu \mathscr{h} \\
%&\ - \bar{g}_{\mu\nu} \left[ \partial_t^2 \mathscr{h} + (n-1) H \partial_t \mathscr{h} - \bar{\partial}_\alpha \bar{\partial}^\alpha \mathscr{h} + \bar{\partial}_\alpha \bar{\partial}_\beta \mathscr{h}^{\alpha\beta} - (n-2) (n-1-2\epsilon+2\delta) H^2 \mathscr{d} \right] = 0 \eqend{.}
%\end{splitequation}
\begin{splitequation}
\label{eq:dynamical_eq_fluid_1st_order_simpler}
&\partial_t^2 \mathscr{h}_{\mu\nu} + (n-5) H \partial_t \mathscr{h}_{\mu\nu} - 2 (n-3-\epsilon) H^2 \mathscr{h}_{\mu\nu} - a^{-2} \laplace \mathscr{h}_{\mu\nu} \\
&\quad+ 2 \bar{\partial}^\alpha \bar{\partial}_{(\mu} \mathscr{h}_{\nu)\alpha} - \bar{\partial}_\mu \bar{\partial}_\nu \mathscr{h} + \bar{g}_{\mu\nu} \left[ H \partial_t \mathscr{h} - 2 (n-1-\epsilon+\delta) H^2 \mathscr{d} \right] = 0 \eqend{.}
\end{splitequation}
We note that the trace of Eq.~\eqref{eq:dynamical_eq_fluid_1st_order_simpler} does not obviously reduce to Eq.~\eqref{eq:dynamical_eq_fluid_1st_order_trace}, but they agree taking into account that the background metric $g_{\mu\nu}$ is time-dependent.

The assumption that the equation of motion for the fluid is satisfied at the background level also implies that the first-order perturbation of the divergence of the fluid's stress tensor is gauge invariant. The linearised equation of motion for the fluid was given in Eq.~\eqref{eq:eom_matter_1st_order}. After performing the substitutions indicated in Eq.~\eqref{eq:substitutions_fluid_1st_order} in that equation, we obtain
\begin{splitequation}
\label{eq:eom_fluid_1st_order}
&\partial_t \left[ (\nu+1) \bar{\mathscr{v}}_\mu \right] - (n-1) \nu (\nu+1) H \bar{\mathscr{v}}_\mu + c_\text{s}^2 \bar{\partial}_\mu \mathscr{d} \\
&\quad+ \frac{1}{2} u_\mu \left[ \partial_t \mathscr{h} (\nu+1) - 2 (n-1) \mathscr{d} H (\nu - c_\text{s}^2) + 2 (\nu+1) \bar{\partial}_\alpha \bar{\mathscr{v}}^\alpha + 2 \partial_t \mathscr{d} \right] = 0 \eqend{,}
\end{splitequation}
where the parameter $\nu$ was defined in Eq.~\eqref{eq:nu}. The temporal and spatial components of Eq.~\eqref{eq:eom_fluid_1st_order} read
\begin{equations}[eq:eom_fluid_1st_order_timespace]
\partial_t \mathscr{d} + (n-1) (c_\text{s}^2 - \nu) H \mathscr{d} + (\nu+1) \left( \partial_i \mathscr{v}^i + \frac{1}{2} \partial_t \mathscr{h} \right) &= 0 \eqend{,} \\
\partial_t \left[ (\nu+1) \mathscr{v}^i \right] + (\nu+1) [2 - (n-1) \nu] H \mathscr{v}^i + a^{-2} c_\text{s}^2 \partial_i \mathscr{d} = 0 \eqend{.}
\end{equations}

To further simplify the linearised invariant Einstein equation and the fluid's equation of motion, it is convenient to use the scalar-vector-tensor (SVT) decomposition~\cite{lifshitz_grg_2017} and express the invariant metric perturbation and the spatial part of the fluid 4-velocity perturbation in terms of irreducible spatial tensors. Hence, we write
\begin{equations}[eq:svt_decomposition]
\mathscr{h}_{ij} &= \mathscr{H}_{ij}^\mathrm{TT} + 2 \partial_{(i} \mathscr{U}^\mathrm{T}_{j)} + \partial_i \partial_j \mathscr{S} + \delta_{ij} \mathscr{T} \eqend{,} \\
\mathscr{v}_i &= \mathscr{V}_i^\mathrm{T} + \partial_i \mathscr{W} \eqend{,}
\end{equations}
where the spatial tensor and vectors above satisfy
\begin{equation} 
\delta^{ij} \mathscr{H}_{ij}^\mathrm{TT} = 0 \eqend{,} \quad \delta^{ij} \partial_i \mathscr{H}_{jk}^\mathrm{TT} = 0 \quad\text{and}\quad \delta^{ij} \partial_i \mathscr{U}^\mathrm{T}_j = \delta^{ij} \partial_i \mathscr{V}^\mathrm{T}_j = 0 \eqend{.}
\end{equation}
In particular, we have
\begin{equation}
\mathscr{h} = a^{-2} \left[ \laplace \mathscr{S} + (n-1) \mathscr{T} \right]
\end{equation}
and
\begin{equation}
\partial_k \mathscr{h}^k{}_i = a^{-2} \laplace \left( \mathscr{U}_i^\mathrm{T} + \partial_i \mathscr{S} + \laplace^{-1} \partial_i \mathscr{T} \right) \eqend{,}
\end{equation}
where $\laplace^{-1}$ denotes the Green's function of the Laplace operator $\laplace$. The substitution of the decomposition~\eqref{eq:svt_decomposition} into the constraints~\eqref{eq:constraint_eq_time_fluid_1st_order} and~\eqref{eq:constraint_eq_space_fluid_1st_order} results in
\begin{equation}
\label{eq:constraint_eq_time_fluid_1st_order_svt}
\laplace \mathscr{T} - a^2 H \partial_t \left[ \laplace \mathscr{S} + (n-1) \mathscr{T} \right] + 2 a^2 H^2 \left[ \laplace \mathscr{S} + (n-1) \mathscr{T} \right] + (n-1) a^4 H^2 \mathscr{d} = 0
\end{equation}
and
\begin{equation}
\label{eq:constraint_eq_space_fluid_1st_order_svt}
\partial_t \left[ (n-2) a^{-2} \partial_i \mathscr{T} - a^{-2} \laplace \mathscr{U}_i^\mathrm{T} \right] - 2 (n-2) \epsilon H^2 \left( \mathscr{V}^\mathrm{T}_i + \partial_i \mathscr{W} \right) = 0 \eqend{.}
\end{equation}
We can now separate the scalar and transverse parts of Eq.~\eqref{eq:constraint_eq_space_fluid_1st_order_svt} into
\begin{equation}
\label{eq:T_eq}
\partial_t \mathscr{T} - 2 H \mathscr{T} - 2 \epsilon a^2 H^2 \mathscr{W} = 0
\end{equation}
and
\begin{equation}
\label{eq:U_eq}
\partial_t \mathscr{U}_i^\mathrm{T} - 2 H \mathscr{U}_i^\mathrm{T} + 2 (n-2) \epsilon a^2 H^2 \laplace^{-1} \mathscr{V}^\mathrm{T}_i = 0 \eqend{.}
\end{equation}
We can then use the Friedmann equation~\eqref{eq:friedmann_fluid} and Eq.~\eqref{eq:T_eq} to simplify the temporal component~\eqref{eq:constraint_eq_time_fluid_1st_order_svt} of the constraint equations. We obtain
\begin{equation}
\label{eq:T_S_eq}
\laplace \mathscr{T} - a^2 H \laplace( \partial_t \mathscr{S} - 2 H \mathscr{S} ) + (n-1) a^4 H^2 \left( \mathscr{d} - 2 \epsilon H \mathscr{W} \right) = 0
\end{equation}
after using Eq.~\eqref{eq:nu}. The constraint equations~\eqref{eq:T_eq}--\eqref{eq:T_S_eq} completely determine the scalars $\mathscr{S}$ and $\mathscr{T}$ and the transverse spatial vector $\mathscr{U}_i$. To obtain the equation for the transverse traceless spatial tensor $\mathscr{H}_{ij}$, we substitute the SVT decomposition~\eqref{eq:svt_decomposition} into the dynamical equation~\eqref{eq:dynamical_eq_fluid_1st_order_simpler}. This substitution results in
\begin{equation}
\label{eq:H_TT_eq}
\partial_t^2 \mathscr{H}_{ij}^\mathrm{TT} + (n-5) H \partial_t \mathscr{H}_{ij}^\mathrm{TT} - a^{-2} \laplace \mathscr{H}_{ij}^\mathrm{TT} - 2 (n-\epsilon-3) H^2 \mathscr{H}_{ij}^\mathrm{TT} = 0 \eqend{,}
\end{equation}
which is the dynamical part of the metric perturbation, and
\begin{equations}[eq:U_i_S_T_eq_second_order]
\partial_t^2 \mathscr{U}^\mathrm{T}_i + (n-5) H \partial_t \mathscr{U}^\mathrm{T}_i - 2 (n-\epsilon-3) H^2 \mathscr{U}^\mathrm{T}_i &= 0 \eqend{,} \label{eq:U_i_eq_second_order} \\
\partial_t^2 \mathscr{S} + (n-5) H \partial_t \mathscr{S} - 2 (n-\epsilon-3) H^2 \mathscr{S} - a^{-2} (n-3) \mathscr{T} &= 0 \eqend{,} \label{eq:S_eq_second_order} \\
\begin{split}
\partial_t^2 \mathscr{T} - 2 (n-3) H \partial_t \mathscr{T} - a^{-2} \laplace \mathscr{T} + 2 (2n-4-\epsilon) H^2 \mathscr{T} &\\
\quad+ H \laplace \left( \partial_t \mathscr{S} - 2 H \laplace \mathscr{S} \right) - 2 a^2 (n-1-\epsilon+\delta) H^2 \mathscr{d} &= 0 \eqend{.} \label{eq:S_T_eq_second_order}
\end{split}
\end{equations}
Eqs.~\eqref{eq:U_i_S_T_eq_second_order} are not independent from the constraint equations~\eqref{eq:T_eq}--\eqref{eq:T_S_eq} and can be obtained from them if the matter perturbations satisfy their equation of motion. The SVT decomposition for the fluid's equation of motion~\eqref{eq:eom_fluid_1st_order_timespace} gives
\begin{equation}
\label{eq:eom_d_SVT}
\partial_t \mathscr{d} - 2 \delta H \mathscr{d} + \frac{\epsilon}{n-1} a^{-2} \left\{ 2 \laplace \mathscr{W} + \left( \partial_t - 2 H \right) [ \laplace \mathscr{S} + (n-1) \mathscr{T} ] \right\} = 0 \eqend{,}
\end{equation}
\begin{equation}
\label{eq:W_eq}
\partial_t \mathscr{W} + \frac{n-1-2\epsilon+2\delta}{2 \epsilon} \left( 2 \epsilon H \mathscr{W} - \mathscr{d} \right) = 0 \eqend{,}
\end{equation}
and
\begin{equation}
\partial_t \mathscr{V}^\mathrm{T}_i + (n-1-2\epsilon+2\delta) H \mathscr{V}^\mathrm{T}_i = 0 \eqend{,}
\end{equation}
after we have expressed the fluid background parameters in terms of the background geometry parameters.

The equations for the metric and fluid perturbations in synchronous gauge were previously studied in Ref.~\cite{matarrese_mollerach_bruni_prd_1998}, assuming an irrotational co-moving fluid and a power-law expansion of the spacetime. We have compared their results for the SVT decomposition at linear order with the ones obtained in this section. Using the dictionary
\begin{splitequation}
&\mathscr{H}_{ij}^\mathrm{T} \to a^2 \chi_\mathrm{S}^{(1) \mathrm{T}} \eqend{,} \quad \mathscr{U}_i^\mathrm{T} \to a^2 \chi_\mathrm{S}^{(1)\bot} \eqend{,} \quad \mathscr{S} \to a^2 \chi_\mathrm{S}^{(1)\|} \eqend{,} \\
&\mathscr{T} \to - a^2 \left( 2 \phi_\mathrm{S}^{(1)} + \frac{1}{3} \laplace \chi_\mathrm{S}^{(1)\|} \right) \eqend{,} \quad \mathscr{v}_i \to 0 \eqend{,}
\end{splitequation}
and expressing everything in conformal time, it is not difficult to show that their equations for the perturbations in the SVT decomposition exactly match ours. This is an example of the fact that, as explained in Sec.~\ref{sec:relational_observables}, for the gauge-fixing condition~\eqref{eq:gauge_condition} the dynamical coordinates $X^{(\mu)}(x)$ coincide with the background coordinates $x^\mu$, and thus the invariant fields coincide with the gauge-fixed ones.

\subsubsection{Second order}
%%%%%%%%%%%%%%%%%%%%%%%%%%%%%%%%%%%%%%%%%%%%%%%%%%%%%%%%%%%%%

Our task now is to express the second-order correction to the invariant Einstein equation~\eqref{eq:invariant_Einstein_eq} in terms of the gauge-invariant perturbation fields $\mathscr{h}_{\mu\nu}$, $\mathscr{v}_\mu$ and $\mathscr{d}$. In order to do that, we first need to express the gauge-variant perturbation fields $h_{\mu\nu}$, $v_\mu$ and $d$ as power series in their invariant parts, up to first order in perturbation theory.

For $h_{\mu\nu}$, we express Eq.~\eqref{eq:invariant_metric_perturbation_expansion} as
\begin{equation}
\label{eq:h_variant_invariant_2nd_order}
h_{\mu\nu} = \mathscr{h}_{\mu\nu} + \lie_{X_{(1)}} g_{\mu\nu} - \kappa \mathscr{g}^{(2)}_{\mu\nu} \big\rvert_{h_{\mu\nu} = \mathscr{h}_{\mu\nu} + \lie_{X_{(1)}} g_{\mu\nu}} + \bigo{\kappa^2} \eqend{,}
\end{equation}
where we have used Eq.~\eqref{eq:invariant_metric_corrections_1} and indicated the replacement of $h_{\mu\nu}$ in all instances it appears explicitly in Eq.~\eqref{eq:invariant_metric_corrections_2b} by its zeroth order expression. For the four-velocity perturbation $v_\mu$, we express its temporal and spatial components in terms of the spatial components of the invariant four-velocity perturbation. The expressions for the temporal component are obtained by substituting $v_i = \mathscr{v}_i - \partial_i t_{(1)}$ --- see Eqs.~\eqref{eq:invariant_4_vel_corrections_components_1_i} and~\eqref{eq:invariant_4_vel_perturbation} --- into Eqs.~\eqref{eq:time_component_4_vel}, and they read
\begin{equations}[eq:v_t_variant_invariant_2nd_order]
v_t^{(1)} &= \frac{1}{2} h_{tt} \eqend{,} \\
v_t^{(2)} &= - \frac{1}{2} \mathscr{v}^i \mathscr{v}_i + \left( \partial^i t_{(1)} + h_t{}^i \right) \mathscr{v}_i - \frac{1}{2} \left( \partial^i t_{(1)} + 2 h_t{}^i \right) \partial_i t_{(1)} + \frac{1}{8} h_{tt}^2 - \frac{1}{2} h_t{}^i h_{ti} + \bigo{\kappa} \eqend{.}
\end{equations}
The expression for the spatial components is obtained by expressing Eq.~\eqref{eq:invariant_4_vel_perturbation} as
\begin{splitequation}
\label{eq:v_i_variant_invariant_2nd_order}
v_i &= \mathscr{v}_i - \partial_i t_{(1)} - \kappa \mathscr{V}_i^{(2)} \big\rvert_{v_i = \mathscr{v}_i - \partial_i t_{(1)}} + \bigo{\kappa^2} \\
&= \mathscr{v}_i - \partial_i t_{(1)} + \kappa \left( t_{(1)} \partial_t \mathscr{v}_i + x_{(1)}^j \partial_j \mathscr{v}_i + \partial_i x_{(1)}^j \mathscr{v}_j - \partial_i t_{(2)}
\right) + \bigo{\kappa^2} \eqend{,}
\end{splitequation}
where we have used Eq.~\eqref{eq:invariant_4_vel_corrections_components_2_t} and replaced $v_i$ by its zeroth-order expression in all instances it appears explicitly in $\mathscr{V}^{(2)}_i$. Finally, we use Eqs.~\eqref{eq:invariant_rho}--\eqref{eq:invariant_d} to express the gauge-variant fractional density perturbation $d$ as
\begin{splitequation}
\label{eq:d_variant_invariant_2nd_order}
d &= \mathscr{d} + \frac{\dot{\rho}}{\rho} t_{(1)} - \kappa \frac{\uprho_{(2)}}{\rho} \Big\rvert_{d \to \mathscr{d} + \frac{\dot{\rho}}{\rho} t_{(1)}} + \bigo{\kappa^2} \\
&= \mathscr{d} + \frac{\dot{\rho}}{\rho} t_{(1)} + \kappa \left( t_{(1)} \partial_t \mathscr{d} + \frac{\dot{\rho}}{\rho} t_{(1)} \mathscr{d} + x_{(1)}^i \partial_i \mathscr{d} + \frac{1}{2} \frac{\ddot{\rho}}{\rho} t_{(1)}^2 + \frac{\dot{\rho}}{\rho} t_{(2)} \right) + \bigo{\kappa^2} \eqend{,}
\end{splitequation}
and as in previous cases we replace all $d$ appearing in the expression for $\uprho_{(2)}$ by its zeroth-order expression.

Our next step is to substitute Eqs.~\eqref{eq:h_variant_invariant_2nd_order}--\eqref{eq:d_variant_invariant_2nd_order} into the perturbative expansion for the invariant Einstein equation~\eqref{eq:invariant_Einstein_eq}. The second-order terms in $\kappa$ are produced by the substitution of the zeroth-order terms in Eqs.~\eqref{eq:h_variant_invariant_2nd_order}--\eqref{eq:d_variant_invariant_2nd_order} into $E_{\mu\nu}^{(2)} - X^{(\rho)}_{(1)} \partial_\rho E_{\mu\nu}^{(1)} - \partial_\mu X^{(\rho)}_{(1)} E^{(1)}_{\rho\nu} - \partial_\nu X^{(\rho)}_{(1)}E^{(1)}_{\rho\mu}$ and by the substitution of their first-order terms into $E^{(1)}_{\mu\nu}$. The tensor algebra involved in the simplifications of the resulting expression is best handled with the help of the tensor algebra package \textsc{xAct}~\cite{xact}. The final expression is very long and given in Eq.~\eqref{eq:invariant_einstein_eq_2nd_order_fluid}.

We can now obtain the second-order correction to the invariant constraint equation. To do so, we first define the invariant part of the observer's 4-velocity in the perturbed spacetime:
\begin{equation}
\label{eq:inv_obs_4_vel}
\mathscr{u}^\mu(X) \equiv \left(\frac{\partial X^{(\mu)}}{\partial x^\nu} \tilde{u}^\nu\right)[x(X)] \eqend{.}
\end{equation}
We then notice that Eqs.~\eqref{eq:perturbed_four_velocity} and~\eqref{eq:synchronous_condition_perturbed} imply that
\begin{equation}
\label{eq:invariant_u}
\mathscr{u}^t = \tilde{u}^\nu \tilde{u}_\nu = -1 \quad\text{and}\quad \mathscr{u}^i = \tilde{u}^\nu \partial_\nu \tilde{x}^i = 0 \eqend{,}
\end{equation}
so $\mathscr{u}^\mu = u^\mu$. The invariant constraint equation at second order can now be obtained from the expansion of
\begin{equation}
\mathscr{E}_\nu \equiv \mathscr{u}^\mu \mathscr{E}_{\mu\nu} = u^\mu \mathscr{E}_{\mu\nu} = \kappa \mathscr{E}^{(1)}_\mu + \kappa^2 \mathscr{E}^{(2)}_\mu + \bigo{\kappa^3} \eqend{.}
\end{equation}
The temporal and spatial components of the second-order correction to the invariant constraint equation are
\begin{splitequation}
\mathscr{E}_t^{(2)} &= - \frac{1}{4} \partial_t \mathscr{h}_{ij} \partial_t \mathscr{h}^{ij} + \frac{1}{4} \left( \partial_t \mathscr{h} \right)^2 - \frac{1}{4} \partial_i \mathscr{h} \partial^i \mathscr{h} + \partial^i \mathscr{h} \partial_j \mathscr{h}_i{}^j - \partial_i \mathscr{h}^{ij} \partial_k \mathscr{h}_j{}^k - \frac{1}{2} \partial_j \mathscr{h}_{ik}\partial^k \mathscr{h}^{ij}\\
&\quad+ \frac{3}{4} \partial_k \mathscr{h}_{ij} \partial^k \mathscr{h}^{ij} + \mathscr{h}^{ij} \partial_i \partial_j \mathscr{h} - 2 \mathscr{h}^{ij} \partial_j \partial_k \mathscr{h}_i{}^k + \mathscr{h}^{ij} \laplace \mathscr{h}_{ij} - (n-3) H \mathscr{h}^{ij} \partial_t \mathscr{h}_{ij} \\
&\quad+ (2n-5) H^2 \mathscr{h}_{ij} \mathscr{h}^{ij} - 2 (n-2) \epsilon H^2 \mathscr{v}_i \mathscr{v}^i \raisetag{1.4em}
\end{splitequation} 
and
\begin{splitequation}
\mathscr{E}_i^{(2)} &= \frac{1}{2} \partial_t\mathscr{h}_{ij} \partial^j\mathscr{h} -  H\mathscr{h}_{ij} \partial^j\mathscr{h} -  \partial_t\mathscr{h}_i{}^j \partial_k\mathscr{h}_j{}^k + 2H \mathscr{h}_i{}^j\partial_k\mathscr{h}_j{}^k + 2 H\mathscr{h}^{jk}\partial_k\mathscr{h}_{ij} - \mathscr{h}^{jk} \partial_k\partial_t\mathscr{h}_{ij}\\
&\quad + \frac{1}{2} \partial_t\mathscr{h}^{jk}\partial_i\mathscr{h}_{jk} - 3 \mathscr{h}^{jk} H\partial_i\mathscr{h}_{jk} + \mathscr{h}^{jk}\partial_i\partial_t\mathscr{h}_{jk}  + 2(n-2)(\epsilon - \delta) H^2 \mathscr{v}_i \mathscr{d} \eqend{.} \raisetag{1.8em}
\end{splitequation}
The second-order correction to the dynamical equation is again obtained by projecting both indices of Eq.~\eqref{eq:invariant_einstein_eq_2nd_order_fluid} on the background spatial section. The result is
\begin{splitequation}
\mathscr{E}^{(2)}_{ij} &= - \partial_t\mathscr{h}_i{}^k \partial_t\mathscr{h}_{jk} + \frac{1}{2} \partial_t\mathscr{h}_{ij} \partial_t\mathscr{h} + 4H \mathscr{h}_{(i}{}^k \partial_t\mathscr{h}_{j)k} - 4H^2 \mathscr{h}_i{}^k \mathscr{h}_{jk} - \frac{1}{2} \partial_k\mathscr{h}_{ij} \partial^k\mathscr{h}\\
&\quad + \partial^k\mathscr{h}_{ij} \partial_l\mathscr{h}_k{}^l - \partial_k\mathscr{h}_{il}\partial^l\mathscr{h}_j{}^k + \partial_k\mathscr{h}_{il} \partial^k\mathscr{h}_j{}^l + \partial^k\mathscr{h} \bar{\partial}_{(i}\mathscr{h}_{j)k} - 2\partial_k\mathscr{h}_l{}^k \partial_{(i}\mathscr{h}_{j) }{}^l\\
&\quad + \frac{1}{2} \partial_i\mathscr{h}^{kl} \partial_j\mathscr{h}_{kl} + \mathscr{h}^{kl} \partial_k\partial_l\mathscr{h}_{ij} -  2\mathscr{h}^{kl}\partial_k \partial_{(i}\mathscr{h}_{j)l} + \mathscr{h}^{kl} \partial_i\partial_j\mathscr{h}_{kl} - 2(n-2)\epsilon H^2 \bar{\mathscr{v}}_i \bar{\mathscr{v}}_j\\
&\quad - \mathscr{h}_{ij} \Big[\partial_t^{2}\mathscr{h} + nH \partial_t\mathscr{h} + \partial_k\partial_l\mathscr{h}^{kl} - a^{-2}\laplace\mathscr{h} - (n-2)(n - 1 - 2\epsilon + 2\delta) H^2 \mathscr{d}\Big] \\
&\quad + \bar{g}_{ij} \bigg[\frac{3}{4} \partial_t\mathscr{h}_{kl} \partial_t\mathscr{h}^{kl} -  \frac{1}{4} \left(\partial_t\mathscr{h}\right)^2 + \mathscr{h}^{kl} \partial_t^2\mathscr{h}_{kl} + \frac{1}{4} \partial_k\mathscr{h}\partial^k\mathscr{h} - \partial_k\mathscr{h}\partial_l\mathscr{h}^{kl} + \partial_k\mathscr{h}^{kl}\partial_m\mathscr{h}_l{}^m\\
&\quad + \frac{1}{2}\partial_k\mathscr{h}_{lm}\partial^m\mathscr{h}^{kl} - \frac{3}{4} \partial_m\mathscr{h}_{kl} \partial^m\mathscr{h}^{kl} -  \mathscr{h}^{kl} \partial_k\partial_l\mathscr{h} + 2 \mathscr{h}^{kl} \partial_k\partial_m\mathscr{h}_l{}^m - a^{-2}\mathscr{h}^{kl}\laplace\mathscr{h}_{kl}\\
&\quad + (n - 8)H \mathscr{h}^{kl} \partial_t\mathscr{h}_{kl} - (2n - 9 - 2\epsilon)H^2 \mathscr{h}_{kl} \mathscr{h}^{kl} + (n-2)\left(1-\frac{\dot{\delta}}{2\epsilon\delta H}\right) \delta H^2\mathscr{d}^2\bigg] \eqend{.}
\end{splitequation}

To obtain the gauge-invariant part of the fluid's equation of motion in terms of the invariant perturbation fields, we must perform the same substitutions as in the case of the Einstein equation. Thus, we use Eqs.~\eqref{eq:h_variant_invariant_2nd_order}--\eqref{eq:d_variant_invariant_2nd_order} to eliminate the variant fields $h_{\mu\nu}$, $v_\mu$ and $d$ in Eq.~\eqref{eq:invariant_eom_matter}. The second order terms are again produced by substituting the zeroth-order terms of Eqs.~\eqref{eq:h_variant_invariant_2nd_order}--\eqref{eq:d_variant_invariant_2nd_order} into $F^{(2)}_\mu - X_{(1)}^{(\sigma)} \partial_\sigma F^{(1)}_\mu - \partial_\mu X_{(1)}^{(\sigma)} F^{(1)}_\sigma$ and the first-order terms in those equations into $F^{(1)}_\mu$, where $F^{(1)}_\mu$ and $F^{(2)}_\mu$ were given in Eqs.~\eqref{eq:eom_matter_1st_order} and~\eqref{eq:eom_matter_2nd_order}. The result is
\begin{splitequation}
\mathscr{F}^{(2)}_\mu &= (n-2)H^2 \bar{\mathscr{v}}_{\mu } \bigg[2 (\epsilon - \delta) \partial_t\mathscr{d} + \epsilon\partial_t\mathscr{h} + 2 \left((n - 1 - 2\epsilon)(\epsilon - \delta) + 2\epsilon\delta - \frac{\dot{\delta}}{H}\right)H\mathscr{d} \\
&\quad+ 2 \epsilon \bar{\partial}_{\alpha }\bar{\mathscr{v}}^{\alpha } \bigg] + \frac{(n-2) H^2}{2}u_{\mu } \bigg[(1 + 2\epsilon - 2\delta) \partial_t\mathscr{h} \mathscr{d} - 2 \epsilon \mathscr{h}^{\alpha \beta } (\partial_t - 2 H)\mathscr{h}_{\alpha \beta } \\
&\quad+ 4 (\epsilon - \delta) \mathscr{d}\bar{\partial}_{\alpha }\bar{\mathscr{v}}^{\alpha } - 4\epsilon \mathscr{h}_{\alpha \beta } \bar{\partial}^{\beta }\bar{\mathscr{v}}^{\alpha } - 2(n-1)\left(1 - \frac{\dot{\delta}}{2\epsilon\delta H}\right)\delta H\mathscr{d}^2 \\
&\quad+ 4 \bar{\mathscr{v}}^{\alpha } \left(2\epsilon \partial_t\bar{\mathscr{v}}_{\alpha } + (n - 2 + 2 \delta - 2 \epsilon) \epsilon H\bar{\mathscr{v}}_{\alpha } + \frac{1}{2}\epsilon \bar{\partial}_{\alpha }\mathscr{h} - (\delta - \epsilon) \bar{\partial}_{\alpha }\mathscr{d} - \epsilon \bar{\partial}_{\beta }\mathscr{h}_{\alpha }{}^{\beta }\right) \bigg] \\
&\quad+ 2(n-2) H^2 \left[ (\epsilon - \delta)\partial_t\bar{\mathscr{v}}_{\mu } \mathscr{d} + \epsilon \bar{\mathscr{v}}^{\alpha }\bar{\partial}_{\alpha }\bar{\mathscr{v}}_{\mu } - \left(1 - \frac{\dot{\delta}}{2 \epsilon\delta H}\right)\delta\mathscr{d} \bar{\partial}_{\mu }\mathscr{d} \right] \eqend{.} \raisetag{2.2em}
\end{splitequation}
The equations that we have obtained in this section extend the second-order analysis of Ref.~\cite{matarrese_mollerach_bruni_prd_1998} to arbitrary cosmological backgrounds and a general flow of the fluid. At this point, we could again use the SVT decomposition~\eqref{eq:svt_decomposition} to express the invariant perturbation fields in terms of irreducible spatial tensors. Unfortunately, this decomposition does not lead to any major simplifications in the second-order corrected invariant Einstein equation nor in the fluid's equation of motion. It is possible that reparametrisations of the invariant fields, such as the ones performed in Ref.~\cite{prokopec_weenink_jcap_2013} in the case of a non-minimally coupled scalar field in the Jordan frame, could simplify our equations. Pursuing this reparametrisation, however, is beyond the scope of this paper.

\subsection{Scalar field model}
\label{sec:scalar_field}
%%%%%%%%%%%%%%%%%%%%%%%%%%%%%%%%%%%%%%%%%%%%%%%%%%%%%%%%%%%%%%%%%%%%%%%%%%%%%%%%%%%%%%%%%%%%%%%%%%%%%%%%%%%%%%%%%%%%%%%%%%%%%%%%%%%%%%%%%%%%%%%%%%%%%

Now, let us consider the perturbed Einstein equation sourced by a perturbed scalar field 
\begin{equation}
\tilde{\phi} = \phi + \kappa \phi^{(1)} \eqend{.}
\end{equation}
The scalar field's stress tensor is given by
\begin{equation}
\tilde{T}_{\mu\nu} = \nabla_\mu \tilde{\phi} \nabla_\nu \tilde{\phi} - \frac{1}{2} \tilde{g}_{\mu\nu} \tilde{g}^{\rho\sigma} \nabla_\rho \tilde{\phi} \nabla_\sigma \tilde{\phi} - \frac{1}{2} \tilde{g}_{\mu\nu} V(\tilde{\phi}) \eqend{,}
\end{equation}
where $V$ is the scalar potential. The perturbative expansion of $\tilde{T}_{\mu\nu}$ in this case reads
\begin{equations}[eq:expansion_field_stress_tensor]
T_{\mu\nu} &= u_\mu u_\nu \dot\phi^2 + \frac{1}{2} g_{\mu\nu} \left[ \dot\phi^2 - V(\phi) \right] \eqend{,} \\
\begin{split}
T_{\mu\nu}^{(1)} &= - \dot\phi \left( u_\mu \partial_\nu \phi^{(1)} + u_\nu \partial_\mu \phi^{(1)} \right) + g_{\mu\nu} \left[ \dot\phi u^\rho \partial_\rho \phi^{(1)} + \frac{1}{2} \dot\phi^2 h_{\rho\sigma} u^\rho u^\sigma - \frac{1}{2} V'(\phi) \phi^{(1)} \right] \\
&\quad+ \frac{1}{2} h_{\mu\nu} \left[ \dot\phi^2 - V(\phi) \right] \eqend{,}
\end{split} \\
\begin{split}
T_{\mu\nu}^{(2)} &= \partial_\mu \phi^{(1)} \partial_\nu \phi^{(1)} + h_{\mu\nu} \left[ \dot\phi u^\rho \partial_\rho \phi^{(1)} + \frac{1}{2} \dot\phi h_{\rho\sigma} u^\rho u^\sigma - \frac{1}{2} V'(\phi) \phi^{(1)} \right] \\
&\quad- \frac{1}{2} g_{\mu\nu} \left[ \partial^\rho \phi^{(1)} \partial_\rho \phi^{(1)} + 2 \dot\phi h^{\rho\sigma} u_\rho \partial_\sigma \phi^{(1)} + \dot\phi^2 h_\rho{}^\lambda h_{\lambda\sigma} u^\rho u^\sigma + \frac{1}{2} V''(\phi) \phi^{(1)2} \right] \eqend{.}
\end{split}
\end{equations}

Next, we consider the perturbed scalar field equation
\begin{equation}
\label{eq:perturbed_field_equation}
\tilde{F} \equiv - \tilde{g}^{\mu\nu} \tilde{\nabla}_\mu \tilde{\nabla}_\nu \tilde{\phi} + \frac{1}{2} V'(\tilde{\phi}) \eqend{.}
\end{equation}
The perturbative expansion of this field equation is
\begin{equation}
\tilde{F} = F + \kappa F^{(1)} + \kappa^2 F^{(2)} + \bigo{\kappa^3}
\end{equation}
where the first-order correction $F^{(1)}$ reads
\begin{equation}
\label{eq:expansion_field_equation}
F^{(1)} = - \nabla^2 \phi^{(1)} + h^{\mu\nu} \nabla_\mu \nabla_\nu\phi + \left( \nabla_\rho h^{\rho\sigma} + \frac{1}{2} \nabla^\sigma h \right) \nabla_\sigma \phi + \frac{1}{2} V''(\phi) \phi^{(1)} \eqend{,}
\end{equation}
and the second-order correction $F^{(2)}$ is given in Eq.~\eqref{eq:expansion_field_equation_2}.

As in the ideal-fluid example, here we also need the invariant part of the scalar field. Since it is a scalar, we define the invariant part of $\tilde{\phi}$ as
\begin{equation}
\label{eq:Phi}
\Phi(X) \equiv \tilde{\phi}[x(X)] = \phi + \kappa \upphi = \phi + \kappa \Phi^{(1)} + \kappa^2 \Phi^{(2)} + \bigo{\kappa^3} \eqend{,}
\end{equation}
where $\upphi$ is the invariant scalar field perturbation and the coefficients of the expansion of the invariant full scalar field are
\begin{equations}[eq:expansion_Phi]
\Phi^{(1)} &= \phi^{(1)} - \dot\phi t_{(1)} \eqend{,} \\
\Phi^{(2)} &= - t_{(1)} \partial_t \phi^{(1)} - x^i_{(1)} \partial_i \phi^{(1)} + \frac{1}{2} \ddot\phi t_{(1)}^2 + \dot\phi \left( x^i_{(1)} \partial_i t_{(1)} - \frac{1}{2} t_{(1)} h_{tt} - t_{(2)} \right) \eqend{,} \label{eq:expansion_Phi_2}
\end{equations}
where we have used Eqs.~\eqref{eq:expansion_invariant_scalar} and~\eqref{eq:coord_correction_first_order} to eliminate the time derivatives of the coordinate corrections. Similarly, the gauge-invariant part of the scalar field equation is
\begin{equation}
\label{eq:invariant_field_equation}
\mathscr{F}(X) \equiv \tilde{F}[x(X)] = \kappa F^{(1)} + \kappa^2 \left( F^{(2)} - X^{(\sigma)}_{(1)} \partial_\sigma F^{(1)} \right) + \bigo{\kappa^3} \eqend{,}
\end{equation}
where we have assumed that the background field equation $F = 0$ is satisfied.

\subsubsection{Background}
%%%%%%%%%%%%%%%%%%%%%%%%%%%%%%%%%%%%%%%%%%%%%%%%%%%%%%%%%%%%%

The Friedmann equations for the scalar fields can be written as
\begin{equations}[eq:Friedmann_eqs_field]
- 2 (n-2) (n-1-2\epsilon) H^2 &= \kappa^2 \left[ \dot\phi^2 - V(\phi) \right] \eqend{,} \\
2 (n-2) \epsilon H^2 &= \kappa^2 \dot\phi^2 \eqend{.}
\end{equations}
The equation of motion for the background scalar field $\phi$ can be obtained from Eqs.~\eqref{eq:Friedmann_eqs_field} simply by taking the derivative of the second equation with respect to time. The result is
\begin{equation}
\label{eq:phi_eq_background}
\ddot\phi + (\epsilon-\delta) H \dot\phi = 0 \eqend{.}
\end{equation}
For later use, we also record the equations satisfied by the scalar potential and its derivatives. We have
\begin{equations}[eq:scalar_potential_derivatives]
V(\phi) &= 2 \kappa^{-2} (n-2) (n-1-\epsilon) H^2 \eqend{,} \\
V'(\phi) &= - 2 (n-1-\epsilon+\delta) H \dot\phi \eqend{,} \\
V''(\phi) &= 2 \left[ (n-1-\epsilon+\delta) (2\epsilon-\delta) + 2 \epsilon \delta - \frac{\dot\delta}{H} \right] H^2 \eqend{.}
\end{equations}

\subsubsection{First order}
%%%%%%%%%%%%%%%%%%%%%%%%%%%%%%%%%%%%%%%%%%%%%%%%%%%%%%%%%%%%%

As for the fluid case, to obtain the invariant Einstein equation at first order it is enough to make the substitutions
\begin{equation}
\label{eq:substitutions_field_1st_order}
h_{\mu\nu} \to \mathscr{h}_{\mu\nu} \quad\text{and}\quad \phi^{(1)} \to \upphi
\end{equation}
into $E^{(1)}_{\mu\nu}$ given in Eqs.~\eqref{eq:expansion_einstein_eq}, with the stress tensor expansion~\eqref{eq:expansion_field_stress_tensor}. We can then write down the constraint and dynamical equations for the perturbations.

In the scalar field case, the constraint equation~\eqref{eq:constraint_eq_linear} reduces to
\begin{splitequation}
\label{eq:constraint_eq_field_1st_order}
&\partial_t \bar{\partial}^\alpha \mathscr{h}_{\alpha\nu} - \partial_t \bar{\partial}_\nu \mathscr{h} - \kappa^2 \dot{\phi} \bar{\partial}_\nu \upphi \\
&\quad- u_\nu \left[ (n-2) H \partial_t \mathscr{h} - \bar{\partial}_\alpha \bar{\partial}^\alpha \mathscr{h} + \bar{\partial}_\alpha \bar{\partial}_\beta \mathscr{h}^{\alpha\beta} - \kappa^2 \dot{\phi} \left( \partial_t \upphi - (n-1-\epsilon+\delta) H \upphi \right) \right] = 0 \eqend{,}
\end{splitequation}
where we have used that $u^\mu \mathscr{h}_{\mu\nu} = 0$ and the $(3+1)$-decomposition of tensors using the induced spatial metric~\eqref{eq:induced_spatial_metric}. Its temporal and spatial components are
\begin{equation}
\label{eq:constraint_eq_field_1st_order_time}
- (n-2) H \partial_t \mathscr{h} + a^{-2} \laplace \mathscr{h} - \partial_i \partial_j \mathscr{h}^{ij} + \kappa^2 \dot{\phi} \left[ \partial_t \upphi - (n-1-\epsilon+\delta) H \upphi \right] = 0
\end{equation}
and
\begin{equation}
\label{eq:constraint_eq_field_1st_order_space}
\partial_t \partial^j \mathscr{h}_{ij} - \partial_t \partial_i \mathscr{h} - \kappa^2 \dot{\phi} \partial_i \upphi = 0 \eqend{.}
\end{equation}
The dynamical equation is again obtained by projecting both indices of $E^{(1)}_{\mu\nu}$ on the spatial section. The result is
\begin{splitequation}
\label{eq:dynamical_eq_field_1st_order}
&\partial_t^2 \mathscr{h}_{\mu\nu} + (n-5) H \partial_t \mathscr{h}_{\mu\nu} - 2(n - 3 - \epsilon) H^2\mathscr{h}_{\mu \nu } -  \bar{\partial}_{\alpha }\bar{\partial}^{\alpha }\mathscr{h}_{\mu \nu } + 2\bar{\partial}_{\alpha }\bar{\partial}_{(\mu }\mathscr{h}_{\nu) }{}^{\alpha } -  \bar{\partial}_{\mu }\bar{\partial}_{\nu }\mathscr{h}\\
&\quad+ \bar{g}_{\mu \nu } \left[- \partial_t^2\mathscr{h} - (n-1) H\partial_t\mathscr{h} + \bar{\partial}_{\alpha }\bar{\partial}^{\alpha }\mathscr{h} -  \bar{\partial}_{\beta }\bar{\partial}_{\alpha }\mathscr{h}^{\alpha \beta } -  \kappa^2\dot{\phi}\left(\partial_t\upphi + (n - 1 - \epsilon + \delta) H \upphi\right)\right]\\
&\quad= 0 \eqend{.} \raisetag{1.4em}
\end{splitequation}
Taking the trace of this equation and then using it to eliminate $\partial_t^2 \mathscr{h}$ reduces Eq.~\eqref{eq:dynamical_eq_field} to
\begin{splitequation}
\label{eq:dynamical_eq_field}
&\partial_t^2 \mathscr{h}_{ij} + (n-5) H \partial_t \mathscr{h}_{ij} - 2 (n-3-\epsilon) H^2 \mathscr{h}_{ij} - a^{-2} \laplace \mathscr{h}_{ij} + 2 \partial^k \partial_{(i} \mathscr{h}_{j)k} \\
&\quad- \partial_i \partial_j \mathscr{h} + \bar{g}_{ij} \left[ H \partial_t \mathscr{h} + \kappa^2 (n-2) \dot{\phi} \left( \partial_t \upphi + (n-1-\epsilon+\delta) H \upphi \right) \right] = 0 \eqend{,}
\end{splitequation}
where we also used the constraint equation~\eqref{eq:constraint_eq_field_1st_order_time}. To obtain the invariant field equation at first order, we perform the substitutions~\eqref{eq:substitutions_field_1st_order} in the first-order correction to the field equation $F^{(1)}$ given in Eq.~\eqref{eq:expansion_field_equation}. The invariant scalar field perturbation $\upphi$ then satisfies
\begin{splitequation}
\label{eq:upphi_eq}
&\partial_t^2 \upphi + (n-1) H \partial_t \upphi - a^{-2} \laplace \upphi \\
&\quad+ \left[ (n-1-\epsilon+\delta) (2 \epsilon - \delta) + 2 \epsilon \delta - \frac{\dot{\delta}}{H} \right] H^2 \upphi + \frac{1}{2} \dot{\phi} \partial_t \mathscr{h} = 0 \eqend{.}
\end{splitequation}

To further simplify the constraint and dynamical equations, we again perform the SVT decomposition~\eqref{eq:svt_decomposition} for the invariant metric perturbation. For the spatial components of the constraint equations~\eqref{eq:constraint_eq_field_1st_order_space}, we obtain
\begin{equation}
\label{eq:T_eq_field}
\partial_t \mathscr{T} - 2 H \mathscr{T} + \frac{\kappa^2}{n-2} a^2 \dot{\phi} \upphi = 0
\end{equation}
and
\begin{equation}
\label{eq:U_eq_field}
\partial_t \mathscr{U}^\text{T}_i - 2 H \mathscr{U}^\text{T}_i = 0 \eqend{,}
\end{equation}
and for the temporal component~\eqref{eq:constraint_eq_field_1st_order_time} we find
\begin{splitequation}
\label{eq:T_S_constraint_field}
\laplace \mathscr{T} - H a^2 \laplace \left( \partial_t \mathscr{S} + 2 H \mathscr{S} \right) + \frac{\kappa^2}{n-2} a^4 \dot{\phi} \left[ \partial_t \upphi + (\epsilon-\delta) H \upphi \right] = 0
\end{splitequation}
after using Eq.~\eqref{eq:T_eq_field}. The SVT decomposition of the dynamical equation~\eqref{eq:dynamical_eq_field} gives again Eqs.~\eqref{eq:H_TT_eq}--\eqref{eq:S_eq_second_order}, while the scalar field equation of motion reads
\begin{splitequation}
\label{eq:eom_upphi}
&\partial_t^2 \upphi + (n-1+\epsilon) H \partial_t \upphi - a^{-2} \laplace \upphi \\
&\quad+ \left[ (n-1-\epsilon+\delta) (\epsilon-\delta) + 2 \epsilon \delta - \frac{\dot\delta}{H} \right] H^2 \upphi + \frac{\dot{\phi}}{2 H a^4} \laplace \mathscr{T} = 0 \eqend{,}
\end{splitequation}
after using the constraint equations~\eqref{eq:T_eq_field} and~\eqref{eq:T_S_constraint_field}.

Note that if we take a time derivative of the constraint equation~\eqref{eq:T_S_constraint_field}, the resulting equation holds by virtue of the other equations of motion. On the other hand, if we express $\upphi$ in terms of $\mathscr{T}$ using Eq.~\eqref{eq:T_eq_field}, this equation turns into a dynamical equation for $\mathscr{T}$:
\begin{equation}
\partial_t^2 \mathscr{T} - 2 (2-\epsilon+\delta) H \partial_t \mathscr{T} - a^{-2} \laplace \mathscr{T} + 2 (2-\epsilon+2\delta) H^2 \mathscr{T} + H \laplace \left( \partial_t \mathscr{S} - 2 H \mathscr{S} \right) = 0 \eqend{,}
\end{equation}
and we could use this equation of motion to replace the equation of motion for $\upphi$. Moreover, the combination
\begin{equation}
\label{eq:sasaki_mukhanov_variable}
Q \equiv \frac{2 H}{\dot\phi} \upphi - a^{-2} \mathscr{T}
\end{equation}
satisfies the source-free equation
\begin{equation}
\label{eq:eom_sasaki_mukhanov_variable}
\partial_t^2 Q + (n-1+2\delta) H \partial_t Q - a^{-2} \laplace Q = 0 \eqend{,}
\end{equation}
and we can identify $Q$ as the Sasaki--Mukhanov variable~\cite{sasaki_ptp_1986,mukhanov_zetf_1988}.

\subsubsection{Second order}
%%%%%%%%%%%%%%%%%%%%%%%%%%%%%%%%%%%%%%%%%%%%%%%%%%%%%%%%%%%%%

We again replace the gauge-variant perturbations $h_{\mu\nu}$ and $\phi^{(1)}$ by their gauge-invariant parts using Eq.~\eqref{eq:h_variant_invariant_2nd_order} and
\begin{equation}
\label{eq:phi_1_variant_invariant_2nd_order}
\phi^{(1)} = \upphi + \dot \phi t_{(1)} - \kappa \Phi^{(2)} \big\rvert_{\phi^{(1)} \to \upphi + \dot\phi t_{(1)}} + \bigo{\kappa^2}
\end{equation}
in Eq.~\eqref{eq:invariant_Einstein_eq}, where $\Phi^{(2)}$ was given in Eq.~\eqref{eq:expansion_Phi_2}. Similarly to the ideal-fluid example, the second-order terms in $\kappa$ are produced by making the substitutions $h_{\mu\nu} \to \mathscr{h}_{\mu\nu} + \lie_{X_{(1)}} g_{\mu\nu}$ and $\phi^{(1)} \to \upphi + \dot\phi t_{(1)}$ in the combination $E_{\mu\nu}^{(2)} - X^{(\rho)}_{(1)} \partial_\rho E^{(1)}_{\mu\nu} - \partial_\mu X^{(\rho)}_{(1)} E^{(1)}_{\rho\nu} - \partial_\nu X^{(\rho)}_{(1)} E^{(1)}_{\rho\mu}$ and the substitutions $h_{\mu\nu} \to - \mathscr{g}^{(2)}_{\mu\nu}$ and $\phi^{(1)} \to - \Phi^{(2)}$ in $E^{(1)}_{\mu\nu}$. The resulting expression is quite long and given in Eq.~\eqref{eq:invariant_einstein_eq_2nd_order_field}.

We can now obtain the second-order correction to the invariant constraint equation $\mathscr{E}^{(2)}_\mu = u^\nu \mathscr{E}^{(2)}_{\mu\nu}$, which is given in Eq.~\eqref{eq:constraint_second_order_scalar}. In terms of its temporal and spatial components, the second-order corrections to the invariant constraint equation read
\begin{splitequation}
\mathscr{E}^{(2)}_t &= - \frac{1}{4} \partial_t\mathscr{h}_{ij } \partial_t\mathscr{h}^{ij } + \frac{1}{4} \left(\partial_t\mathscr{h}\right)^2 - (n - 3)H \mathscr{h}^{ij } \partial_t\mathscr{h}_{ij } + (2n - 5)H^2 \mathscr{h}_{ij } \mathscr{h}^{ij } \\
&\quad-  \frac{1}{4} \partial_{i }\mathscr{h} \partial^{i }\mathscr{h}  + \partial^{i }\mathscr{h} \partial_{j }\mathscr{h}_{i }{}^{j } -  \partial_{i }\mathscr{h}^{ij } \partial_{k }\mathscr{h}_{j }{}^{k } -  \frac{1}{2} \partial_{j }\mathscr{h}_{i k } \partial^{k }\mathscr{h}^{ij } + \frac{3}{4} \partial_{k }\mathscr{h}_{ij } \partial^{k }\mathscr{h}^{ij } \\
&\quad+ \mathscr{h}^{ij } \partial_i\partial_j \mathscr{h} - 2 \mathscr{h}^{ij } \partial_j \partial_k\mathscr{h}_{i }{}^{k } + \mathscr{h}^{ij } \partial^{k }\partial_{k }\mathscr{h}_{ij } - \frac{1}{2}\kappa^2 \left[ \left(\partial_t\upphi\right)^2 + \partial_{i }\upphi \partial^{i }\upphi \right] \\
&\quad- \frac{1}{2}\kappa^2 \left( (n - 1 - \epsilon + \delta) (2\epsilon-\delta) + 2\epsilon\delta - \frac{\dot{\delta}}{H}\right) H^2 \upphi^2
\end{splitequation}
and
\begin{splitequation}
\mathscr{E}^{(2)}_i &= - \partial_t\mathscr{h}_{i }{}^{j } \partial_{k }\mathscr{h}_{j }{}^{k } + \frac{1}{2} \partial_t\mathscr{h}_{i j } \partial^{j }\mathscr{h} - H\mathscr{h}_{i j } \partial^{j }\mathscr{h} + 2H \mathscr{h}_{i }{}^{j } \partial_{k }\mathscr{h}_{j }{}^{k } + 2H \mathscr{h}^{j k } \partial_{k }\mathscr{h}_{i j }\\
&\quad - \mathscr{h}^{j k } \partial_{k }\partial_t\mathscr{h}_{i j } + \frac{1}{2} \partial_t\mathscr{h}^{j k } \partial_{i }\mathscr{h}_{j k } - 3 H\mathscr{h}^{j k } \partial_{i }\mathscr{h}_{j k } +  \mathscr{h}^{j k } \partial_{i }\partial_t\mathscr{h}_{j k } - \kappa^2 \partial_t\upphi \partial_{i }\upphi \eqend{.}
\end{splitequation}
Finally, the second-order correction to the equation of motion reads
\begin{splitequation}
\mathscr{E}^{(2)}_{ij}
& = - \partial_t\mathscr{h}_{i }{}^{k } \partial_t\mathscr{h}_{j k } + \frac{1}{2} \partial_t\mathscr{h}_{i j } \partial_t\mathscr{h} -  \mathscr{h}_{i j } \partial_t^2\mathscr{h} + 2H \mathscr{h}_{(i }{}^{k } \partial_t\mathscr{h}_{j) k } -  n H\mathscr{h}_{i j } \partial_t\mathscr{h} \\
&\quad- 4H^2 \mathscr{h}_{i }{}^{k } \mathscr{h}_{j k }  -  \frac{1}{2} \partial_{k }\mathscr{h}_{i j } \partial^{k }\mathscr{h} + \partial^{k }\mathscr{h}_{i j } \partial_{l }\mathscr{h}_{k }{}^{l } -  \partial_{k }\mathscr{h}_{j l } \partial^{l }\mathscr{h}_{i }{}^{k } + \partial_{l }\mathscr{h}_{j k } \partial^{l }\mathscr{h}_{i }{}^{k } \\
&\quad+ \partial^{k }\mathscr{h} \partial_{(i }\mathscr{h}_{j) k } -  \partial_{l }\mathscr{h}_{k }{}^{l } \partial_{(i }\mathscr{h}_{j) }{}^{k } + \frac{1}{2} \partial_{i }\mathscr{h}^{k l } \partial_{j }\mathscr{h}_{k l } -  2\mathscr{h}^{k l } \partial_l\partial_{(i}\mathscr{h}_{j) k } + \mathscr{h}^{k l } \partial_i \partial_j\mathscr{h}_{k l } \\
&\quad-  \mathscr{h}_{i j } \partial_k \partial_l \mathscr{h}^{k l } + \mathscr{h}^{k l } \partial_k \partial_l \mathscr{h}_{i j } + \mathscr{h}_{i j } \partial^k\partial_k\mathscr{h} + g_{i j } \bigg[\frac{3}{4} \partial_t\mathscr{h}_{k l } \partial_t\mathscr{h}^{k l } + \mathscr{h}^{k l } \partial_t^2\mathscr{h}_{k l } \\
&\quad -  \frac{1}{4} \left(\partial_t\mathscr{h}\right)^2 + (n - 8) H\mathscr{h}^{k l } \partial_t\mathscr{h}_{k l } - (2n - 9 - 2\epsilon)H^2 \mathscr{h}_{k l } \mathscr{h}^{k l } + \frac{1}{4} \partial_{k }\mathscr{h} \partial^{k }\mathscr{h} \\
&\quad -  \partial^{k }\mathscr{h} \partial_{l }\mathscr{h}_{k }{}^{l } + \partial_{k }\mathscr{h}^{k l } \partial_{m }\mathscr{h}_{l }{}^{m } + \frac{1}{2} \partial_{l }\mathscr{h}_{k m } \partial^{m }\mathscr{h}^{k l } -  \frac{3}{4} \partial_{m }\mathscr{h}_{k l } \partial^{m }\mathscr{h}^{k l } + 2 \mathscr{h}^{k l } \partial_l \partial_m \mathscr{h}_{k }{}^{m } \\
&\quad -  \mathscr{h}^{k l } \partial^m \partial_m\mathscr{h}_{k l } -  \mathscr{h}^{k l } \partial_k \partial_l \mathscr{h} \bigg] - \kappa^2 \partial_{i }\upphi \partial_{j }\upphi + \kappa^2 \dot{\phi}\mathscr{h}_{i j }\Big[\partial_t\upphi + (n - 1 - \epsilon + \delta) H \upphi \Big] \\
&\quad- \frac{1}{2} \kappa^2 g_{i j } \bigg[\left(\partial_t\upphi\right)^2 - \partial_{k }\upphi \partial^{k }\upphi - \bigg((n - 1 - \epsilon + \delta) (2\epsilon - \delta) + 2\epsilon\delta - \frac{\dot{\delta}}{H} \bigg)H^2 \upphi^2 \bigg] \eqend{.}
\end{splitequation}

The second-order correction to the invariant field equation is obtained by performing the substitutions given in Eqs.~\eqref{eq:h_variant_invariant_2nd_order} and~\eqref{eq:phi_1_variant_invariant_2nd_order} in Eq.~\eqref{eq:invariant_field_equation}. Similarly to the invariant Einstein equation, the second-order terms in $\kappa$ are produced by the substitution of the zeroth-order terms of Eqs.~\eqref{eq:h_variant_invariant_2nd_order} and~\eqref{eq:phi_1_variant_invariant_2nd_order} in $F^{(2)} - X^{(\sigma)}_{(1)} \partial_\sigma F^{(1)}$, and the substitution of the first-order terms in those equations into $F^{(1)}$. The resulting expression is
\begin{splitequation}
\mathscr{F}_{(2)} &= \frac{1}{2} \partial_t\mathscr{h} \partial_t\upphi -  \frac{1}{2} \partial_{i }\upphi \partial^{i }\mathscr{h} + \partial^{i }\upphi \partial_{j }\mathscr{h}_{i }{}^{j } + \mathscr{h}^{i j } \partial_i\partial_j\upphi -  \frac{1}{2}\dot{\phi} \mathscr{h}^{i j } \partial_t\mathscr{h}_{i j } + \dot{\phi}H\mathscr{h}_{i j } \mathscr{h}^{i j }\\
&- \frac{\kappa^2\dot{\phi}}{2 (n-2)}\left[ (n - 1 - \epsilon + 2 \delta)(2\epsilon - \delta) - 2 \epsilon \delta + (n - 1 + 2 \delta - 6 \epsilon)\frac{\dot{\delta}}{2\epsilon H} + \frac{\ddot{\delta}}{2\epsilon H^2}\right]H\upphi^2\eqend{.}
\end{splitequation}

%%%%%%%%%%%%%%%%%%%%%%%%%%%%%%%%%%%%%%%%%%%%%%%%%%%%%%%%%%%%%%%%%%%%%%%%%%%%%%%%%%%%%%%%%%%%%%%%%%%%%%%%%%%%%%%%%%%%%%%%%%%%%%%%%%%%%%%%%%%%%%%%%%%%%
\section{Gauge-invariant Hubble rate}                           									                                                %
\label{sec:hubble_rate}                                                                                                                             %
%%%%%%%%%%%%%%%%%%%%%%%%%%%%%%%%%%%%%%%%%%%%%%%%%%%%%%%%%%%%%%%%%%%%%%%%%%%%%%%%%%%%%%%%%%%%%%%%%%%%%%%%%%%%%%%%%%%%%%%%%%%%%%%%%%%%%%%%%%%%%%%%%%%%%

Our aim now is to obtain an observable that corresponds to a local measurement of the expansion rate of the universe, the local Hubble rate. There are different ways to define this rate, depending on how its measurement is performed. As a first example, let us consider the local Hubble rate defined by the expansion of the observer's spatial section. This spatial section can be defined by its normal vector, the observer's perturbed four velocity $\tilde{u}^\mu$ defined in Eq.~\eqref{eq:perturbed_four_velocity}. The corresponding (gauge-dependent) local Hubble rate is proportional to its divergence:
\begin{equation}
\label{eq:H_u_tilde}
\tilde{H}_u(x) \equiv \frac{1}{n-1} \tilde{\nabla}_\mu \tilde{u}^\mu(x) \eqend{.}
\end{equation}
So far, to obtain an invariant observable from a gauge-dependent expression, we have been using the map $\mathcal{X}$ from the background to the field-dependent synchronous coordinates, which in the case of Eq.~\eqref{eq:H_u_tilde} gives the following invariant local Hubble rate:
\begin{equation}
\label{eq:H_U_inv_def}
\mathscr{H}_u(X) \equiv \tilde{H}_u[x(X)] \eqend{.}
\end{equation}
The explicitly gauge-invariant expression is then obtained by expanding Eq.~\eqref{eq:H_U_inv_def} as a power series in the gauge-dependent perturbation fields and then eliminating them in favour of their gauge-invariant part.

In this section, however, we will take a different approach. As we are mainly interested in the final expression for the Hubble rate, we will first transform the tensor quantities appearing on the right-hand side of Eq.~\eqref{eq:H_U_inv_def} to the field-dependent synchronous coordinates, expressing them in terms of the invariant perturbed fields and only then expand the result as a power series in the perturbations. This method will result in the same final expression, but is somewhat easier to use in practice. Hence, we use the invariant covariant derivative $\bnabla_\mu$ defined in Eq.~\eqref{eq:inv_derivative} and the invariant four-velocity~\eqref{eq:inv_obs_4_vel} to express Eq.~\eqref{eq:H_U_inv_def} as
\begin{equation}
\label{eq:H_U_inv_div}
\mathscr{H}_u(X) = \frac{1}{n-1} \bnabla_\mu \mathscr{u}^\mu(X) \eqend{.}
\end{equation}
To find the perturbative expansion of Eq.~\eqref{eq:H_U_inv_div}, we first expand the derivative operator. This yields
\begin{splitequation}
\label{eq:div_U_inv_expansion_1}
\bnabla_\mu \mathscr{u}^\mu &= \nabla_\mu \mathscr{u}^\mu + \mathscr{C}_{\rho\mu}^\rho \mathscr{u}^\mu \\
&= \nabla_\mu \mathscr{u}^\mu + \frac{\kappa}{2} \mathscr{u}^\mu \nabla_\mu \mathscr{h} - \frac{\kappa^2}{4} \mathscr{u}^\mu \nabla_\mu \left( \mathscr{h}^{\alpha\beta} \mathscr{h}_{\alpha\beta} \right) + \bigo{\kappa^3} \eqend{,}
\end{splitequation}
where the invariant tensor $\mathscr{C}^\rho_{\mu\nu}$ was given in Eq.~\eqref{eq:C_invariant_tensor}, and we recall that $\mathscr{h}_{\mu\nu}$ is the invariant metric perturbation defined by~\eqref{eq:invariant_metric_perturbation}. Recalling that $\mathscr{u}^\mu = u^\mu$~\eqref{eq:invariant_u}, up to second order in perturbation theory Eq.~\eqref{eq:H_U_inv_div} can be cast in the form
\begin{equation}
\label{eq:inv_H_u_final}
\mathscr{H}_u = H + \frac{\kappa}{2 (n-1)} \partial_t \mathscr{h} - \frac{\kappa^2}{4 (n-1)} \partial_t \left( \mathscr{h}^{ij} \mathscr{h}_{ij} \right) + \bigo{\kappa^3} \eqend{.}
\end{equation}

A second possible definition for the local Hubble rate is the expansion rate measured by the elements of fluid. This is given by
\begin{equation}
\label{H_V_inv_def}
\mathscr{H}_V \equiv \frac{1}{n-1} \bnabla_\mu \mathscr{V}^\mu(X) \eqend{,}
\end{equation}
where $\mathscr{V}_\mu$ was defined in Eq.~\eqref{eq:inv_4_vel_fluid}. The fluid's four-velocity can be expressed as
\begin{equation}
\mathscr{V}^\mu = \mathscr{g}^{\mu\nu} \left( u_\nu + \kappa \mathscr{v}_\nu \right) = \mathscr{u}^\mu + \kappa \mathscr{v}^\mu - \kappa^2 \mathscr{h}^\mu{}_\nu \mathscr{v}^\nu + \bigo{\kappa^3} \eqend{,}
\end{equation}
where the indices in the right-hand side of the expression were raised with the background metric to be consistent with the perturbative expansion performed in Sec.~\ref{sec:perturbed_einstein_eq}. We can then compute the divergence of $\mathscr{V}^\mu$. It reads
\begin{splitequation}
&\bnabla_\mu \mathscr{V}^\mu = \bnabla_\mu \mathscr{u}^\mu + \kappa \nabla_\mu \mathscr{v}^\mu + \frac{1}{2} \kappa^2 \mathscr{v}^\mu \nabla_\mu \mathscr{h} - \kappa^2 \nabla_\mu \left( \mathscr{h}^\mu{}_\nu \mathscr{v}^\nu \right) + \bigo{\kappa^3} \\
&\quad= \bnabla_\mu \mathscr{u}^\mu + \kappa \left[ - \partial_t \mathscr{v}_t + \partial_i \mathscr{v}^i - (n+1) H \mathscr{v}_t \right] + \frac{1}{2} \kappa^2 \left[ \mathscr{v}^i \partial_i \mathscr{h} - 2 \partial_i \left( \mathscr{h}^i{}_j \mathscr{v}^j \right) \right] + \bigo{\kappa^3} \\
&\quad= \bnabla_\mu \mathscr{u}^\mu + \kappa \partial_i \mathscr{v}^i + \frac{1}{2} \kappa^2 \left[ \left( \partial_t + (n+1) H \right) \left( \mathscr{v}^i \mathscr{v}_i \right) + \mathscr{v}^i \partial_i \mathscr{h} - 2 \partial_i \left( \mathscr{h}^i{}_j \mathscr{v}^j \right) \right] + \bigo{\kappa^3} \eqend{,}
\end{splitequation}
where we have used Eq.~\eqref{eq:C_invariant_tensor} in the first equality and the expansion for the temporal component of $\mathscr{v}_\mu$ given in Eqs.~\eqref{eq:time_components_invariant_4_vel} in the third equality. Finally, the invariant Hubble rate defined by the expansion of the fluid elements reads
\begin{splitequation}
\label{eq:inv_H_V_final}
\mathscr{H}_V &= \mathscr{H}_u + \frac{\kappa}{n-1} \partial_i \mathscr{v}^i \\
&\quad+ \frac{\kappa^2}{2 (n-1)} \left[ \left( \partial_t + (n+1) H \right) \left( \mathscr{v}^i \mathscr{v}_i \right) + \mathscr{v}^i \partial_i \mathscr{h} - 2 \partial_i \left( \mathscr{h}^i{}_j \mathscr{v}^j \right) \right] + \bigo{\kappa^3} \eqend{,}
\end{splitequation}
which in general differs from $\mathscr{H}_u$. However, they agree if the fluid is non-expanding as seen by the observer, which at lowest order is the condition that $\partial_i \mathscr{v}^i = 0$. We also note that we have kept the observer's four-velocity arbitrary. Hence, the magnitude of the extra terms in Eq.~\eqref{eq:inv_H_V_final} with respect to Eq.~\eqref{eq:inv_H_u_final} is highly model-dependent; in particular, for an observer co-moving with the fluid the extra terms will vanish.

Finally, we can consider the local Hubble rate defined by the expansion of the hypersurfaces on which the scalar field is constant. The normal vector field defining this foliation in the perturbed spacetime is
\begin{equation}
\label{eq:n}
\tilde{n}_\mu(x) \equiv \frac{\partial_\mu \tilde{\phi}(x)}{\sqrt{ - \tilde{g}^{\alpha\beta}(x) \partial_\alpha \tilde{\phi}(x) \partial_\beta \tilde{\phi}(x) }} \eqend{.}
\end{equation}
We recall that at the background level the observers with four-velocity $u^\mu$ are co-moving with the scalar field, and hence on the background we have $n_\mu = u_\mu$. The local Hubble rate for observers co-moving with the perturbed scalar field is then defined by its expansion
\begin{equation}
\label{eq:H_phi_def}
\tilde{H}_\phi(x) \equiv \frac{1}{n-1} \tilde{\nabla}_\mu \tilde{n}^\mu(x) \eqend{,}
\end{equation}
and the corresponding invariant observable is given by
\begin{equation}
\label{eq:H_phi_inv_def}
\mathscr{H}_\phi(\tilde{X}) \equiv \tilde{H}_\phi[x(X)] \eqend{.}
\end{equation}
As in the examples above, we now transform the normal vector field~\eqref{eq:n} to the field-dependent synchronous coordinates $\tilde{X}^{(\mu)}$ and obtain
\begin{equation}
\label{eq:N}
\mathscr{N}_\mu(X) \equiv \frac{\partial_\mu \Phi(X)}{\sqrt{ - \mathscr{g}^{\alpha\beta}(X) \partial_\alpha \Phi(X) \partial_\beta \Phi(X) }} \eqend{,}
\end{equation}
where we recall that $\Phi$ is the invariant scalar field defined by Eq.~\eqref{eq:Phi}. Using the invariant normal vector $\mathscr{N}^\mu = \mathscr{g}^{\mu\nu} \mathscr{N}_\nu$, we can write Eq.~\eqref{eq:H_phi_inv_def} in the form
\begin{equation}
\label{eq:H_phi_div}
\mathscr{H}_\phi(X) = \frac{1}{n-1} \bnabla_\mu \mathscr{N}^\mu(X) \eqend{.}
\end{equation}

To obtain the perturbative expansion of $\mathscr{H}_\phi$~\eqref{eq:H_phi_div}, we first expand the expression for the invariant normal vector $\mathscr{N}^\mu$:
\begin{splitequation}
\mathscr{N}^\mu &= \mathscr{u}^\mu - \frac{\kappa}{\dot{\phi}} \left( g^{\mu\nu} + u^\mu u^\nu \right) \nabla_\nu \left( \Phi^{(1)} + \kappa \Phi^{(2)} \right) + \frac{\kappa^2}{\dot{\phi}}\mathscr{h}^{\mu\rho} \nabla_\rho \Phi^{(1)} \\
&\quad+ \frac{\kappa^2}{2 \dot\phi^2} \left[ u^\mu \left( g^{\alpha\beta} + 3 u^\alpha u^\beta \right) \nabla_\alpha \Phi^{(1)} \nabla_\beta \Phi^{(1)} + 2 \nabla^\mu \Phi^{(1)} u^\rho \nabla_\rho \Phi^{(1)} \right] + \bigo{\kappa^3} \eqend{,}
\end{splitequation}
where $\Phi^{(1)}$ and $\Phi^{(2)}$ were given in Eq.~\eqref{eq:expansion_Phi}. The invariant divergence of the invariant normal vector $\mathscr{N}^\mu$ then reads
\begin{splitequation}
\bnabla_\mu \mathscr{N}^\mu &= \bnabla_\mu \mathscr{u}^\mu - \frac{\kappa}{\dot{\phi}} \nabla_\mu \left[ \left( g^{\mu\nu} + u^\mu u^\nu \right) \nabla_\nu \left(\Phi^{(1)} + \kappa \Phi^{(2)} \right) \right] \\
&\quad- \frac{\kappa^2}{2 \dot{\phi}} \nabla_\mu \mathscr{h} \left( g^{\mu\nu} + u^\mu u^\nu \right) \nabla_\nu \Phi^{(1)} + \kappa^2 \nabla_\mu \left( \frac{1}{\dot\phi} \mathscr{h}^{\mu\rho} \nabla_\rho \Phi^{(1)} \right) \\
&\quad+ \frac{\kappa^2}{2} \nabla_\mu \left[ \frac{1}{\dot\phi^2} \left( u^\mu g^{\alpha\beta} + 3 u^\mu u^\alpha u^\beta + 2 g^{\mu\alpha} u^\beta \right) \nabla_\alpha \Phi^{(1)} \nabla_\beta \Phi^{(1)} \right] + \bigo{\kappa^3} \\
&= \bnabla_\mu \mathscr{u}^\mu - \frac{\kappa}{\dot{\phi}} \laplace \left( \Phi^{(1)} + \kappa \Phi^{(2)} \right) - \frac{\kappa^2}{2\dot{\phi}} \partial^i \mathscr{h} \partial_i \Phi^{(1)} + \frac{\kappa^2}{\dot{\phi}} \partial_i \left(\mathscr{h}^{ij} \partial_j \Phi^{(1)} \right) \\
&\quad + \frac{\kappa^2}{2} \nabla_\mu \left[ \frac{1}{\dot\phi^2} \left(u^\mu\bar{g}^{\nu\rho}\partial_\rho\Phi^{(1)} + 2\bar{g}^{\mu\rho}\partial_\rho\Phi^{(1)} u^\nu \right) \partial_\nu \Phi^{(1)} \right] + \bigo{\kappa^3} \eqend{.}
\end{splitequation}
Finally we insert this expression back into the expression for the invariant Hubble rate~\eqref{eq:H_phi_div}, and use the background equation~\eqref{eq:phi_eq_background} for $\phi$ to simplify it. Expressing everything in terms of the invariant scalar field perturbation $\upphi$ defined in Eq.~\eqref{eq:Phi}, we then obtain
\begin{splitequation}
\label{eq:inv_H_phi_final}
\mathscr{H}_\phi &= \mathscr{H}_u - \frac{\kappa}{(n-1) a^2 \dot{\phi}} \laplace \upphi + \frac{\kappa^2}{(n-1) \dot{\phi}^2} \bigg[ \dot{\phi} \partial_i \left( \mathscr{h}^{ij} \partial_j \upphi \right) - \dot{\phi} \partial^i \mathscr{h} \partial_i \upphi \\
&\quad+ \left[ \partial_t + \left( \frac{n-1}{2} + \epsilon - \delta \right) H \right] \left( \partial^i \upphi \partial_i \upphi \right) + \frac{1}{a^2} \laplace \upphi \partial_t \upphi \bigg] + \bigo{\kappa^3} \eqend{.}
\end{splitequation}
We see that the invariant Hubble parameters $\mathscr{H}_u$ and $\mathscr{H}_\phi$ also differ from each other in general. However, the difference vanishes whenever the spatial derivative of the invariant scalar perturbation vanishes, \ie, whenever the observer sees spatially homogeneous constant scalar field hypersurfaces.

%%%%%%%%%%%%%%%%%%%%%%%%%%%%%%%%%%%%%%%%%%%%%%%%%%%%%%%%%%%%%%%%%%%%%%%%%%%%%%%%%%%%%%%%%%%%%%%%%%%%%%%%%%%%%%%%%%%%%%%%%%%%%%%%%%%%%%%%%%%%%%%%%%%%%
\section{Linearised quantum gravity}                                                                                                                %
\label{sec:linearised_qg}                                                                                                                           %
%%%%%%%%%%%%%%%%%%%%%%%%%%%%%%%%%%%%%%%%%%%%%%%%%%%%%%%%%%%%%%%%%%%%%%%%%%%%%%%%%%%%%%%%%%%%%%%%%%%%%%%%%%%%%%%%%%%%%%%%%%%%%%%%%%%%%%%%%%%%%%%%%%%%%

We now consider the linearised theory in the absence of matter but including a cosmological constant, \ie, we expand the Einstein--Hilbert action $S_\text{EH} = \kappa^{-2} \int \left( \tilde{R} - 2 \Lambda \right) \sqrt{-\tilde{g}} \total^n x$ around a fixed background to order $\kappa^0$. Assuming that the background fulfills the background Einstein equation $G_{\mu\nu} + \Lambda g_{\mu\nu} = 0$, this results in
\begin{splitequation}
\label{eq:eh_action_secondorder}
S_\text{EH} &= \frac{1}{4} \int \bigg[ h^{\mu\nu} \left( \nabla^2 h_{\mu\nu} - 2 \nabla^\rho \nabla_\mu h_{\nu\rho} + 2 \nabla_\mu \nabla_\nu h \right) - h \nabla^2 h \\
&\qquad+ \frac{2}{n-2} \left( 2 h^{\mu\nu} h_{\mu\nu} - h^2 \right) \Lambda \bigg] \sqrt{-g} \total^n x + \text{const} + \bigo{\kappa} \eqend{,}
\end{splitequation}
where the constant term does not depend on the metric perturbation $h_{\mu\nu}$. We then add a suitable gauge-fixing term and quantise the resulting theory, which gives us the two-point function of the quantised $\hat{h}_{\mu\nu}$ in a suitable (vacuum) state $\ket{0}$. This is the primary ingredient in perturbative quantum gravity, the effective quantum field theory of General Relativity~\cite{burgess_lrr_2004}. Starting from this two-point function of the gauge-fixed metric perturbation
\begin{equation}
\label{eq:hmunu_2_point_function_def}
G_{\mu\nu\rho\sigma}(x,x') \equiv - \mathi \bra{0} \hat{h}_{\mu\nu}(x) \hat{h}_{\rho\sigma}(x') \ket{0} \eqend{,}
\end{equation}
our aim is then to construct the two-point function of the quantised gauge-invariant metric perturbation $\hat{\mathscr{h}}_{\mu\nu}$:
\begin{equation}
\label{eq:inv_2_point_function_def}
\mathscr{G}_{\mu\nu\rho\sigma}(x,x') \equiv - \mathi \bra{0} \hat{\mathscr{h}}_{\mu\nu}(x) \hat{\mathscr{h}}_{\rho\sigma}(x') \ket{0} \eqend{.}
\end{equation}
Here, the quantised gauge-invariant metric perturbation $\hat{\mathscr{h}}_{\mu\nu}$ is simply obtained by taking the classical expression $\mathscr{h}_{\mu\nu}$~\eqref{eq:invariant_metric_perturbation_expansion} and replacing all metric perturbations $h_{\mu\nu}$ by their quantum counterparts $\hat{h}_{\mu\nu}$, including in the coordinate corrections $X^{(\mu)}_{(k)}$. In terms of the two-point function~\eqref{eq:hmunu_2_point_function_def}, we thus obtain
\begin{splitequation}
\label{eq:inv_2_point_function}
\mathscr{G}_{\mu\nu\rho\sigma}(x,x') &= G_{\mu\nu\rho\sigma}(x,x') + 2 \mathi g_{\gamma(\rho} \nabla^{x'}_{\sigma)} \expect{ h_{\mu\nu}(x) X^{(\gamma)}_{(1)}(x') } + 2 \mathi g_{\gamma(\mu} \nabla^x_{\nu)} \expect{ X^{(\gamma)}_{(1)}(x) h_{\rho\sigma}(x') } \\
&\quad- 4 \mathi g_{\gamma(\mu} \nabla^x_{\nu)} g_{\delta(\rho} \nabla^{x'}_{\sigma)} \expect{ X^{(\gamma)}_{(1)}(x) X^{(\delta)}_{(1)}(x') } \eqend{,} \raisetag{1.6em}
\end{splitequation}
where we have dropped all terms of order $\kappa$ or higher since we only consider the linearised theory. Moreover, in Eq.~\eqref{eq:inv_2_point_function} we have re-expressed the derivative terms appearing in Eq.~\eqref{eq:invariant_metric_corrections_1} using the covariant derivative of the background metric. We also recall that the first-order correction to the field-dependent coordinates $X^{(\mu)}_{(1)}$ is given by Eqs.~\eqref{eq:coord_correction_first_order_integral}, and that they are linear functionals of the metric perturbation. For simplicity, we restrict to four dimensions in the remainder of this section.

\subsection{Minkowski spacetime}
%%%%%%%%%%%%%%%%%%%%%%%%%%%%%%%%%%%%%%%%%%%%%%%%%%%%%%%%%%%%%%%%%%%%%%%%%%%%%%%%%%%%%%%%%%%%%%%%%%%%%%%%%%%%%%%%%%%%%%%%%%%%%%%%%%%%%%%%%%%%%%%%%%%%%

As a first concrete example, we consider a Minkowski background spacetime with metric $g_{\mu\nu} = \eta_{\mu\nu}$ and use the gauge-fixing action
\begin{equation}
\label{eq:general_linear_gauge}
S_\text{GF} \equiv - \frac{1}{2 \alpha} \int G_\mu G^\mu \total^4 x \quad\text{with}\quad G_\mu \equiv \partial^\nu h_{\mu\nu} - \frac{1+\beta}{\beta} \partial_\mu h \eqend{,}
\end{equation}
where $\alpha$ and $\beta$ are real parameters. The two-point function in the Minkowski vacuum state $\ket{0}$ in the general linear gauge determined by Eq.~\eqref{eq:general_linear_gauge} can be computed easily by inverting the differential operator that appears in the full action $S_\text{EH} + S_\text{GF}$ [with the Einstein--Hilbert action given in Eq.~\eqref{eq:eh_action_secondorder}], and reads
\begin{splitequation}
\label{eq:gauge_fixed_2_point_function}
G_{\mu\nu\rho\sigma}(x,x') &= \left( 2 \eta_{\mu(\rho} \eta_{\sigma)\nu} - \eta_{\mu\nu} \eta_{\rho\sigma} \right) G_0(x,x') + 4 (\alpha-1) \frac{\partial_{(\mu} \eta_{\nu)(\rho} \partial_{\sigma)}}{\partial^2} G_0(x,x') \\
&\quad+ (2+\beta) \left( \eta_{\mu\nu} \frac{\partial_\rho \partial_\sigma}{\partial^2} + \eta_{\rho\sigma} \frac{\partial_\mu \partial_\nu}{\partial^2} \right) G_0(x,x') \\
&\quad- (2+\beta) \left[ 2 (2+\beta) + (\alpha-1) (2-\beta) \right] \frac{\partial_\mu \partial_\nu \partial_\rho \partial_\sigma}{\left( \partial^2 \right)^2} G_0(x,x') \eqend{,}
\end{splitequation}
where $\partial^{-2}$ denotes the Green's function of the d'Alembertian and
\begin{equation}
\label{eq:Fourier_rep_2point_function}
G_0(x,x') = - \mathi \int \frac{\mathe^{- \mathi \abs{\vec{p}} (t-t')}}{2 \abs{\vec{p}}} \mathe^{\mathi \vec{p} (\vec{x}-\vec{x}')} \frac{\total^3 p}{(2\pi)^3}
\end{equation}
is the two-point function of a massless scalar field in Fourier space.

We have already shown in Sec.~\ref{subsec:gauge_invariant_metric} that our gauge-invariant metric perturbation satisfies $\mathscr{h}_{t\nu} = 0$, which is expected from the synchronous condition. For the same reason, after quantisation any temporal component of the invariant two-point function~\eqref{eq:inv_2_point_function_def} must vanish. Indeed, by using the equations of motion~\eqref{eq:coord_correction_first_order} for the coordinate corrections $X_{(1)}^{(\mu)}$ in Eq.~\eqref{eq:inv_2_point_function}, which also hold after quantisation, it is easy to check that
\begin{equation}
\mathscr{G}_{tt\rho\sigma}(x,x') = \mathscr{G}_{\mu\nu tt}(x,x') = \mathscr{G}_{ti\rho\sigma}(x,x') = \mathscr{G}_{\mu\nu ti}(x,x') = 0 \eqend{.}
\end{equation}
Hence, here we can focus only on the purely spatial components of Eq.~\eqref{eq:inv_2_point_function_def}.

The explicit expression for the purely spatial components of the invariant two-point function is obtained after inserting Eqs.~\eqref{eq:coord_correction_first_order_integral} into Eq.~\eqref{eq:inv_2_point_function}. This results in
\begin{splitequation}
\label{eq:inv_2_point_function_spatial}
&\mathscr{G}_{ijkl}(t,\vec{x},t',\vec{x}') = G_{ijkl}(t,\vec{x},t',\vec{x}') \\
&+ \int_{-\infty}^{t'} \left[ \partial_k \partial_l \int_{-\infty}^{s'} G_{ijtt}(t,\vec{x},u',\vec{x}') \total u' - 2 \partial'_{(k} G_{|ij|l)t}(t,\vec{x},s',\vec{x}') \right] \total s' \\
&+ \int_{-\infty}^t \left[ \partial_i \partial_j \int_{-\infty}^s G_{ttkl}(u,\vec{x},t',\vec{x}') \total u - 2 \partial_{(i} G_{j)tkl}(s,\vec{x},t',\vec{x}') \right] \total s \\
&+ \partial_i \partial_j \int_{-\infty}^{t'} \int_{-\infty}^t \int_{-\infty}^s \left[ \partial'_{k} \partial'_{l} \int_{-\infty}^{s'} G_{tttt}(u,\vec{x},u',\vec{x}') \total u' - 2 \partial'_{(k} G_{|tt|l)t}(u,\vec{x},s',\vec{x}') \right] \total u \total s \total s' \\
&- 2 \int_{-\infty}^{t'} \int_{-\infty}^t \left[ \partial'_k \partial'_l \int_{-\infty}^{s'} \partial_{(i} G_{j)ttt}(s,\vec{x},u',\vec{x}') \total u' - 2 \partial'_{(k} \partial_{|(i} G_{j)t|l)t}(s,\vec{x},s',\vec{x}') \right] \total s \total s' \eqend{,}
\end{splitequation}
where primed derivatives act on $\vec{x}'$. It remains to insert the two-point function of the gauge-fixed metric perturbation~\eqref{eq:gauge_fixed_2_point_function} and perform the integrals. We first notice that the derivative terms in Eq.~\eqref{eq:gauge_fixed_2_point_function} are symmetrised in the indices $\mu\nu$ and $\rho\sigma$, and thus have the form of a two-point function of operator-valued diffeomorphisms. Since we have already shown that $\mathscr{h}_{\mu\nu}$ (and thus $\hat{\mathscr{h}}_{\mu\nu}$) is gauge invariant, those terms in the gauge-fixed propagator do not contribute to Eq.~\eqref{eq:inv_2_point_function_spatial}. To check this, let us consider explicitly the term $\partial_{(\mu} \eta_{\nu)(\rho} \partial_{\sigma)} \partial^{-2} G_0$ of Eq.~\eqref{eq:gauge_fixed_2_point_function}. Its contribution to Eq.~\eqref{eq:inv_2_point_function_spatial} reduces to
\begin{splitequation}
\label{eq:inv_2_point_function_spatial_gaugecontr}
&\frac{\partial_{(i} \eta_{j)(k} \partial_{l)}}{\partial^2} G_0(t,\vec{x},t',\vec{x}') - \int_{-\infty}^{t'} \frac{\partial_{(i} \eta_{j)(k} \partial_{l)}}{\partial^2} \partial_{s'} G_0(t,\vec{x},s',\vec{x}') \total s' \\
&- \int_{-\infty}^t \partial_s \frac{\partial_{(i} \eta_{j)(k} \partial_{l)}}{\partial^2} G_0(s,\vec{x},t',\vec{x}') \total s + \int_{-\infty}^{t'} \int_{-\infty}^t \partial_s \partial_{s'} \frac{\partial_{(i}\eta_{j)(k}\partial_{l)}}{\partial^2} G_0(s,\vec{x},s',\vec{x}') \total s \total s' \\
&+ \partial_i \partial_j \partial_k \partial_l \int_{-\infty}^{t'} \int_{-\infty}^t \int_{-\infty}^s \left[ \int_{-\infty}^{s'} \frac{\partial_{u'} \partial_u}{\partial^2} G_0(u,\vec{x},u',\vec{x}') \total u' - \frac{\partial_u}{\partial^2} G_0(u,\vec{x},s',\vec{x}') \right] \total u \total s \total s' \\
&- \partial_i \partial_j \partial_k \partial_l \int_{-\infty}^{t'} \int_{-\infty}^t \left[ \int_{-\infty}^{s'} \frac{\partial_{u'}}{\partial^2} G_0(s,\vec{x},u',\vec{x}') \total u' - \frac{1}{\partial^2} G_0(s,\vec{x},s',\vec{x}') \right] \total s \total s' \eqend{,} \raisetag{2.2em}
\end{splitequation}
where we have traded primed for unprimed derivatives and vice versa, using the fact that $G_0$ only depends on the difference of the coordinates. Since the integrand always contains a time derivative, the integrals are trivial, and using that the scalar propagator~\eqref{eq:Fourier_rep_2point_function} vanishes as one of the arguments goes to $-\infty$, we see that all terms cancel and the whole expression~\eqref{eq:inv_2_point_function_spatial_gaugecontr} vanishes. The same happens with the other gauge-dependent terms of the two-point function~\eqref{eq:gauge_fixed_2_point_function}.

We thus only need to care about the derivative-free part of the two-point function of the gauge-fixed metric perturbation~\eqref{eq:gauge_fixed_2_point_function} involving $2 \eta_{\mu(\rho} \eta_{\sigma)\nu} - \eta_{\mu\nu} \eta_{\rho\sigma}$. Its contribution to the invariant two-point function~\eqref{eq:inv_2_point_function_spatial} is
\begin{splitequation}
\mathscr{G}_{ijkl}(t,\vec{x},t',\vec{x}') &= \left( 2 \eta_{i(k} \eta_{l)j} - \eta_{ij} \eta_{kl} \right) G_0(t,\vec{x},t',\vec{x}') \\
&\quad+ \eta_{ij} \partial_k \partial_l \int_{-\infty}^{t'} \int_{-\infty}^{s'} G_0(t,\vec{x},u',\vec{x}') \total u' \total s' \\
&\quad+ \eta_{kl} \partial_i \partial_j \int_{-\infty}^t \int_{-\infty}^s G_0(u,\vec{x},t',\vec{x}') \total u \total s \\
&\quad+ \partial_i \partial_j \partial_k \partial_l \int_{-\infty}^{t'} \int_{-\infty}^t \int_{-\infty}^s \int_{-\infty}^{s'} G_0(u,\vec{x},u',\vec{x}') \total u' \total u \total s \total s' \\
&\quad+ 4 \partial_{(i} \eta_{j)(k} \partial_{l)} \int_{-\infty}^{t'} \int_{-\infty}^t G_0(s,\vec{x},s',\vec{x}') \total s \total s' \eqend{,}
\end{splitequation}
where again we have changed all the primed derivatives into unprimed ones. The integrals are easy to compute in Fourier space, introducing a convergence factor $\mathe^{\epsilon \abs{\vec{p}} s}$ and taking the limit $\epsilon \to 0^+$ after integration.\footnote{This prescription selects the interacting vacuum of the theory~\cite{peskinschroeder}.} Thus, using the explicit Fourier space expression~\eqref{eq:Fourier_rep_2point_function} for the scalar two-point function we obtain
\begin{equation}
\label{eq:inv_2point_function_final_flat}
\mathscr{G}_{ijkl}(x,x') = - \mathi \int K_{ijkl}(\vec{p}) \frac{\mathe^{- \mathi \abs{\vec{p}} (t-t')}}{2 \abs{\vec{p}}} \mathe^{\mathi \vec{p} (\vec{x}-\vec{x}')} \frac{\total^3 p}{(2\pi)^3} \eqend{,}
\end{equation}
where we have defined the tensor
\begin{equation}
\label{eq:K_ijkl}
K_{ijkl}(\vec{p}) \equiv 2 \left[ \delta_{i(k} - \frac{\vec{p}_i \vec{p}_{(k}}{\vec{p}^2} \right] \left[ \delta_{l)j} - \frac{\vec{p}_{l)} \vec{p}_j}{\vec{p}^2} \right] - \left[ \delta_{ij} - \frac{\vec{p}_i \vec{p}_j}{\vec{p}^2} \right] \left[ \delta_{kl} - \frac{\vec{p}_k \vec{p}_l}{\vec{p}^2} \right] \eqend{.}
\end{equation}
The matrices within the square brackets are positive semi-definite (of rank 2). Moreover, they only contain the transverse and traceless tensor modes. Indeed, contracting Eq.~\eqref{eq:K_ijkl} with $\delta^{ik} \delta^{jl}$ to extract the tensor mode and with $\delta^{ij} \delta^{kl}$ to extract the scalar mode we find
\begin{equation}
\delta^{ik} \delta^{jl} K_{ijkl}(\vec{p}) = 4 \quad\text{and}\quad \delta^{ij} \delta^{kl} K_{ijkl}(\vec{p}) = 0 \eqend{,}
\end{equation}
respectively. That is, the invariant two-point function $\mathscr{G}_{\mu\nu\rho\sigma}(x,x')$ contains only the propagating tensor modes. It is also positive definite (and therefore has a spectral representation): we write~\cite{fordparker1977}
\begin{splitequation}
K_{ijkl}(\vec{p}) &= \left[ e^1_i(\vec{p}) e^2_j(\vec{p}) + e^2_i(\vec{p}) e^1_j(\vec{p}) \right] \left[ e^1_k(\vec{p}) e^2_l(\vec{p}) + e^2_k(\vec{p}) e^1_l(\vec{p}) \right] \\
&\quad+ \left[ e^1_i(\vec{p}) e^1_j(\vec{p}) - e^2_i(\vec{p}) e^2_j(\vec{p}) \right] \left[ e^1_k(\vec{p}) e^1_l(\vec{p}) - e^2_k(\vec{p}) e^2_l(\vec{p}) \right] \eqend{,}
\end{splitequation}
where $e^A_i(\vec{p})$ with $A \in \{1,2,3\}$ is a right-handed orthogonal set of real polarisation vectors with $e^3_i(\vec{p}) = \vec{p}_i/\abs{\vec{p}}$, $e^1_i(-\vec{p}) = e^2_i(\vec{p})$ and $e^2_i(-\vec{p}) = e^1_i(\vec{p})$. It follows that
\begin{splitequation}
&\iint f^{*ij}(x) \left[ \mathi \mathscr{G}_{ijkl}(x,x') \right] f^{kl}(x') \total^4 x \total^4 x' \\
&= \int \frac{1}{2 \abs{\vec{p}}} \abs{ \left[ e^1_k(\vec{p}) e^2_l(\vec{p}) + e^2_k(\vec{p}) e^1_l(\vec{p}) \right] \tilde{f}^{kl}(\abs{\vec{p}},\vec{p}) }^2 \frac{\total^3 p}{(2\pi)^3} \\
&\quad+ \int \frac{1}{2 \abs{\vec{p}}} \abs{ \left[ e^1_k(\vec{p}) e^1_l(\vec{p}) - e^2_k(\vec{p}) e^2_l(\vec{p}) \right] \tilde{f}^{kl}(\abs{\vec{p}},\vec{p}) }^2 \frac{\total^3 p}{(2\pi)^3} \geq 0
\end{splitequation}
with the Fourier transform
\begin{equation}
\tilde{f}^{ij}(p) = \int f^{ij}(x) \, \mathe^{- \mathi p x} \frac{\total^4 p}{(2\pi)^4} \eqend{.}
\end{equation}

\subsection{De~Sitter spacetime}
%%%%%%%%%%%%%%%%%%%%%%%%%%%%%%%%%%%%%%%%%%%%%%%%%%%%%%%%%%%%%%%%%%%%%%%%%%%%%%%%%%%%%%%%%%%%%%%%%%%%%%%%%%%%%%%%%%%%%%%%%%%%%%%%%%%%%%%%%%%%%%%%%%%%%

As a second example, let us consider metric perturbations around a de~Sitter background $g_{\mu\nu} = a^2(t) \eta_{\mu\nu}$ with the scale factor $a(t) = \mathe^{Ht}$ in cosmological time. We now follow Ref.~\cite{tsamis_woodard_cmp_1994} and use the gauge-fixing action
\begin{equation}
\label{eq:gauge_Lagrangian_dS}
S_\text{GF} = - \frac{1}{2} \int G_\mu G^\mu \total^4 x \quad\text{with}\quad G_\mu \equiv \partial_\nu h^\nu{}_\mu - \frac{1}{2} \partial_\mu h + 2 H h^t{}_\mu \eqend{,}
\end{equation}
where we have expressed their gauge condition using the cosmological time. The gauge condition~\eqref{eq:gauge_Lagrangian_dS} corresponds to the Feynman-gauge version of the harmonic gauge in four dimensions, see for example Ref.~\cite{froeb_lima_cqg_2018}. With this choice of gauge, the two-point function of the gauge-fixed metric perturbation in the Euclidean or Bunch--Davies vacuum assumes a particularly simple form:
\begin{splitequation}
\label{eq:dS_gauge_fixed_2_point_function_conformal_time}
G_{\mu\nu\rho\sigma}(x,x') &= 2 a^2(t) a^2(t') \left( \bar{\eta}_{\mu(\rho} \bar{\eta}_{\sigma)\nu} - \bar{\eta}_{\mu\nu} \bar{\eta}_{\rho\sigma} \right) G_0(x,x') \\
&\quad+ \left[ \left( a^2(t) \bar{\eta}_{\mu\nu} + \delta_\mu^t \delta_\nu^t \right) \left( a^2(t') \bar{\eta}_{\rho\sigma} + \delta_\rho^t \delta_\sigma^t \right) - 4 a(t) a(t') \delta^t_{(\mu} \bar{\eta}_{\nu)(\rho} \delta_{\sigma)}^t \right] G_1(x,x') \eqend{,}
\end{splitequation}
where the purely spatial part of the Minkowski metric $\eta_{\mu\nu}$ is defined by
\begin{equation}
\bar{\eta}_{\mu\nu} \equiv \eta_{\mu\nu} + \delta_\mu^t \delta_\nu^t \eqend{.}
\end{equation}
The remaining scalar two-point functions $G_s(x,x')$ with $s = 0,1$ are given by
\begin{equations}
G_s(x,x') &= \int \tilde{G}_s(\eta,\eta',\vec{p}) \mathe^{\mathi \vec{p} (\vec{x} - \vec{x}')} \frac{\total^3 p}{(2\pi)^3} \eqend{,} \\
\tilde{G}_0(\eta,\eta',\vec{p}) &\equiv - \frac{\mathi}{a(\eta) a(\eta')} \frac{( \mathi H a(\eta) + \abs{\vec{p}} ) ( - \mathi H a(\eta') + \abs{\vec{p}} )}{2 \abs{\vec{p}}^3} \mathe^{-\mathi \abs{\vec{p}} (\eta - \eta')} \eqend{,} \\
\tilde{G}_1(\eta,\eta',\vec{p}) &\equiv - \frac{\mathi}{a(\eta) a(\eta')} \frac{1}{2 \abs{\vec{p}}} \mathe^{- \mathi \abs{\vec{p}} (\eta - \eta')}
\end{equations}
in Fourier space. In these expressions, $\eta \equiv - H^{-1} \mathe^{H t}$ is the conformal time, and using this time coordinate the scale factor reads $a(\eta) = - 1/(H \eta)$.

As before, all the temporal components of the two-point function of the invariant metric perturbation vanish, and we only need to be concerned with its purely spatial components. These are given by
\begin{splitequation}
\label{eq:inv_2point_function_dS_spatial}
\mathscr{G}_{ijkl}(x,x') &= G_{ijkl}(t,\vec{x},t',\vec{x}') + H a^2(t') \eta_{kl} \int_{-\infty}^{t'} G_{ijtt}(t,\vec{x},s',\vec{x}') \total s' \\
&\quad+ H a^2(t) \eta_{ij} \int_{-\infty}^t G_{ttkl}(s,\vec{x},t',\vec{x}') \total s \\
&\quad+ H^2 a^2(t) a^2(t') \eta_{ij} \eta_{kl} \int_{-\infty}^t \int_{-\infty}^{t'} G_{tttt}(s,\vec{x},s',\vec{x}') \total s' \total s \\
&\quad+ \int_{-\infty}^{t'} \frac{a^2(t')}{a^2(s)} \left[ \partial'_k \partial'_l \int_{-\infty}^s G_{ijtt}(t,\vec{x},u,\vec{x}') \total u - 2 \partial'_{(k} G_{|ij|l)t}(t,\vec{x},s,\vec{x}') \right] \total s \\
&\quad+ \int_{-\infty}^t \frac{a^2(t)}{a^2(s)} \left[ \partial_i \partial_j \int_{-\infty}^s G_{ttkl}(u,\vec{x},t',\vec{x}') \total u - 2 \partial_{(i} G_{j)tkl}(s,\vec{x},t',\vec{x}') \right] \total s \\
&\quad+ H a^2(t) \eta_{ij} \int_{-\infty}^t \int_{-\infty}^{t'} \frac{a^2(t')}{a^2(s')} \bigg[ \partial'_k \partial'_l \int_{-\infty}^{s'} G_{tttt}(s,\vec{x},u,\vec{x}') \total u \\
&\hspace{13em}- 2 \partial'_{(k} G_{|tt|l)t}(s,\vec{x},s',\vec{x}') \bigg] \total s' \total s \\
&\quad+ H a^2(t') \eta_{kl} \int_{-\infty}^{t'} \int_{-\infty}^t \frac{a^2(t)}{a^2(s)} \bigg[ \partial_i \partial_j \int_{-\infty}^s G_{tttt}(u,\vec{x},s',\vec{x}') \total u \\
&\hspace{13em}- 2 \partial_{(i} G_{j)ttt}(s,\vec{x},s',\vec{x}') \bigg] \total s \total s' \\
&\quad- 2 \int_{-\infty}^t \frac{a^2(t)}{a^2(s)} \int_{-\infty}^{t'} \frac{a^2(t')}{a^2(s')} \bigg[ \partial'_k \partial'_l \int_{-\infty}^{s'} \partial_{(i} G_{j)ttt}(s,\vec{x},u,\vec{x}') \total u \\
&\hspace{13em}- 2 \partial'_{(k} \partial_{|(i} G_{j)|t|l)t}(s,\vec{x},s',\vec{x}') \bigg] \total s' \total s \\
&\quad+ \partial_i \partial_j \int_{-\infty}^t \frac{a^2(t)}{a^2(s)} \int_{-\infty}^s \int_{-\infty}^{t'} \frac{a^2(t')}{a^2(s')} \bigg[ \partial_{k'} \partial_{l'} \int_{-\infty}^{s'} G_{tttt}(u,\vec{x},u',\vec{x}') \total u' \\
&\hspace{13em}- 2 \partial'_{(k} G_{|tt|l)t}(u,\vec{x},s',\vec{x}') \bigg] \total s' \total u \total s \eqend{.} \raisetag{2.2em}
\end{splitequation}
We now substitute Eq.~\eqref{eq:dS_gauge_fixed_2_point_function_conformal_time} into Eq.~\eqref{eq:inv_2point_function_dS_spatial} and obtain
\begin{splitequation}
&a^{-2}(t) a^{-2}(t')\mathscr{G}_{ijkl}(x,x') = 2 \left( \eta_{i(k} \eta_{l)j} - \eta_{ij} \eta_{kl} \right) G_0(x,x') + \eta_{ij} \eta_{kl} G_1(x,x') \\
&\quad+ H \eta_{ij} \eta_{kl} \int_{-\infty}^{t'} G_1(t,\vec{x},s,\vec{x}') \total s + H \eta_{ij} \eta_{kl} \int_{-\infty}^t G_1(s,\vec{x},t',\vec{x}') \total s \\
&\quad+ H^2 \eta_{ij} \eta_{kl} \int_{-\infty}^t \int_{-\infty}^{t'} G_1(s,\vec{x},s',\vec{x}') \total s' \total s + \eta_{ij} \partial_k \partial_l \int_{-\infty}^{t'} \frac{1}{a^2(s)} \int_{-\infty}^s G_1(t,\vec{x},u,\vec{x}') \total u \total s \\
&\quad+ \eta_{kl} \partial_i \partial_j \int_{-\infty}^t \frac{1}{a^2(s)} \int_{-\infty}^s G_1(u,\vec{x},t',\vec{x}') \total u \total s \\
&\quad+ H \eta_{ij} \partial_k \partial_l \int_{-\infty}^t \int_{-\infty}^{t'} \frac{1}{a^2(s')} \int_{-\infty}^{s'} G_1(s,\vec{x},u,\vec{x}') \total u \total s' \total s \\
&\quad+ H \eta_{kl} \partial_i \partial_j \int_{-\infty}^{t'} \int_{-\infty}^t \frac{1}{a^2(s)} \int_{-\infty}^s G_1(u,\vec{x},s',\vec{x}') \total u \total s \total s' \\
&\quad+ 4 \partial_{(i} \eta_{j)(k} \partial_{l)} \int_{-\infty}^t \frac{1}{a(s)} \int_{-\infty}^{t'} \frac{1}{a(s')} G_1(s,\vec{x},s',\vec{x}') \total s' \total s \\
&\quad+ \partial_i \partial_j \partial_k \partial_l \int_{-\infty}^t \frac{1}{a^2(s)} \int_{-\infty}^s \int_{-\infty}^{t'} \frac{1}{a^2(s')} \int_{-\infty}^{s'} G_1(u,\vec{x},u',\vec{x}') \total u' \total s' \total u \total s \eqend{.} \raisetag{2.1em}
\end{splitequation}
To perform the time integrals above, it is convenient to change integration variables to the conformal time $\eta$. As in the Minkowski example, we also need to introduce convergence factors $\mathe^{\epsilon \abs{\vec{p}} \eta}$ and take the limit $\epsilon \to 0^+$ after integration. This results in
\begin{splitequation}
\label{eq:inv_2point_function_final_dS}
\mathscr{G}_{ijkl}(x,x') &= - \mathi H^2 a^2(\eta) a^2(\eta') \int K_{ijkl}(\vec{p})\frac{( 1 + \mathi \abs{\vec{p}} \eta ) ( 1 - \mathi \abs{\vec{p}} \eta' )}{2 \abs{\vec{p}}^3}  \mathe^{- \mathi \abs{\vec{p}} (\eta-\eta')} \mathe^{\mathi \vec{p} (\vec{x}-\vec{x}')} \frac{\total^3 p}{(2\pi)^3} \eqend{,}
\end{splitequation}
where we have expressed the functional time-dependence of the two-point function in terms of the conformal time $\eta$, and the tensor $K_{ijkl}(\vec{p})$ was defined in Eq.~\eqref{eq:K_ijkl}. It is thus clear that the two-point function $\mathscr{G}_{\mu\nu\rho\sigma}$ of the invariant metric perturbation also in the de~Sitter case only contains the propagating transverse and traceless tensor modes. Moreover, as in flat space it is positive definite (apart from the usual infrared issues of massless fields in de~Sitter space) and thus has a spectral representation. The flat-space result~\eqref{eq:inv_2point_function_final_flat} can in fact be easily obtained from Eq.~\eqref{eq:inv_2point_function_final_dS}, expressing the conformal time $\eta$ in terms of cosmological time $t$ and then taking the limit $H \to 0$.

%%%%%%%%%%%%%%%%%%%%%%%%%%%%%%%%%%%%%%%%%%%%%%%%%%%%%%%%%%%%%%%%%%%%%%%%%%%%%%%%%%%%%%%%%%%%%%%%%%%%%%%%%%%%%%%%%%%%%%%%%%%%%%%%%%%%%%%%%%%%%%%%%%%%%
\section{Conclusions}                                                                                                                               %
\label{sec:conclusions}                                                                                                                             %
%%%%%%%%%%%%%%%%%%%%%%%%%%%%%%%%%%%%%%%%%%%%%%%%%%%%%%%%%%%%%%%%%%%%%%%%%%%%%%%%%%%%%%%%%%%%%%%%%%%%%%%%%%%%%%%%%%%%%%%%%%%%%%%%%%%%%%%%%%%%%%%%%%%%%

We have given a construction of field-dependent synchronous coordinates in the relational approach to observables in perturbative (quantum) gravity. At linear order, the invariant metric perturbation, constructed as a relational observable using these coordinates, equals the gauge-fixed metric perturbation in synchronous gauge. Our construction provides thus an extension of this widely used gauge to higher orders, and moreover clarifies its interpretation. Namely, using it corresponds to measurements made in the coordinate system defined by Eqs.~\eqref{eq:perturbed_four_velocity} and~\eqref{eq:synchronous_condition_perturbed}, where the time is the proper time of a given observer and the spatial coordinates are rectangular and orthogonal to the observer's four-velocity, both defined in the full perturbed geometry.

We have then considered the perturbed Einstein equation for a perturbed FLRW spacetime, sourced by either a perfect fluid or a scalar field (the inflaton). Using the relational approach, we have obtained the gauge-invariant part of these equations and made their gauge invariance explicit by expressing them in terms of the invariant metric and matter fields perturbations, up to second order in perturbation theory. The first-order results~\eqref{eq:constraint_eq_fluid_1st_order} and~\eqref{eq:dynamical_eq_fluid_1st_order_simpler} for the fluid as well as~\eqref{eq:constraint_eq_field_1st_order} and~\eqref{eq:dynamical_eq_field_1st_order} for the scalar field are well known~\cite{dodelson_cosmology_book, peebles_cosmology_book}, but the second-order results~\eqref{eq:invariant_einstein_eq_2nd_order_fluid} and~\eqref{eq:invariant_einstein_eq_2nd_order_field} are (to the best of our knowledge) new. An important point in the construction is that at second order one obtains additional quadratic contributions from first-order terms that arise when expressing the gauge-dependent perturbation fields in terms of the gauge-invariant ones. Only when taking those contributions into account does one obtain a gauge-invariant result also at second order, and at higher orders one needs to take into account contributions from all lower orders. As a check on our results, we have also compared the results for a fluid source and the inflaton field as source, and found complete agreement when using the equation of state~\eqref{eq:del_t_upphi} for the inflaton.

Gauge-invariant perturbations of the Einstein equation on cosmological backgrounds were also studied in Ref.~\cite{nakamura_ptp_2007} for the ideal fluid and scalar field models, but using a different method to produce gauge-invariant perturbation fields. In their case, $X^{(\mu)}_{(1)}$ is determined at linear order by imposing that the scalar modes correspond to the Bardeen potentials~\cite{bardeen_prd_1980} and the SVT decomposition for the gauge-invariant parts of $h_{ti}$ and $h_{ij}$. This leads to elliptic equations for $X^{(\mu)}_{(1)}$. The higher-order corrections $X^{(\mu)}_{(n)}$ to the field-dependent coordinates are then obtained recursively by imposing these same conditions to the higher-order corrections $\mathscr{g}_{\mu\nu}^{(n)}$ to the invariant metric and working out the gauge-dependent terms. Although this method produces gauge-invariant perturbation fields at every order, their relation to the gauge-dependent perturbation fields is non-causal, and their physical meaning at higher orders is not very transparent.

An important observable in cosmology, both in inflationary cosmology and today, is the Hubble rate, the local expansion rate of the universe. However, the status of perturbative corrections to it is not yet fully clear. In particular, the issue of back-reaction, \ie, the effect of fluctuations on the average expansion rate has not been completely solved. While there is a vast literature involving computations done in the last decades (see, \eg, Refs.~\cite{mukhanov_abramo_brandenberger_1997,tsamis_woodard_1997,abramo_woodard_prd_1999,nambu_prd_2000,ghosh_madden_veneziano_npb_2000,zibin_brandenberger_scott_prd_2001,abramo_woodard_2002,geshnizjani_brandenberger_2002,nambu_prd_2002,finelli_marozzi_vacca_venturi_prd_2004,geshnizjani_brandenberger_jcap_2005,rasanen_jcap_2004,martineau_brandenberger_prd_2005,geshnizjani_afshordi_jcap_2005,rasanen_cqg_2006,sloth_npb_2007,marozzi_prd_2007,pereznadal_roura_verdaguer_cqg_2008,pereznadal_roura_verdaguer_prd_2008,seery_cqg_2010,finelli_marozzi_vacca_venturi_prl_2011,koivisto_prokopec_prd_2011,buchert_rasanen_arnps_2012,marozzi_vacca_cqg_2012,marozzi_vacca_brandenberger_jcap_2012,levasseur_mcdonough_prd_2015,brandenberger_graef_marozzi_vacca_prd_2018}), it seems to us that no conclusive result has been obtained. We believe that one of the reasons for this is that it is difficult to find a gauge-invariant observable that properly describes the expansion rate that is actually measured in observations. The relational approach not only furnishes a concrete and systematic way to construct such observables, but also gives their interpretation: they correspond to measurements made in the coordinate system that is used to define them. We have thus computed the invariant relational Hubble rate in synchronous coordinates to second order. Apart from the expansion rate that the observer experiences~\eqref{eq:inv_H_u_final}, \ie, the one defined from his four-velocity, we have also computed a) the expansion rate experienced by the fluid elements~\eqref{eq:inv_H_V_final} and b) the one defined by the four-velocity obtained from the gradient of the scalar field~\eqref{eq:inv_H_phi_final}. As expected, these differ in general, but agree if a) the fluid is non-expanding or b) the perturbed inflaton is spatially homogeneous, both as seen in the observer's frame.

With a proper definition of a gauge-invariant Hubble rate, the issue of back-reaction can now be tackled anew. In previous work~\cite{froeb_lima_cqg_2018,froeb_cqg_2019,lima_cqg_2021}, we have already computed one-loop quantum corrections to the invariant Hubble rate in generalised harmonic coordinates. While these coordinates do not have such a clear physical interpretation as the synchronous one that we studied in this work (or the geodesic lightcone coordinates~\cite{gasperini_et_al_jcap_2011,fanizza_etal_jcap_2013,fanizza_etal_jcap_2015,nugier_mg_2015,fleury_nugier_fanizza_jcap_2016,fanizza_etal_jcap_2021,mitsou_etal_cqg_2021,froeb_lima_jcap_2022,fanizza_marozzi_medeiros_jcap_2023}), the computation done there shows that back-reaction exists for a fully gauge-invariant observable and, thus, is not merely a gauge effect. Moreover, the results obtained there agree with the physical intuition: the accelerated expansion of the background spacetime creates particles and in particular gravitons, whose mutual attraction then slows down the expansion. In the future, we would like to compute the back-reaction also for the invariant Hubble rate in synchronous coordinates, and verify that it persists also for measurements done in this coordinate system.

Last but not least, we have considered quantum fluctuations of the metric around Minkowski and de~Sitter spacetimes. We have shown that the invariant metric perturbation, defined as a relational observable using synchronous coordinates, is not only gauge-invariant as required, but also that its correlator only contains the propagating tensor modes. That is, for both flat and de~Sitter backgrounds, the correlator of the invariant metric perturbation using synchronous coordinates captures exactly the physical content of the metric fluctuations. It then follows that this correlator is positive definite and consequently has a spectral representation, as it must be for a physical observable.

While we have restricted in this work to second order, the extension of both the coordinate system and the relational invariant observables to higher orders is straightforward but lengthy, and best left to computer algebra (such as the tensor algebra package \textsc{xAct}~\cite{xact}). As discussed in the introduction, relational observables clearly have applications in all approaches to quantum gravity, and can be used to disentangle gauge effects from physical contributions. In particular, it would be most interesting to evaluate the results of Refs.~\cite{bonanno_et_al_2021,fehre_et_al_2021,braunetal_2022} regarding the renormalisation group flow of the graviton propagator in this light, and check whether the graviton spectral function that is found there corresponds to the physical correlator of the gauge-invariant metric perturbations in synchronous coordinates.

\ack

The authors thank Kevin Falls, Renata Ferrero and Manuel Reichert for discussions on relational observables at the early stages of this work, as well as the anonymous referees for comments, references and suggestions that helped to improve the manuscript. M.B.F.\ acknowledges the support by the Deutsche Forschungsgemeinschaft (DFG, German Research Foundation), project No.~396692871 within the Emmy Noether grant~CA1850/1-1 and project No.~406116891 within the Research Training Group RTG~2522/1.

\appendix

%%%%%%%%%%%%%%%%%%%%%%%%%%%%%%%%%%%%%%%%%%%%%%%%%%%%%%%%%%%%%%%%%%%%%%%%%%%%%%%%%%%%%%%%%%%%%%%%%%%%%%%%%%%%%%%%%%%%%%%%%%%%%%%%%%%%%%%%%%%%%%%%%%%%%
\section{Comparison between the fluid and scalar field models}  									                                                %
\label{sec:comparison_fluid_field}                                                                                                                  %
%%%%%%%%%%%%%%%%%%%%%%%%%%%%%%%%%%%%%%%%%%%%%%%%%%%%%%%%%%%%%%%%%%%%%%%%%%%%%%%%%%%%%%%%%%%%%%%%%%%%%%%%%%%%%%%%%%%%%%%%%%%%%%%%%%%%%%%%%%%%%%%%%%%%%

In this appendix, we compare the equations for the perturbations in the perfect fluid and scalar field models, as to map them into each other.

We start by comparing the fluid's background stress tensor with the one for the scalar field in the full spacetime. They read
\begin{equation}
\tilde{T}_{\mu\nu} = ( \tilde{\rho} + \tilde{p} ) \tilde{V}_\mu \tilde{V}_\nu + \tilde{p} \tilde{g}_{\mu\nu}
\end{equation}
and
\begin{equation}
\tilde{T}_{\mu\nu} = \partial_\mu \tilde{\phi} \partial_\nu \tilde{\phi} - \frac{1}{2} \tilde{g}_{\mu\nu} \left[ \tilde{g}^{\rho\sigma} \partial_\rho \tilde{\phi} \partial_\sigma \tilde{\phi} + V(\tilde{\phi}) \right] \eqend{,}
\end{equation}
respectively. Hence, we have the following identifications
\begin{equations}[eq:fluid_field_identifications]
\tilde{V}_\mu &\longleftrightarrow \tilde{n}_\mu = \frac{\partial_\mu \tilde{\phi}}{\sqrt{- \tilde{g}^{\rho\sigma} \partial_\rho \tilde{\phi} \partial_\sigma \tilde{\phi}}} \eqend{,}\label{eq:fluid_field_identification_V_N} \\
\tilde{\rho} &\longleftrightarrow - \frac{1}{2} \left[ \tilde{g}^{\rho\sigma} \partial_\rho \tilde{\phi} \partial_\sigma \tilde{\phi} - V(\tilde{\phi}) \right] \eqend{,} \\
\tilde{p} &\longleftrightarrow - \frac{1}{2} \left[ \tilde{g}^{\rho\sigma} \partial_\rho \tilde{\phi} \partial_\sigma \tilde{\phi} + V(\tilde{\phi}) \right] \eqend{.}
\end{equations}

An important difference between perfect fluid and scalar field models is that the former have an equation of state that relates the energy density to the pressure. In the case of scalar fields, we can see from the expressions above that energy density and pressure are in general independent since there is no functional relation between $\tilde{\phi}$ and $\partial_t \tilde{\phi}$, \ie, their Cauchy data can be prescribed independently. In order to compare these models, we therefore have to assume that such functional relation exists for the scalar field as well. That is, we shall assume that there is a function $f$ for which
\begin{equation}
- \frac{1}{2} \left[ \tilde{g}^{\rho\sigma} \partial_\rho \tilde{\phi} \partial_\sigma \tilde{\phi} + V(\tilde{\phi}) \right] = f\left( - \frac{1}{2} \left[ \tilde{g}^{\rho\sigma} \partial_\rho \tilde{\phi} \partial_\sigma \tilde{\phi} - V(\tilde{\phi}) \right] \right) \eqend{,}
\end{equation}
and which thus constrains the Cauchy data of $\tilde{\phi}$ and $\partial_t \tilde{\phi}$.

We can now proceed with the expansion of the scalar field normal vector, energy density and pressure. We find (using that $\dot{\phi} < 0$)
\begin{equation}
\tilde{n}_\mu = u_\mu + \kappa n_\mu^{(1)} + \kappa^2 n_\mu^{(2)} + \bigo{\kappa^3}
\end{equation}
with
\begin{equations}
n_\mu^{(1)} &= - \frac{1}{2} \left( h_{tt} + \frac{2 \partial_t \phi^{(1)}}{\dot{\phi}} \right) u_\mu - \frac{1}{\dot{\phi}} \partial_\mu \phi^{(1)} \eqend{,} \\
\begin{split}
n_\mu^{(2)} &= \frac{1}{2} \left[ \frac{1}{\dot{\phi}^2} \partial_\alpha \phi^{(1)} \partial^\alpha \phi^{(1)} + h_t{}^\gamma h_{t\gamma} + \frac{2}{\dot{\phi}} h_{t\beta} \partial^\beta \phi^{(1)} + \frac{3}{4} \left( h_{tt} + \frac{2 \partial_t \phi^{(1)}}{\dot{\phi}} \right)^2 \right] u_\mu \\
&\quad+ \frac{1}{2 \dot{\phi}} \left( h_{tt} + \frac{2 \partial_t \phi^{(1)}}{\dot{\phi}} \right) \partial_\mu \phi^{(1)}
\end{split}
\end{equations}
for the normal vector,
\begin{equation}
\tilde{\rho} = \rho + \kappa \rho_{(1)} + \kappa^2 \rho_{(2)} + \bigo{\kappa^3}
\end{equation}
with
\begin{equations}
\rho &= \frac{1}{2} \dot{\phi}^2 + \frac{1}{2} V(\phi) \eqend{,} \label{eq:rho_phi_background} \\
\rho_{(1)} &= \dot{\phi} \partial_t \phi^{(1)} + \frac{1}{2} \dot{\phi}^2 h_{tt} + \frac{1}{2} V'(\phi) \phi^{(1)} \eqend{,} \\
\rho_{(2)} &= - \frac{1}{2} \partial_\alpha \phi^{(1)} \partial^\alpha \phi^{(1)} - \frac{1}{2} \dot{\phi}^2 h_t{}^\alpha h_{t\alpha} - \dot{\phi} h_{t\alpha} \partial^\alpha \phi^{(1)} + \frac{1}{4} V''(\phi) \phi^{(1)2}
\end{equations}
for the energy density and
\begin{equation}
\tilde{p} = p + \kappa p_{(1)} + \kappa^2 p_{(2)} + \bigo{\kappa^3}
\end{equation}
with
\begin{equations}
p &= \frac{1}{2} \dot{\phi}^2 - \frac{1}{2} V(\phi) \eqend{,} \label{eq:p_phi_background} \\
p_{(1)} &= \dot{\phi} \partial_t \phi^{(1)} + \frac{1}{2} \dot{\phi}^2 h_{tt} - \frac{1}{2} V'(\phi) \phi^{(1)} \eqend{,} \\
p_{(2)} &= - \frac{1}{2} \partial_\mu \phi^{(1)} \partial^\mu \phi^{(1)} - \frac{1}{2} \dot{\phi}^2 h_t{}^\alpha h_{t\alpha} - \dot{\phi} h_{t\alpha} \partial^\alpha \phi^{(1)} - \frac{1}{4} V''(\phi) \phi^{(1)2}
\end{equations}
for the pressure. It is not difficult now to write down the expressions for the gauge-invariant fields. We find 
\begin{equation}
\mathscr{N}_\mu = u_\mu + \kappa \mathscr{N}_\mu^{(1)} + \kappa^2 \mathscr{N}_\mu^{(2)} + \bigo{\kappa^3}
\end{equation}
with
\begin{equations}[eq:expansion_N_V_comparison]
\mathscr{N}_\mu^{(1)} &= - \frac{1}{\dot{\phi}} \bar{\partial}_\mu \upphi \eqend{,} \\
\mathscr{N}_\mu^{(2)} &= \frac{1}{2 \dot{\phi}^2} \bar{\partial}_\alpha \upphi \bar{\partial}^\alpha \upphi u_\mu + \frac{1}{\dot{\phi}^2} \partial_t \upphi \bar{\partial}_\mu \upphi \eqend{,}
\end{equations}
where $\bar{\partial}_\mu$ is the covariant derivative of the induced spatial metric $\bar{g}_{\mu\nu}$~\eqref{eq:induced_spatial_metric}, and $\mathscr{N}_\nu$ was defined in Eq.~\eqref{eq:N}. Eqs.~\eqref{eq:expansion_N_V_comparison} agree with what we have found in Eqs.~\eqref{eq:time_components_invariant_4_vel} for the time component of the fluid four-velocity using the fact that this four-vector is normalised. Indeed, we see from Eqs.~\eqref{eq:expansion_N_V_comparison} that
\begin{equation}
\mathscr{N}_t^{(1)} = 0 \quad\text{and}\quad \mathscr{N}_t^{(2)} = - \frac{1}{2} \mathscr{N}^i_{(1)} \mathscr{N}_i^{(1)}
\end{equation}
also as a consequence of the normalisation of the scalar field's normal vector. Moreover, the scalar field's invariant energy density and pressure are given by
\begin{equation}
\uprho = \rho + \kappa \uprho_{(1)} + \kappa^2 \uprho_{(2)} + \bigo{\kappa^3}
\end{equation}
with
\begin{equations}[eq:inv_energy_density_corrections]
\uprho_{(1)} &= \dot{\phi} \left[ \partial_t \upphi - (n-1-\epsilon+\delta) H \upphi \right] \eqend{,} \\
\uprho_{(2)} &= \frac{1}{2} ( \partial_t \upphi )^2 - \frac{1}{2} \bar{\partial}_\mu \upphi \bar{\partial}^\mu \upphi + \frac{1}{2} \left[ (n-1-\epsilon+\delta) (2\epsilon-\delta) + 2 \epsilon \delta - \frac{\dot\delta}{H} \right] H^2 \upphi^2
\end{equations}
and
\begin{equation}
\mathscr{p} = p + \kappa \mathscr{p}_{(1)} + \kappa^2 \mathscr{p}_{(2)} + \bigo{\kappa^3}
\end{equation}
with
\begin{equations}[eq:inv_pressure_corrections]
\mathscr{p}_{(1)} &= \dot{\phi} \left[ \partial_t \upphi + (n-1-\epsilon+\delta) H \upphi \right] \eqend{,} \\
\mathscr{p}_{(2)} &= \frac{1}{2} ( \partial_t \upphi )^2 - \frac{1}{2} \bar{\partial}_\mu \upphi \bar{\partial}^\mu \upphi - \frac{1}{2} \left[ (n-1-\epsilon+\delta) (2\epsilon-\delta) + 2 \epsilon \delta - \frac{\dot\delta}{H} \right] H^2 \upphi^2 \eqend{,}
\end{equations}
respectively. In the expressions above, we have used the Friedmann equations to write the derivatives of the scalar potential in terms of the background-geometry parameters~\eqref{eq:scalar_potential_derivatives}. For latter use, we also compute the invariant fractional energy density for the scalar field. It reads
\begin{equation}
\label{eq:d_scalar_field}
\mathscr{d} = \mathscr{d}_{(1)} + \kappa \mathscr{d}_{(2)} + \bigo{\kappa^2}
\end{equation}
with
\begin{equations}[eq:d_scalar_field_12]
\mathscr{d}_{(1)} &= \frac{2 \epsilon}{(n-1) \dot{\phi}} \left[ \partial_t \upphi - (n-1-\epsilon+\delta) H \upphi \right] \eqend{,} \label{eq:d_scalar_field_1} \\
\mathscr{d}_{(2)} &= \frac{\epsilon}{(n-1) \dot{\phi}^2} \bigg\{ ( \partial_t \upphi )^2 - \bar{\partial}_\alpha \upphi \bar{\partial}^\alpha \upphi + \frac{1}{2} \left[ (n-1-\epsilon+\delta) (2\epsilon-\delta) + 2 \epsilon \delta - \frac{\dot\delta}{H} \right] H^2 \upphi^2 \bigg\} \eqend{.} \label{eq:d_scalar_field_2}
\end{equations}

We can now find the relation between $\upphi$ and $\partial_t \upphi$ if we have an equation of state for the scalar field given by the function $f$. In that case, we know from the perfect-fluid model that invariant pressure and energy density are related by~\eqref{eq:invariant_d}, \eqref{eq:invariant_p}
\begin{equation}
\label{eq:inv_eos_field}
\mathscr{p} = p + \kappa c_\text{s}^2 \uprho_{(1)} + \kappa^2\left( c_\text{s}^2 \uprho_{(2)} + \frac{1}{2} \frac{\total c_\text{s}^2}{\total \rho} \uprho_{(1)}^2 \right) + \bigo{\kappa^3} \eqend{.}
\end{equation}
Thus, by substituting Eqs.~\eqref{eq:inv_energy_density_corrections} into Eq.~\eqref{eq:inv_eos_field}, we obtain
\begin{splitequation}
\label{eq:del_t_uphi_full}
\partial_t \upphi &= - \frac{1 + c_\text{s}^2}{1 - c_\text{s}^2} (n-1-\epsilon+\delta) H \upphi - \kappa \frac{1}{\dot{\phi}} \bigg\{ \frac{1}{2} \left( 1 - \frac{1}{1 - c_\text{s}^2} \frac{\total c_\text{s}^2}{\total \rho} \dot{\phi}^2 \right) ( \partial_t \upphi )^2 \\
&\qquad- \frac{1}{2} \bar{\partial}_\mu \upphi \bar{\partial}^\mu \upphi - \frac{1 + c_\text{s}^2}{1 - c_\text{s}^2} \bigg[ \frac{1}{2} (n-1-\epsilon+\delta) (2\epsilon-\delta) + \epsilon \delta - \frac{\dot\delta}{2 H} \\
&\qquad+ \frac{1}{1 + c_\text{s}^2} \left( \frac{1}{2} + \frac{1 + c_\text{s}^2}{1 - c_\text{s}^2} \right) \frac{\total c_\text{s}^2}{\total \rho} \dot{\phi}^2 (n-1-\epsilon+\delta)^2 \bigg] H^2 \upphi^2 \bigg\} + \bigo{\kappa^2} \eqend{.}
\end{splitequation}
To obtain an explicit expression for $\partial_t \upphi$, we can use its zeroth order term to eliminate the term $( \partial_t \upphi )^2$ appearing at first order. The resulting equation can then be further simplified if we employ the following relations for the speed of sound $c_\mathrm{s}$ and background scalar field $\phi$:
\begin{equations}[eq:cs2_phi_background]
1 + c_\text{s}^2 &= \frac{2}{n-1} (\epsilon-\delta) \eqend{,} \\
1 - c_\text{s}^2 &= \frac{2}{n-1} (n-1-\epsilon+\delta) \eqend{,} \\
\frac{\total c_\text{s}^2}{\total \rho} \dot{\phi}^2 &= - \frac{4}{(n-1)^2} \left( \epsilon \delta - \frac{\dot{\delta}}{2 H} \right) \eqend{.}
\end{equations}
Substituting these relations in Eq.~\eqref{eq:del_t_uphi_full} yields
\begin{equation}
\label{eq:del_t_upphi}
\partial_t \upphi = - (\epsilon-\delta) H \upphi + \frac{\kappa}{2 \dot{\phi}} \left[ \bar{\partial}_\mu \upphi \bar{\partial}^\mu \upphi + \left( \epsilon - 3 \delta + \frac{\dot{\delta}}{\epsilon H} \right) \epsilon H^2 \upphi^2 \right] + \bigo{\kappa^2} \eqend{,}
\end{equation}
and the invariant fractional energy density~\eqref{eq:d_scalar_field} reads
\begin{equation}
\label{eq:d_scalar_field_del_t_upphi}
\mathscr{d} = - \frac{2 \epsilon}{\dot{\phi}} H \upphi + \frac{\kappa \epsilon}{2 (n-1) \dot{\phi}^2} \left[ (n-1+\epsilon-2\delta) (2\epsilon-\delta) - \delta^2 + \frac{\dot\delta}{H} \right] H^2 \upphi^2 + \bigo{\kappa^2} \eqend{.}
\end{equation}
We now check how these expansions imply the correspondence of the fluid and scalar field energy-momentum tensor.

\subsection{Background}
%%%%%%%%%%%%%%%%%%%%%%%%%%%%%%%%%%%%%%%%%%%%%%%%%%%%%%%%%%%%%%%%%%%%%%%%%%%%%%%%%%%%%%%%%%%%%%%%%%%

As a quick check for the background equations, let us consider the continuity equation for the perfect fluid. The identifications~\eqref{eq:fluid_field_identifications} for the background fields yield
\begin{equation}
0 = \dot{\rho} + (n-1) H (\rho + p) = \left[ \ddot{\phi} + (n-1) H \dot{\phi} + \frac{1}{2} V'(\phi) \right] \dot{\phi} \eqend{,}
\end{equation}
which corresponds to the equation of motion for the background scalar field $\phi$. The Friedmann equations can also be easily checked using these identifications.

\subsection{First order}
%%%%%%%%%%%%%%%%%%%%%%%%%%%%%%%%%%%%%%%%%%%%%%%%%%%%%%%%%%%%%%%%%%%%%%%%%%%%%%%%%%%%%%%%%%%%%%%%%%%

At linear order, it is interesting to establish the comparison between the perfect fluid and scalar field models using their SVT decompositions. The identification between the fluid's four-velocity and the scalar field's normal vector~\eqref{eq:fluid_field_identification_V_N} implies that
\begin{equation}
\label{eq:N_V_identification_SVT}
\mathscr{V}^\mathrm{T}_i = 0 \quad\text{and}\quad \mathscr{W} = - \frac{\upphi}{\dot{\phi}} \eqend{.}
\end{equation}
Furthermore, we can relate the invariant fractional energy density $\mathscr{d}$ to the Sasaki-Mukhanov variable~\eqref{eq:sasaki_mukhanov_variable}. Indeed, from Eq.~\eqref{eq:d_scalar_field} we have that
\begin{equation}
\label{eq:d_phi_comparison}
\mathscr{d} = \frac{\epsilon}{(n-1) H} \left[ \partial_t Q - (n-1) H \left( Q + a^{-2} \mathscr{T} \right) \right] \eqend{,}
\end{equation}
where we have used Eq.~\eqref{eq:sasaki_mukhanov_variable} to eliminate $\upphi$ and the constraint equation~\eqref{eq:T_eq} for $\mathscr{T}$, together with the identification~\eqref{eq:N_V_identification_SVT}. We can now check the equation of motion for $\mathscr{d}$~\eqref{eq:eom_d_SVT}. Indeed, if we substitute Eq.~\eqref{eq:d_phi_comparison} into the left-hand side of Eq.~\eqref{eq:eom_d_SVT}, then the equation of motion for $\mathscr{d}$ is satisfied if the equation of motion for $Q$~\eqref{eq:eom_sasaki_mukhanov_variable} is fulfilled (or, alternatively, if the equation of motion for $\upphi$~\eqref{eq:eom_upphi} holds).

Finally, we can also check the equation of motion for $\mathscr{W}$~\eqref{eq:W_eq}. The substitution of Eqs.~\eqref{eq:N_V_identification_SVT} and~\eqref{eq:d_phi_comparison} into the left-hand side of that equation yields
\begin{equation}
\partial_t \mathscr{W} + \frac{n-1-2\epsilon+\delta}{2\epsilon} \left( 2 \epsilon H \mathscr{W} - \mathscr{d} \right) = - \frac{2}{(n-1) \dot{\phi}} (\epsilon-\delta) \left[ \partial_t \upphi + (\epsilon-\delta) H \upphi \right] \eqend{.}
\end{equation} 
The right-hand side of this equation then vanishes if the (invariant) scalar field and its derivative are related by Eq.~\eqref{eq:del_t_upphi}, \ie, if there is an equation of state that constrains the Cauchy data for the scalar field.

\subsection{Second order}
%%%%%%%%%%%%%%%%%%%%%%%%%%%%%%%%%%%%%%%%%%%%%%%%%%%%%%%%%%%%%%%%%%%%%%%%%%%%%%%%%%%%%%%%%%%%%%%%%%%

At second order, it is more convenient to just compare the respective invariant stress tensors for each model. The invariant stress tensor is defined as
\begin{equation}
\mathscr{T}_{\mu\nu}(X) \equiv \frac{\partial x^\alpha}{\partial X^{(\mu)}}(X) \frac{\partial x^\beta}{\partial X^{(\nu)}}(X) \tilde{T}_{\alpha\beta} \eqend{,}
\end{equation}
where $\tilde{T}_{\mu\nu}$ is the perturbed stress tensor of the matter. In the case of a perfect fluid, the perturbative expansion for $\mathscr{T}_{\mu\nu}$ is given by
\begin{equation}
\mathscr{T}_{\mu\nu} = T_{\mu\nu} + \kappa \mathscr{T}^{(1)}_{\mu\nu} + \kappa^2 \mathscr{T}^{(2)}_{\mu\nu} + \bigo{\kappa^3}
\end{equation}
with
\begin{equations}
\mathscr{T}^{(1)}_{\mu\nu} &\equiv \rho u_\mu u_\nu \mathscr{d} + 2 (\rho + p) u_{(\mu} \bar{\mathscr{v}}_{\nu)} + c_\text{s}^2 \rho \bar{g}_{\mu\nu} \mathscr{d} + p \mathscr{h}_{\mu\nu} \eqend{,} \\
\begin{split}
\mathscr{T}^{(2)}_{\mu\nu} &\equiv \frac{1}{2} (\rho + p) \left[ u_\mu u_\nu \bar{\mathscr{v}}^\alpha \bar{\mathscr{v}}_\alpha + 2 \bar{\mathscr{v}}_\mu \bar{\mathscr{v}}_\nu \right] + 2 \rho (1 + c_\text{s}^2) u_{(\mu} \bar{\mathscr{v}}_{\nu)} \mathscr{d} \\
&\quad+ \frac{1}{2} \rho^2 \frac{\total c_\text{s}^2}{\total \rho} \bar{g}_{\mu\nu} \mathscr{d}^2 + c_\text{s}^2 \rho \mathscr{d} \mathscr{h}_{\mu\nu} \eqend{.}
\end{split}
\end{equations}

To obtain the second-order correction to the invariant stress tensor for the scalar field, we perform the substitutions
\begin{equation}
\bar{\mathscr{v}}_\mu \to \bar{\mathscr{N}}_\mu^{(1)} + \kappa \bar{\mathscr{N}}_\mu^{(2)} \eqend{,} \quad \mathscr{d} \to \mathscr{d}_{(1)} + \kappa \mathscr{d}_{(2)}
\end{equation}
in $\mathscr{T}_{\mu\nu}^{(1)}$ and the substitutions
\begin{equation}
\bar{\mathscr{v}}_\mu \to \bar{\mathscr{N}}_\mu^{(1)} \eqend{,} \quad \mathscr{d} \to \mathscr{d}_{(1)}
\end{equation}
into $\mathscr{T}_{\mu\nu}^{(2)}$, where the expansion of $\bar{\mathscr{N}}_\mu$ is given by~\eqref{eq:expansion_N_V_comparison} and the one of $\mathscr{d}$ by~\eqref{eq:d_scalar_field_12}. After performing these substitutions and using Eqs.~\eqref{eq:Friedmann_eqs_field}, \eqref{eq:scalar_potential_derivatives}, \eqref{eq:rho_phi_background}, \eqref{eq:p_phi_background}, \eqref{eq:cs2_phi_background} and~\eqref{eq:del_t_upphi}, we obtain
\begin{splitequation}
&\left. \mathscr{T}^{(1)}_{\mu\nu} + \kappa \mathscr{T}^{(2)}_{\mu\nu} \right\rvert_{\bigo{\kappa}} = \partial_\mu \upphi \partial_\nu \upphi + \dot{\phi} \mathscr{h}_{\mu\nu} \left[ \partial_t \upphi + (n-1-\epsilon+\delta) H \upphi \right] \\
&\qquad+ \frac{1}{2} g_{\mu\nu} \left\{ ( \partial_t \upphi )^2 - \bar{\partial}^\alpha \upphi \bar{\partial}_\alpha \upphi - \left[ (n-1-\epsilon+\delta) (2\epsilon-\delta) + 2 \epsilon \delta - \frac{\dot{\delta}}{H} \right] H^2 \upphi^2 \right\} \eqend{,}
\end{splitequation}
which is the second-order correction to the invariant stress tensor of the scalar field.

% \mathscr{d} = - \frac{2 \epsilon}{\dot{\phi}} H \upphi + \kappa \frac{\epsilon}{2 (n-1) \dot{\phi}^2} \left[ (n-1+\epsilon-2\delta) (2\epsilon-\delta) - \delta^2 + \frac{\dot\delta}{H} \right] H^2 \upphi^2

%%%%%%%%%%%%%%%%%%%%%%%%%%%%%%%%%%%%%%%%%%%%%%%%%%%%%%%%%%%%%%%%%%%%%%%%%%%%%%%%%%%%%%%%%%%%%%%%%%%%%%%%%%%%%%%%%%%%%%%%%%%%%%%%%%%%%%%%%%%%%%%%%%%%%
\section{Second-order expressions}  									                                                %
\label{sec:second_order}                                                                                                                  %
%%%%%%%%%%%%%%%%%%%%%%%%%%%%%%%%%%%%%%%%%%%%%%%%%%%%%%%%%%%%%%%%%%%%%%%%%%%%%%%%%%%%%%%%%%%%%%%%%%%%%%%%%%%%%%%%%%%%%%%%%%%%%%%%%%%%%%%%%%%%%%%%%%%%%

In this appendix, we list expressions for various quantities at second order which are too long to fit in the main text. These are:
\begin{itemize}
\item The explicit second-order correction to the invariant perturbed metric~\eqref{eq:invariant_metric_corrections}
\begin{splitequation}
\label{eq:invariant_metric_corrections_2b}
\mathscr{g}_{\mu\nu}^{(2)} &= - \left[ X^{(\rho)}_{(2)} - X^{(\sigma)}_{(1)} \partial_\sigma X^{(\rho)}_{(1)} \right] \partial_\rho g_{\mu\nu} + \frac{1}{2} X^{(\rho)}_{(1)} X^{(\sigma)}_{(1)} \partial_\rho \partial_\sigma g_{\mu\nu} - X^{(\rho)}_{(1)} \partial_\rho h_{\mu\nu} \\
&\quad- \partial_\mu X^{(\rho)}_{(1)} \left[ h_{\rho\nu} - X^{(\sigma)}_{(1)} \partial_\sigma g_{\rho\nu} \right] - \partial_\nu X^{(\rho)}_{(1)} \left[ h_{\rho\mu} - X^{(\sigma)}_{(1)} \partial_\sigma g_{\rho\mu} \right] + \partial_\mu X^{(\rho)}_{(1)} \partial_\nu X^{(\sigma)}_{(1)} g_{\rho\sigma} \\
&\quad- \partial_\mu \left[ X^{(\rho)}_{(2)} - X^{(\sigma)}_{(1)} \partial_\sigma X^{(\rho)}_{(1)} \right] g_{\rho\nu} - \partial_\nu \left[ X^{(\rho)}_{(2)} - X^{(\sigma)}_{(1)} \partial_\sigma X^{(\rho)}_{(1)} \right] g_{\rho\mu} \eqend{.} \raisetag{1.6em}
\end{splitequation}
\item The explicit second-order correction to the invariant co-vector observable~\eqref{eq:covector}
\begin{splitequation}
\label{eq:expansion_invariant_covector_2}
\mathscr{W}^{(2)}_\mu &= W^{(2)}_\mu - X_{(1)}^{(\rho)} \partial_\rho W^{(1)}_\mu - \left[ X_{(2)}^{(\sigma)} - X_{(1)}^{(\rho)} \partial_\rho X_{(1)}^{(\sigma)} \right] \partial_\sigma W_\mu + \frac{1}{2} X_{(1)}^{(\rho)} X_{(1)}^{(\sigma)} \partial_\rho \partial_\sigma W_\mu \\
&\quad- \partial_\mu X_{(1)}^{(\sigma)} \left[ W^{(1)}_\sigma - X_{(1)}^{(\rho)} \partial_\rho W_\sigma \right] - \partial_\mu \left[ X_{(2)}^{(\sigma)} - X_{(1)}^{(\rho)} \partial_\rho X_{(1)}^{(\sigma)} \right] W_\sigma \eqend{.} \raisetag{1.6em}
\end{splitequation}
\item The change of the temporal second-order correction~\eqref{eq:coord_correction_second_order_integral_t} under a diffeomorphism
\begin{splitequation}
\label{eq:coord_correction_deltaxi_t2}
\delta_\xi t_{(2)}(t,\vec{x}) &= \int_{-\infty}^t \left( \partial^\mu \delta_\xi t_{(1)} \partial_\mu t_{(1)} + \delta_\xi h_t{}^\mu \partial_\mu t_{(1)} + h_t{}^\mu \partial_\mu \delta_\xi t_{(1)} + \delta_\xi h_t{}^\mu h_{t\mu} \right)(s,\vec{x}) \total s \\
&= \int_{-\infty}^t \left( \partial_t \xi_\mu \partial^\mu t_{(1)} - 2 \Gamma^k_{ti} \xi_k \partial^i t_{(1)} + \partial_t \xi_\mu h_t{}^\mu - 2 \Gamma^k_{ti} \xi_k h_t{}^i \right)(s,\vec{x}) \total s + \bigo{\kappa} \\
&= \int_{-\infty}^t \left( \partial_t \xi_\mu \partial^\mu t_{(1)} - 2 H \xi^i \partial_i t_{(1)} + \partial_t \xi_\mu h_t{}^\mu - 2 H \xi^i h_{ti} \right)(s,\vec{x}) \total s + \bigo{\kappa} \\
&= \int_{-\infty}^t \left( \partial_t \xi^\mu \partial_\mu t_{(1)} + \partial_t \xi^\mu h_{t\mu} \right)(s,\vec{x}) \total s + \bigo{\kappa} \\
&= \int_{-\infty}^t \left[ \partial_t \left( \xi^\mu \partial_\mu t_{(1)} \right) - \xi^\mu \partial_\mu \partial_t t_{(1)} + \partial_t \xi^\mu h_{t\mu} \right](s,\vec{x}) \total s + \bigo{\kappa} \\
&= \xi^\mu \partial_\mu t_{(1)}(t,\vec{x}) + \frac{1}{2} \int_{-\infty}^t \left[ \xi^\mu \partial_\mu h_{tt} + 2 \partial_t \xi^\mu h_{t\mu} \right](s,\vec{x}) \total s + \bigo{\kappa} \eqend{,} \raisetag{1.8em}
\end{splitequation}
where as for change~\eqref{eq:coord_correction_deltaxi_t1} of the first-order correction we have used the assumption that the diffeomorphism is localised.
\item The temporal component of the second-order correction to the invariant metric, Eq.~\eqref{eq:invariant_metric_corrections_2a} or Eq.~\eqref{eq:invariant_metric_corrections_2b}
\begin{splitequation}
\label{eq:synchronous_metric_second_order}
\mathscr{g}_{t\nu}^{(2)} &= - t_{(1)} \partial_t h_{t\nu} - x^i_{(1)} \partial_i h_{t\nu} - \partial_t t_{(1)} h_{t\nu} - \partial_t x^i_{(1)} \left[ h_{i\nu} - t_{(1)} \partial_t g_{i\nu} \right] \\
&\quad- \partial_\nu t_{(1)} h_{tt} - \partial_\nu x^i_{(1)} h_{ti} - \partial_t t_{(1)} \partial_\nu t_{(1)} + \partial_t x^i_{(1)} \partial_\nu x^j_{(1)} g_{ij} \\
&\quad- \partial_t \left[ t_{(2)} - t_{(1)} \partial_t t_{(1)} - x^i_{(1)} \partial_i t_{(1)} \right] g_{t\nu} - \partial_t \left[ x^i_{(2)} - t_{(1)} \partial_t x^i_{(1)} - x^j_{(1)} \partial_j x^i_{(1)} \right] g_{i\nu} \\
&\quad+ \partial_\nu \left[ t_{(2)} - t_{(1)} \partial_t t_{(1)} - x^i_{(1)} \partial_i t_{(1)} \right] \\
&= - \Big[ 2 \partial_t t_{(2)} - \partial^\mu t_{(1)} \partial_\mu t_{(1)} - 2 h_t{}^\mu \partial_\mu t_{(1)} - h_t{}^\mu h_{t\mu} - \left( 2 \partial_t t_{(1)} + h_{tt} \right)^2 \\
&\qquad- \partial_i t_{(1)} \left( \partial_t x^i_{(1)} - \partial^i t_{(1)} - h_t{}^i \right) - \left( t_{(1)} \partial_t + x^i_{(1)} \partial_i \right) \left( 2 \partial_t t_{(1)} + h_{tt} \right) \Big] g_{t\nu} \\
&\quad- \Big[ \partial_t x^i_{(2)} - \partial^i t_{(2)} - \partial^\mu t_{(1)} \partial_\mu x^i_{(1)} + h^{i\mu} \partial_\mu t_{(1)} - h_t{}^\mu \partial_\mu x^i_{(1)} + h_{t\mu} h^{i\mu} \\
&\qquad- \left( 2 \partial_t t_{(1)} + h_{tt} \right) \left( \partial_t x^i_{(1)} - \partial^i t_{(1)} - h_t{}^i \right) + \partial^i t_{(1)} \left( 2 \partial_t t_{(1)} + h_{tt} \right) \\
&\qquad- \left( \partial_t x^j_{(1)} - \partial^j t_{(1)} - h_t{}^j \right) \partial_j x^i_{(1)} - t_{(1)} \partial_t \left( \partial_t x^i_{(1)} - \partial^i t_{(1)} - h_t{}^i \right) \\
&\qquad- x^j_{(1)} \partial_j \left( \partial_t x^i_{(1)} - \partial^i t_{(1)} - h_t{}^i \right) \Big] g_{i\nu} = 0 \eqend{,} \raisetag{1.8em}
\end{splitequation}
where we have used Eqs.~\eqref{eq:coord_correction_first_order} and~\eqref{eq:coord_correction_second_order}.
\item The second-order correction to the perturbed Einstein tensor~\eqref{eq:perturbed_einstein_tensor}
\begin{splitequation}
\label{eq:perturbed_einstein_tensor_2}
G_{\mu\nu}^{(2)} &= - \frac{1}{4} \nabla_{\alpha }h \nabla^{\alpha }h_{\mu \nu } + \frac{1}{2}\nabla^{\alpha }h_{\mu \nu } \nabla_{\beta }h_{\alpha }{}^{\beta } - \frac{1}{2}h_{\mu \nu }\nabla_{\alpha }\nabla_{\beta }h^{\alpha \beta } + \frac{1}{2}h^{\alpha \beta } \nabla_{\alpha }\nabla_{\beta }h_{\mu \nu }\\
&\quad+ \frac{1}{2}h_{\mu \nu } \nabla_{\beta }\nabla^{\beta }h - a^2h^{\alpha \beta } \nabla_{\alpha }\nabla_{(\mu }h_{\nu) \beta } - \frac{1}{2}\nabla_{\alpha }h_{\nu \beta } \nabla^{\beta }h_{\mu }{}^{\alpha } + \frac{1}{2}\nabla_{\beta }h_{\nu \alpha } \nabla^{\beta }h_{\mu }{}^{\alpha }\\
&\quad+ \frac{1}{2}\nabla_{\alpha }h \nabla_{(\mu }h_{\nu) }{}^{\alpha } - \nabla_{\beta }h_{\alpha }{}^{\beta } \nabla_{(\mu }h_{\nu) }{}^{\alpha } + \frac{1}{2}a^2h^{\alpha \beta } \nabla_{\mu }\nabla_{\nu }h_{\alpha \beta } + \frac{1}{4} \nabla_{\mu }h^{\alpha \beta } \nabla_{\nu }h_{\alpha \beta } \\
&\quad+ \frac{1}{2} g_{\mu\nu} \bigg( h^{\alpha \beta } \nabla_{\beta }\nabla_{\gamma }h_{\alpha }{}^{\gamma } - h^{\alpha \beta } \nabla_{\alpha }\nabla_{\beta }h + \frac{1}{4} \nabla_{\beta }h \nabla^{\beta }h + \nabla_{\alpha }h^{\alpha \beta } \nabla_{\gamma }h_{\beta }{}^{\gamma }\\
&\qquad- \nabla^{\alpha }h \nabla_{\beta }h_{\alpha }{}^{\beta } + h^{\alpha \beta } \nabla_{\gamma }\nabla_{\beta }h_{\alpha }{}^{\gamma }  - h^{\alpha \beta } \nabla^2h_{\alpha \beta } + \frac{1}{2} \nabla_{\beta }h_{\alpha \gamma } \nabla^{\gamma }h^{\alpha \beta }\\
&\qquad- \frac{3}{4} \nabla_{\gamma }h_{\alpha \beta } \nabla^{\gamma }h^{\alpha \beta } - (n - 1 - \epsilon)H^2 h_{\alpha \beta } h^{\alpha \beta } - (n-2) \epsilon H^2 u^{\alpha } u^{\beta }h_{\alpha }{}^{\gamma } h_{\beta \gamma } \bigg) \\
&\quad+ \frac{1}{2} H^2 [ (n-1-\epsilon) h + (n-2) \epsilon u^{\alpha } u^{\beta } h_{\alpha \beta } ] h_{\mu \nu } \eqend{.} \raisetag{1.8em}
\end{splitequation}
\item The second-order correction to the stress tensor divergence~\eqref{eq:divergence_stress_tensor}
\begin{splitequation}
\label{eq:eom_matter_2nd_order}
F^{(2)}_\nu &\equiv \nabla^\mu T^{(2)}_{\mu\nu} - h^{\mu\rho} \nabla_\mu T^{(1)}_{\rho\nu} + h^{\mu\rho} h_\rho{}^\sigma T^{(1)}_{\sigma\nu} - \left( \nabla_\mu h^{\mu\rho} - \frac{1}{2} \nabla^\rho h \right) T^{(1)}_{\rho\nu} - \frac{1}{2} \nabla_\nu h^{\mu\rho} T^{(1)}_{\mu\rho} \\
&\quad+ h^{\mu\rho} \left( \nabla_\mu h_\rho{}^\sigma - \frac{1}{2} \nabla^\sigma h_{\mu\rho} \right) T_{\sigma\nu} + \frac{1}{2} h^{\mu\rho} \left( \nabla_\mu h_\nu{}^\sigma + \nabla_\nu h_\mu{}^\sigma - \nabla^\sigma h_{\mu\nu} \right) T_{\rho\sigma} \\
&\quad+ h^{\sigma\lambda} \left( \nabla_\mu h^\mu{}_\lambda - \frac{1}{2} \nabla_\lambda h \right) T_{\sigma\nu} + \frac{1}{2} h^{\sigma\lambda} \left( \nabla^\rho h_{\nu\lambda} + \nabla_\nu h^\rho{}_\lambda - \nabla_\lambda h^\rho{}_\nu \right) T_{\rho\sigma} \eqend{.} \raisetag{1.8em}
\end{splitequation}
\item The second-order terms of the invariant Einstein equation~\eqref{eq:invariant_Einstein_eq} for the ideal fluid model
\begin{splitequation}
\label{eq:invariant_einstein_eq_2nd_order_fluid}
\mathscr{E}^{(2)}_{\mu\nu} &= - \partial_t\mathscr{h}_{\mu }{}^{\alpha } \partial_t\mathscr{h}_{\nu \alpha } + \frac{1}{2} \partial_{t}{}\mathscr{h}_{\mu \nu } \partial_t\mathscr{h} + 4H \mathscr{h}_{(\mu }{}^{\alpha } \partial_t\mathscr{h}_{\nu) \alpha } - 4H^2 \mathscr{h}_{\mu }{}^{\alpha } \mathscr{h}_{\nu \alpha } - \frac{1}{2} \bar{\partial}_{\alpha }\mathscr{h}_{\mu \nu } \bar{\partial}^{\alpha }\mathscr{h} \\
&\quad+ \bar{\partial}^{\alpha }\mathscr{h}_{\mu \nu } \bar{\partial}_{\beta }\mathscr{h}_{\alpha }{}^{\beta } -  \bar{\partial}_{\alpha }\mathscr{h}_{\nu \beta } \bar{\partial}^{\beta }\mathscr{h}_{\mu }{}^{\alpha } + \bar{\partial}_{\beta }\mathscr{h}_{\nu \alpha } \bar{\partial}^{\beta }\mathscr{h}_{\mu }{}^{\alpha } + \bar{\partial}^{\alpha }\mathscr{h} \bar{\partial}_{(\mu }\mathscr{h}_{\nu) \alpha }  + \mathscr{h}^{\alpha \beta } \bar{\partial}_\mu\bar{\partial}_\nu\mathscr{h}_{\alpha \beta } \\
&\quad- 2\bar{\partial}_{\beta }\mathscr{h}_{\alpha }{}^{\beta } \bar{\partial}_{(\mu }\mathscr{h}_{\nu) }{}^{\alpha } + \frac{1}{2} \bar{\partial}_{\mu }\mathscr{h}^{\alpha \beta } \bar{\partial}_{\nu }\mathscr{h}_{\alpha \beta } + \mathscr{h}^{\alpha \beta } \bar{\partial}_\alpha\bar{\partial}_\beta\mathscr{h}_{\mu \nu } -  2\mathscr{h}^{\alpha \beta }\bar{\partial}_\beta \bar{\partial}_{(\mu}\mathscr{h}_{\nu) \alpha }\\
&\quad + 2u_{(\mu } \bigg[\partial_t\mathscr{h}_{\nu) }{}^{\alpha } \bar{\partial}_{\beta }\mathscr{h}_{\alpha }{}^{\beta } - \frac{1}{2} \partial_t\mathscr{h}_{\nu) \alpha } \bar{\partial}^{\alpha }\mathscr{h} + \mathscr{h}^{\alpha \beta } \bar{\partial}_{\beta }\partial_t\mathscr{h}_{\nu) \alpha } -  \frac{1}{2} \partial_t\mathscr{h}^{\alpha \beta } \bar{\partial}_{\nu) }\mathscr{h}_{\alpha \beta } \\
&\quad -  \mathscr{h}^{\alpha \beta } \bar{\partial}_{\nu) }\partial_t\mathscr{h}_{\alpha \beta } + H\big[ \mathscr{h}_{\nu) \alpha }\bar{\partial}^{\alpha }\mathscr{h} - 2\mathscr{h}_{\nu) }{}^{\alpha }\bar{\partial}_{\beta }\mathscr{h}_{\alpha }{}^{\beta } - 2\mathscr{h}^{\alpha \beta } \bar{\partial}_{\beta }\mathscr{h}_{\nu) \alpha } + 3\mathscr{h}^{\alpha \beta } \bar{\partial}_{\nu) }\mathscr{h}_{\alpha \beta } \big] \\
&\quad+ 2 (n-2)(\delta - \epsilon) H^2 \bar{\mathscr{v}}_{\nu) } \mathscr{d}\bigg] + u_{\mu } u_{\nu } \bigg[\frac{1}{4} \left(\partial_t\mathscr{h}\right)^2 - \frac{1}{4} \partial_t\mathscr{h}_{\alpha \beta } \partial_t\mathscr{h}^{\alpha \beta } - \frac{1}{4} \bar{\partial}_{\alpha }\mathscr{h} \bar{\partial}^{\alpha }\mathscr{h} \\
&\quad + \bar{\partial}^{\alpha }\mathscr{h} \bar{\partial}_{\beta }\mathscr{h}_{\alpha }{}^{\beta } -  \bar{\partial}_{\alpha }\mathscr{h}^{\alpha \beta } \bar{\partial}_{\gamma }\mathscr{h}_{\beta }{}^{\gamma } -  \frac{1}{2} \bar{\partial}_{\beta }\mathscr{h}_{\alpha \gamma } \bar{\partial}^{\gamma }\mathscr{h}^{\alpha \beta } + \frac{3}{4} \bar{\partial}_{\gamma }\mathscr{h}_{\alpha \beta } \bar{\partial}^{\gamma }\mathscr{h}^{\alpha \beta }\\
&\quad + \mathscr{h}^{\alpha \beta } \bar{\partial}_\alpha\bar{\partial}_\beta\mathscr{h} - 2 \mathscr{h}^{\alpha \beta } \bar{\partial}_\beta\bar{\partial}_\gamma\mathscr{h}_{\alpha }{}^{\gamma } + \mathscr{h}^{\alpha \beta } \bar{\partial}^\gamma\bar{\partial}_\gamma\mathscr{h}_{\alpha \beta } - (n - 3)H \mathscr{h}^{\alpha \beta } \partial_t\mathscr{h}_{\alpha \beta }\\
&\quad + (2 n - 5)H^2 \mathscr{h}_{\alpha \beta } \mathscr{h}^{\alpha \beta } - 2(n-2)\epsilon H^2 \bar{\mathscr{v}}_{\alpha } \bar{\mathscr{v}}^{\alpha }\bigg] - 2(n-2)\epsilon H^2 \bar{\mathscr{v}}_{\mu } \bar{\mathscr{v}}_{\nu }\\
&\quad - \mathscr{h}_{\mu \nu } \Big[\partial_t^{2}\mathscr{h} + nH \partial_t\mathscr{h} + \bar{\partial}_\alpha\bar{\partial}_\beta\mathscr{h}^{\alpha \beta } - \bar{\partial}^\alpha\bar{\partial}_\alpha\mathscr{h} - (n-2)(n - 1 - 2\epsilon + 2\delta) H^2 \mathscr{d}\Big] \\
&\quad + \bar{g}_{\mu \nu } \bigg[\frac{3}{4} \partial_t\mathscr{h}_{\alpha \beta } \partial_t\mathscr{h}^{\alpha \beta } -  \frac{1}{4} \left(\partial_t\mathscr{h}\right)^2 + \mathscr{h}^{\alpha \beta } \partial_t^2\mathscr{h}_{\alpha \beta } + \frac{1}{4} \bar{\partial}_{\alpha }\mathscr{h} \bar{\partial}^{\alpha }\mathscr{h} -  \bar{\partial}^{\alpha }\mathscr{h} \bar{\partial}_{\beta }\mathscr{h}_{\alpha }{}^{\beta }\\
&\quad + \bar{\partial}_{\alpha }\mathscr{h}^{\alpha \beta } \bar{\partial}_{\gamma }\mathscr{h}_{\beta }{}^{\gamma } + \frac{1}{2} \bar{\partial}_{\beta }\mathscr{h}_{\alpha \gamma } \bar{\partial}^{\gamma }\mathscr{h}^{\alpha \beta } -  \frac{3}{4} \bar{\partial}_{\gamma }\mathscr{h}_{\alpha \beta } \bar{\partial}^{\gamma }\mathscr{h}^{\alpha \beta } -  \mathscr{h}^{\alpha \beta } \bar{\partial}_\alpha\bar{\partial}_\beta\mathscr{h} \\
&\quad+ 2 \mathscr{h}^{\alpha \beta } \bar{\partial}_\beta\bar{\partial}_\gamma\mathscr{h}_{\alpha }{}^{\gamma } - \mathscr{h}^{\alpha \beta } \bar{\partial}^\gamma\bar{\partial}_{\gamma }\mathscr{h}_{\alpha \beta } + (n - 8)H \mathscr{h}^{\alpha \beta } \partial_t\mathscr{h}_{\alpha \beta } \\
&\quad- (2n - 9 - 2\epsilon)H^2 \mathscr{h}_{\alpha \beta } \mathscr{h}^{\alpha \beta } + (n-2)\left(1-\frac{\dot{\delta}}{2\epsilon\delta H}\right) \delta H^2\mathscr{d}^2\bigg] \eqend{,} \raisetag{2em}
\end{splitequation}
where we recall that $\bar{g}_{\mu\nu}$ is the induced spatial metric and $\bar{\mathscr{v}}_\mu$ is the projection of the four-velocity perturbation on the background spatial section, see Eqs.~\eqref{eq:induced_spatial_metric} and~\eqref{eq:spatial_projection}.
\item The second-order terms in the expansion of the perturbed scalar field equation~\eqref{eq:perturbed_field_equation}
\begin{splitequation}
\label{eq:expansion_field_equation_2}
F^{(2)} &= \left( \frac{1}{2} h^{\beta \gamma } \nabla_{\alpha }h_{\beta \gamma } + \frac{1}{2}h_{\alpha\beta } \nabla^{\beta }h - h^{\beta \gamma } \nabla_{\gamma }h_{\alpha \beta } -  h_{\alpha\beta } \nabla_{\gamma }h^{\beta\gamma } \right) \nabla^{\alpha }\phi + h^{\alpha \beta } \nabla_{\alpha }\nabla_{\beta }\phi^{(1)} \\
&\quad- h_{\alpha }{}^{\gamma } h^{\alpha \beta } \nabla_{\gamma }\nabla_{\beta }\phi + \nabla^{\alpha }\phi^{(1)} \left( \nabla_{\beta }h_{\alpha }{}^{\beta } - \frac{1}{2} \nabla_{\alpha }h \right) + \frac{1}{4}V'''(\phi) \phi^{(1)2} \eqend{.} \raisetag{1.8em}
\end{splitequation}
\item The second-order terms of the invariant Einstein equation~\eqref{eq:invariant_Einstein_eq} for the scalar field model
\begin{splitequation}
\label{eq:invariant_einstein_eq_2nd_order_field}
\mathscr{E}^{(2)}_{\mu\nu} &= - \partial_t\mathscr{h}_{\mu }{}^{\alpha } \partial_t\mathscr{h}_{\nu \alpha } + \frac{1}{2} \partial_t\mathscr{h}_{\mu \nu } \partial_t\mathscr{h} -  \mathscr{h}_{\mu \nu } \partial_t^2\mathscr{h} + 2H \mathscr{h}_{(\mu }{}^{\alpha } \partial_t\mathscr{h}_{\nu) \alpha } -  n H\mathscr{h}_{\mu \nu } \partial_t\mathscr{h}\\
&\quad- 4H^2 \mathscr{h}_{\mu }{}^{\alpha } \mathscr{h}_{\nu \alpha } -  \frac{1}{2} \bar{\partial}_{\alpha }\mathscr{h}_{\mu \nu } \bar{\partial}^{\alpha }\mathscr{h} + \bar{\partial}^{\alpha }\mathscr{h}_{\mu \nu } \bar{\partial}_{\beta }\mathscr{h}_{\alpha }{}^{\beta } -  \bar{\partial}_{\alpha }\mathscr{h}_{\nu \beta } \bar{\partial}^{\beta }\mathscr{h}_{\mu }{}^{\alpha } + \bar{\partial}_{\beta }\mathscr{h}_{\nu \alpha } \bar{\partial}^{\beta }\mathscr{h}_{\mu }{}^{\alpha }\\
&\quad+ \bar{\partial}^{\alpha }\mathscr{h} \bar{\partial}_{(\mu }\mathscr{h}_{\nu) \alpha } -  \bar{\partial}_{\beta }\mathscr{h}_{\alpha }{}^{\beta } \bar{\partial}_{(\mu }\mathscr{h}_{\nu) }{}^{\alpha } + \frac{1}{2} \bar{\partial}_{\mu }\mathscr{h}^{\alpha \beta } \bar{\partial}_{\nu }\mathscr{h}_{\alpha \beta } -  2\mathscr{h}^{\alpha \beta } \bar{\partial}_\beta\bar{\partial}_{(\mu}\mathscr{h}_{\nu) \alpha } \\
&\quad+ \mathscr{h}^{\alpha \beta } \bar{\partial}_\mu \bar{\partial}_\nu\mathscr{h}_{\alpha \beta } - \mathscr{h}_{\mu \nu } \bar{\partial}_\alpha \bar{\partial}_\beta \mathscr{h}^{\alpha \beta } + \mathscr{h}^{\alpha \beta } \bar{\partial}_\alpha \bar{\partial}_\beta \mathscr{h}_{\mu \nu } + \mathscr{h}_{\mu \nu } \bar{\partial}^\alpha\bar{\partial}_\alpha\mathscr{h} \\
&\quad+ 2u_{(\mu } \bigg[\partial_t\mathscr{h}_{\nu) }{}^{\alpha } \bar{\partial}_{\beta }\mathscr{h}_{\alpha }{}^{\beta } - \frac{1}{2} \partial_t\mathscr{h}_{\nu) \alpha } \bar{\partial}^{\alpha }\mathscr{h} + \mathscr{h}^{\alpha \beta } \bar{\partial}_{\beta }\partial_t\mathscr{h}_{\nu) \alpha } - \frac{1}{2} \partial_t\mathscr{h}^{\alpha \beta } \bar{\partial}_{\nu) }\mathscr{h}_{\alpha \beta } \\
&\quad- \mathscr{h}^{\alpha \beta } \bar{\partial}_{\nu) }\partial_t\mathscr{h}_{\alpha \beta } + H\left[ \mathscr{h}_{\nu) \alpha } \bar{\partial}^{\alpha }\mathscr{h} - 2\mathscr{h}_{\nu) }{}^{\alpha } \bar{\partial}_{\beta }\mathscr{h}_{\alpha }{}^{\beta } - 2\mathscr{h}^{\alpha \beta } \bar{\partial}_{\beta }\mathscr{h}_{\nu) \alpha } + 3\mathscr{h}^{\alpha \beta } \bar{\partial}_{\nu) }\mathscr{h}_{\alpha \beta }\right] \bigg] \\
&\quad+ \bar{g}_{\mu \nu } \bigg[\frac{3}{4} \partial_t\mathscr{h}_{\alpha \beta } \partial_t\mathscr{h}^{\alpha \beta } -  \frac{1}{4} \left(\partial_t\mathscr{h}\right)^2 + \mathscr{h}^{\alpha \beta } \partial_t^2\mathscr{h}_{\alpha \beta } + (n - 8) H\mathscr{h}^{\alpha \beta } \partial_t\mathscr{h}_{\alpha \beta } \\
&\quad+ \frac{1}{4} \bar{\partial}_{\alpha }\mathscr{h} \bar{\partial}^{\alpha }\mathscr{h} - (2n - 9 - 2\epsilon)H^2 \mathscr{h}_{\alpha \beta } \mathscr{h}^{\alpha \beta } + \bar{\partial}_{\alpha }\mathscr{h}^{\alpha \beta } \bar{\partial}_{\gamma }\mathscr{h}_{\beta }{}^{\gamma } -  \frac{3}{4} \bar{\partial}_{\gamma }\mathscr{h}_{\alpha \beta } \bar{\partial}^{\gamma }\mathscr{h}^{\alpha \beta } \\
&\quad- \bar{\partial}^{\alpha }\mathscr{h} \bar{\partial}_{\beta }\mathscr{h}_{\alpha }{}^{\beta } + \frac{1}{2} \bar{\partial}_{\beta }\mathscr{h}_{\alpha \gamma } \bar{\partial}^{\gamma }\mathscr{h}^{\alpha \beta } - \mathscr{h}^{\alpha \beta } \bar{\partial}_\alpha \bar{\partial}_\beta \mathscr{h} + 2 \mathscr{h}^{\alpha \beta } \bar{\partial}_\beta \bar{\partial}_\gamma \mathscr{h}_{\alpha }{}^{\gamma } - \mathscr{h}^{\alpha \beta } \bar{\partial}^2 \mathscr{h}_{\alpha \beta }\bigg] \\
&\quad+ u_{\mu } u_{\nu } \bigg[ - \frac{1}{4} \partial_t\mathscr{h}_{\alpha \beta } \partial_t\mathscr{h}^{\alpha \beta } + \frac{1}{4} \left(\partial_t\mathscr{h}\right)^2 - (n - 3)H \mathscr{h}^{\alpha \beta } \partial_t\mathscr{h}_{\alpha \beta } - \bar{\partial}_{\alpha }\mathscr{h}^{\alpha \beta } \bar{\partial}_{\gamma }\mathscr{h}_{\beta }{}^{\gamma } \\
&\quad+ (2n - 5)H^2 \mathscr{h}_{\alpha \beta } \mathscr{h}^{\alpha \beta } - \frac{1}{4} \bar{\partial}_{\alpha }\mathscr{h} \bar{\partial}^{\alpha }\mathscr{h} + \bar{\partial}^{\alpha }\mathscr{h} \bar{\partial}_{\beta }\mathscr{h}_{\alpha }{}^{\beta } - \frac{1}{2} \bar{\partial}_{\beta }\mathscr{h}_{\alpha \gamma } \bar{\partial}^{\gamma }\mathscr{h}^{\alpha \beta } \\
&\quad+ \frac{3}{4} \bar{\partial}_{\gamma }\mathscr{h}_{\alpha \beta } \bar{\partial}^{\gamma }\mathscr{h}^{\alpha \beta } + \mathscr{h}^{\alpha \beta } \bar{\partial}_{\alpha}\bar{\partial}_{\beta }\mathscr{h} - 2 \mathscr{h}^{\alpha \beta } \bar{\partial}_{\beta}\bar{\partial}_{\gamma }\mathscr{h}_{\alpha }{}^{\gamma } + \mathscr{h}^{\alpha \beta } \bar{\partial}^{\gamma }\bar{\partial}_{\gamma }\mathscr{h}_{\alpha \beta }\bigg] \\
&\quad- \kappa^2 \bigg\{\bar{\partial}_{\mu }\upphi \bar{\partial}_{\nu }\upphi - 2u_{(\mu}\partial_t\upphi \bar{\partial}_{\nu) }\upphi + \frac{1}{2} g_{\mu\nu} \bigg[ \left(\partial_t\upphi\right)^2 - \bar{\partial}_{\alpha }\upphi \bar{\partial}^{\alpha }\upphi \\
&\qquad+ \bigg( (n - 1 - \epsilon + \delta) (\delta-2\epsilon) - 2\epsilon\delta + \frac{\dot{\delta}}{H} \bigg) H^2 \upphi^2 \bigg] + u_\mu u_\nu \left(\partial_t\upphi\right)^2 \\
&\qquad+ \dot{\phi}\mathscr{h}_{\mu \nu }\Big[\partial_t\upphi + (n - 1 - \epsilon + \delta) H \upphi\Big] \bigg\} \eqend{.} \raisetag{2em}
\end{splitequation}
\item The second-order correction to the invariant constraint equation for the scalar field model
\begin{splitequation}
\label{eq:constraint_second_order_scalar}
\mathscr{E}^{(2)}_\mu &= \partial_t\mathscr{h}_{\mu }{}^{\alpha } \bar{\partial}_{\beta }\mathscr{h}_{\alpha }{}^{\beta } + \frac{1}{2} \partial_t\mathscr{h}_{\mu \alpha } \bar{\partial}^{\alpha }\mathscr{h} - H\mathscr{h}_{\mu \alpha } \bar{\partial}^{\alpha }\mathscr{h} + 2H \mathscr{h}_{\mu }{}^{\alpha } \bar{\partial}_{\beta }\mathscr{h}_{\alpha }{}^{\beta } + 2H \mathscr{h}^{\alpha \beta } \bar{\partial}_{\beta }\mathscr{h}_{\mu \alpha }\\
&\quad - \mathscr{h}^{\alpha \beta } \bar{\partial}_{\beta }\partial_t\mathscr{h}_{\mu \alpha } + \frac{1}{2} \partial_t\mathscr{h}^{\alpha \beta } \bar{\partial}_{\mu }\mathscr{h}_{\alpha \beta } - 3 H\mathscr{h}^{\alpha \beta } \bar{\partial}_{\mu }\mathscr{h}_{\alpha \beta } + \mathscr{h}^{\alpha \beta } \bar{\partial}_{\mu }\partial_t\mathscr{h}_{\alpha \beta }\\
&\quad + u_{\mu }\bigg[\frac{1}{4} \partial_t\mathscr{h}_{\alpha \beta } \partial_t\mathscr{h}^{\alpha \beta } - \frac{1}{4} \left(\partial_t\mathscr{h}\right)^2 + (n - 3)H \mathscr{h}^{\alpha \beta } \partial_t\mathscr{h}_{\alpha \beta } - (2n - 5)H^2 \mathscr{h}_{\alpha \beta } \mathscr{h}^{\alpha \beta }\\
&\quad + \frac{1}{4} \bar{\partial}_{\alpha }\mathscr{h} \bar{\partial}^{\alpha }\mathscr{h} - \bar{\partial}^{\alpha }\mathscr{h} \bar{\partial}_{\beta }\mathscr{h}_{\alpha }{}^{\beta } + \bar{\partial}_{\alpha }\mathscr{h}^{\alpha \beta } \bar{\partial}_{\gamma }\mathscr{h}_{\beta }{}^{\gamma } + \frac{1}{2} \bar{\partial}_{\beta }\mathscr{h}_{\alpha \gamma } \bar{\partial}^{\gamma }\mathscr{h}^{\alpha \beta } - \frac{3}{4} \bar{\partial}_{\gamma }\mathscr{h}_{\alpha \beta } \bar{\partial}^{\gamma }\mathscr{h}^{\alpha \beta }\\
&\quad - \mathscr{h}^{\alpha \beta } \bar{\partial}_\alpha \bar{\partial}_\beta\mathscr{h} + 2 \mathscr{h}^{\alpha \beta } \bar{\partial}_\beta \bar{\partial}_\gamma\mathscr{h}_{\alpha }{}^{\gamma } - \mathscr{h}^{\alpha \beta } \bar{\partial}^\gamma \bar{\partial}_{\gamma }\mathscr{h}_{\alpha \beta }\bigg] - \kappa^2 \partial_t\upphi \bar{\partial}_{\mu }\upphi \\
&\quad+ \frac{\kappa^2}{2} u_\mu \bigg[ \bar{\partial}_{\alpha }\upphi \bar{\partial}^{\alpha }\upphi + \left(\partial_t\upphi\right)^2 - \left( (n - 1 - \epsilon + \delta) (\delta-2\epsilon) - 2\epsilon\delta + \frac{\dot{\delta}}{H} \right) H^2 \upphi^2\bigg] \eqend{.}
\end{splitequation}
\end{itemize}

% the iopart definition
\providecommand\newblock{\ }
\bibliography{references}

\end{document}